\documentclass{aa}  
\usepackage{graphicx}
\usepackage[normalem]{ulem}
\usepackage{txfonts}
\usepackage[hyperindex=true,colorlinks=true,citecolor=blue,linkcolor=blue,breaklinks=true]{hyperref}
\usepackage{amsmath}
\usepackage{lscape}
\usepackage{natbib,twoopt}
\usepackage{color}
\usepackage{multirow}
\usepackage{subfig}

\definecolor{red}{rgb}{1.,0.0,0.}

\definecolor{orange}{rgb}{1.,.65,0.}

\definecolor{vert}{rgb}{.0,.65,0.}

\bibpunct{(}{)}{;}{a}{}{,} 
\newcommandtwoopt{\citeads}[3][][]{\href{http://adsabs.harvard.edu/abs/#3}%
	{\citealp[#1][#2]{#3}}} 
\newcommandtwoopt{\citepads}[3][][]{\href{http://adsabs.harvard.edu/abs/#3}%
	{\citep[#1][#2]{#3}}} 
\newcommandtwoopt{\citetads}[3][][]{\href{http://adsabs.harvard.edu/abs/#3}%
	{\citet[#1][#2]{#3}}}
\newcommandtwoopt{\citeyearads}[3][][]%
{\href{http://adsabs.harvard.edu/abs/#3}{\citeyear[#1][#2]{#3}}}

\usepackage{etoolbox}

\begin{document}

	\title{Extended envelopes around Galactic Cepheids}
	\subtitle{V.~Multi-wavelength and time-dependent analysis of IR excess\thanks{Based on observations collected at the European Southern Observatory under ESO programme 097.D-305(A)}}
	
	\author{A.~Gallenne\inst{1,2,3},
		A.~M\'erand\inst{4},
		P.~Kervella\inst{5},
		G.~Pietrzy\'nski\inst{1},
		W.~Gieren\inst{2},
		V.~Hocd\'e\inst{6},
		L.~Breuval\inst{5},
		N.~Nardetto\inst{6},
		\and E.~Lagadec\inst{6}
	}

	\authorrunning{A. Gallenne et al.}
	
	\institute{
		Nicolaus Copernicus Astronomical Centre, Polish Academy of Sciences,  Bartycka 18, 00-716 Warszawa, Poland
		\and Universidad de Concepci\'on, Departamento de Astronom\'ia, Casilla 160-C, Concepci\'on, Chile
		\and Unidad Mixta Internacional Franco-Chilena de Astronom\'ia (CNRS UMI 3386), Departamento de Astronom\'ia, Universidad de Chile, Camino El Observatorio 1515, Las Condes, Santiago, Chile
		\and European Southern Observatory, Karl-Schwarzschild-Str. 2, 85748 Garching, Germany
		\and LESIA, Observatoire de Paris, Universit\'e PSL, CNRS, Sorbonne Universit\'e, Univ. Paris Diderot, Sorbonne Paris Cit\'e, 5 place Jules Janssen, 92195 Meudon, France
		\and Laboratoire Lagrange, UMR7293, Universit\'e C\^ote d'Azur, CNRS, Observatoire de la C\^ote d'Azur, Nice, France
	}
	
	\offprints{A. Gallenne} \mail{alexandre.gallenne@gmail.com}
	
	   \date{Received January 15, 2021; accepted April 30, 2021}
	
	
	\abstract
	{}
	{We aim to investigate the infrared excess of 45 Milky Way (MW) Cepheids combining different observables in order to constrain the presence of circumstellar envelopes (CSEs).}
	{We used the SpectroPhoto-Interferometry of Pulsating Stars (SPIPS) algorithm, a robust implementation of the parallax-of-pulsation method that combines photometry, angular diameter, stellar effective temperature, and radial velocity measurements in a global modelling of the pulsation of the Cepheid. We obtained new photometric measurements at mid-infrared (mid-IR) with the VISIR instrument at the Very Large Telescope complemented with data gathered from the literature. We then compared the mean magnitude of the Cepheids from 0.5\,$\mu$m to 70\,$\mu$m with stellar atmosphere models to infer the IR excess, which we attribute to the presence of a circumstellar envelope.}
	{We report that at least 29\,\% of the Cepheids of our sample have a detected IR excess ($> 3\sigma$). We estimated a mean excess of $0.08\pm0.04$\,mag at 2.2\,$\mu$m and $0.13\pm0.06$\,mag at 10\,$\mu$m. Other Cepheids possibly also have IR excess, but they were rejected due to their low detection level compared to a single-star model. We do not see any correlation between the IR excess and the pulsation period as previously suspected for MW Cepheids, but a rather constant trend at a given wavelength. We also do not find any correlation between the CO absorption and the presence of a CSE, but rather with the stellar effective temperature, which confirms that the CO features previously reported are mostly photospheric. No bias caused by the presence of the circumstellar material is detected on the average distance estimates from a SPIPS analysis with a fitted colour excess. We also do not find correlation between the presence of IR excess and the evolution stage of the Cepheids.}
	{We report a fraction of 29\,\% of Cepheids with an IR excess likely produced by the circumstellar envelope surrounding the stars. Longer period Cepheids do not exhibit greater excess than short periods as previously suspected from observations and theoretical dusty-wind models. Other mechanisms such as free-free emission, among others, may be at the origin of the formation of the CSEs. We also show that not fitting the colour excess leads to a bias on the distance estimates in our Galaxy.}
	
	
	\maketitle
	%
	
	\section{Introduction}
	
	A significant fraction of Cepheid stars are known to exhibit extended infrared emission, possibly linked to a circumstellar shell. Near- and mid-IR interferometric observations revealed the first circumstellar envelope (CSE) surrounding $\ell$~Car \citep{Kervella_2006_03_0}, then followed by similar detections around other Cepheids \citep{Merand_2006_07_0,Merand_2007_08_0,Gallenne_2013_10_0}. Near- and mid-IR photometry also showed an IR excess around several Cepheids \citep{Gallenne_2010_12_0,Barmby_2011_11_0,Gallenne_2012_02_0,Schmidt_2015_11_0,Gallenne_2017_11_0}, leading to the hypothesis that perhaps all Cepheids are surrounded by a CSE. Theoretical modelling using pulsation-driven mass loss supports the presence of IR excess \citep[see e.g.][]{Neilson_2008_09_0,Neilson_2010_06_0}. 
	
	These envelopes are interesting from several perspectives. First, they might be tracers of past or ongoing stellar mass loss. Mass loss for Cepheids is still poorly understood and is important for our understanding of stellar evolution. It is still unknown whether the presence of these CSEs and IR excess are global or local phenomena. It has been proposed that mass loss can account for the Cepheid mass discrepancy problem \citep[i.e. the disagreement of $\sim 10$\,\% between masses calculated from stellar evolution and pulsation models, see e.g.][]{Caputo_2005_08_0}, but this still needs to be investigated further. It is also suggested that the circumstellar material may be induced by shock waves propagating in the Cepheid atmosphere during the pulsation cycle \citep[see e.g.][]{Neilson_2008_09_0,Neilson_2010_06_0,Engle_2014_10_0,Hocde_2020_09_0}, but there is still no clear evidence. Second, IR excess may affect the IR period-luminosity relations (PLRs) and their calibration. Indeed, the presence of a CSE can affect both the photometric and interferometric measurements because the Cepheids would appear brighter and larger. The effect of the CSEs on the PLRs has not been quantified yet, but the near-IR relations are likely less affected than in the mid-IR as the CSE contribution is lower, going from a few percent in the $K$ band \citep{Merand_2006_07_0,Merand_2007_08_0} to tens of percent in the $N$ band \citep{Gallenne_2012_02_0,Gallenne_2013_10_0}. Interestingly, there might be a correlation between the CSE flux contribution and the period of the Cepheids both at near- and mid-IR wavelengths, with long-period Cepheids exhibiting relatively brighter CSEs than the short-period ones \citep{Merand_2006_07_0,Merand_2007_08_0,Gallenne_2012_02_0,Gallenne_2013_10_0}. Cepheids with long periods have higher masses and larger radii; therefore if we assume that the CSE IR brightness is an indicator of the mass-loss rate, this would mean that heavier stars experience higher mass-loss rates. Although the sample used to study this correlation is still too small to make firm conclusions, this would be consistent with the previously mentioned theoretical predictions that the mass-loss mechanism is linked to a pulsation-driven mass-loss mechanism. This scenario is also supported by the observations, with a stronger velocity field detected at certain pulsation phases \citep{Nardetto_2006_07_0,Nardetto_2008_10_0,Hocde_2020_01_0}.
	
	Physical origin and composition are still poorly studied, probably because of a lack of specific observations. \citet{Gallenne_2013_10_0} probed the close environment of two Cepheids using mid-IR interferometric observations. We used a numerical radiative transfer code to simultaneously fit the spectral energy distribution (SED) and visibility profile with a dust shell model to determine physical parameters of the CSEs. Although the dust composition was restricted to typical composition for circumstellar environment, an optically thin envelope with an internal dust shell radius in the $15-20$\,mas range provided a satisfying fit. Recent work of \citet{Hocde_2020_01_0} analysed the SED of five other Cepheids and proposed that IR excess is caused by free-free emission produced by a thin shell of ionised gas. \citet{Groenewegen_2020_03_0} also investigated the IR excess of 477 Cepheids by gathering photometric data from the literature. He fitted the average SEDs with a dust radiative transfer code and only detected IR excess for a few Cepheids.
	
	In this work we investigated the IR excess of 45 bright Cepheids using the SPIPS algorithm \citep[SpectroPhoto-Interferometry of Pulsating Stars,][]{Merand_2015_12_0}, which is an implementation of the parallax of pulsation (PoP) method using all available observables (photometry, radial velocities, effective temperature, ...). We did not aim to estimate distances which is usually done with the PoP technique, but instead we analysed the phase-dependent SED in a more global way taking into account the pulsation of the star and various observables. In Sect.~\ref{section__observations_and_data_reduction}, we present new mid-IR photometric observations and list the literature data we gathered for our global PoP analysis. The SPIPS algorithm and its use to our Cepheids are detailed in Sect.~\ref{section__spips_analysis}. We then discuss our results and conclusions in Sects.~\ref{section__discussion} and \ref{section__conclusion}.
	
	\section{Observations and data reduction}
	\label{section__observations_and_data_reduction}
	
	In this section we present the Cepheids we used in our SPIPS analysis, mainly selected from the \emph{Spitzer} light curves published by \citet{Monson_2012_11_0}. We report new single-epoch mid-IR photometric observations for 19 of them, and list the additional data taken from the literature which are used in the global fit. 
	
	\begin{figure*}[!h]
		\centering
		\resizebox{\hsize}{!}{\includegraphics{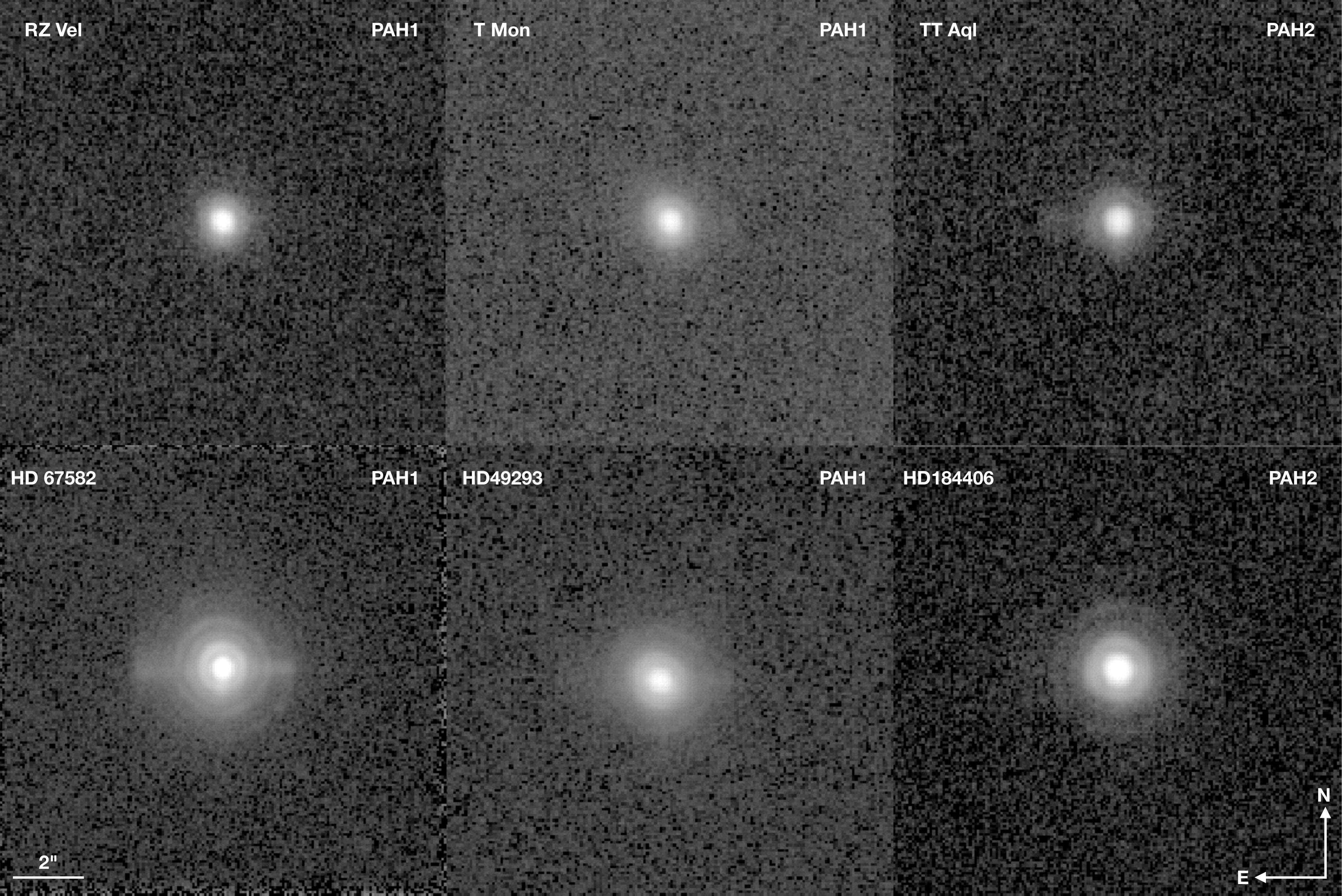}}
		\caption{Sample of the final VISIR images for three Cepheids (top) and their calibrator (bottom). The scale is logarithmic for all images.}
		\label{fig__images}
	\end{figure*}
	
	\subsection{VISIR data}
	
	We obtained new mid-IR observations with the VLT/VISIR instrument \citep{Lagage_2004_09_0} installed at the Cassegrain focus of UT3. It provides imaging and long-slit spectroscopy in the spectral bands M ($3-7\,\mu$m), N ($8-13\,\mu$m), and Q ($16-24\,\mu$m). Our observations were performed with the new AQUARIUS detector providing a field of view of $38\arcsec \times 38\arcsec$ with a plate scale of 45\,mas/pixel. Images were acquired with the PAH1 ($\lambda_\mathrm{c}=8.59\,\mu$m, $\Delta \lambda = 0.42\,\mu$m) and PAH2 ($\lambda_\mathrm{c}=11.25\,\mu$m, $\Delta \lambda = 0.59\,\mu$m) filters with a total on-source integration time of 180\,s. To remove the sky and telescope thermal background, chopping and nodding sequences were used during the acquisitions with a perpendicular amplitude of 8\,\arcsec. To obtain an absolute flux calibration, we observed a nearby standard star from the \citet{Cohen_1999_04_0} catalogue before or after each Cepheid. The observing log is listed in Table~\ref{table__log}. Our sample was selected from three main criteria: 1) they are bright ($V < 8$), to avoid a significant observing time; 2) their pulsation period ranges from 3 to 39\,days to provide a sufficient coverage in period; and 3) they have at least $B, V, J, H,$ and $K$-band photometric light curves, which will allow us to better match the measured mid-IR excess for a given pulsation phase.
	
	Data were reduced with the ESO data reduction pipeline\footnote{\url{https://www.eso.org/sci/software/pipelines/visir/visir- pipe- recipes.html}}, taking into account the chopping and nodding corrections to obtain a final average image. Examples of the images obtained are shown in Fig.\ref{fig__images}. We then extracted flux densities by using classical aperture photometry with an aperture radius of 50\,pixels (3.75\arcsec) and a background annulus with inner radius of 55\,pixels and a thickness of 10\,pixels ($4.13\arcsec-4.88\arcsec$). No aperture correction was applied as we used the same for all. Absolute flux calibration was obtained taking into account the filter transmission\footnote{Filter transmission profiles available at \url{http://www.eso.org/sci/facilities/paranal/instruments/visir/inst.html}}. We also applied an airmass correction such as
	\begin{displaymath}
		F_\mathrm{corr}(\lambda) = F_\mathrm{meas}(\lambda,AM) \times C(\lambda,AM),
	\end{displaymath}
	with $\lambda$ being the wavelength, $AM$ the airmass, and the correction factor $C(\lambda,AM)$ taken from \citet{Schutz_2005__0}:
	\begin{displaymath}
		C(\lambda,AM) = 1 + \left[0.220 - \dfrac{0.104}{3} (\lambda - 8.6) \right] (AM - 1)
	\end{displaymath}
	where $\lambda$ is expressed in $\mu$m. Our absolute flux measurements are listed in Table~\ref{table__log}. Errors were estimated using the formalism of \citet{Laher_2012_07_0}. In addition, we added a 5\,\% error to account for possible additional mid-IR sky variability.
	
	\begin{table*}[!ht]
		\centering
		\caption{Log of the VISIR observations with our photometric measurements.}
		\begin{tabular}{cccccccc}
			\hline
			\hline
			Star 							    & MJD   & Seeing & Airmass & Flux $\mathrm{\pm \sigma_{stat}\pm \sigma_{sys}}$ & mag $\mathrm{\pm \sigma_{stat}\pm \sigma_{sys}}$ & Cal. & Rej. \\
			&                   & (\arcsec) &            & ($\times 10^{-14}\, \mathrm{W/m^2/\mu m}$) &  \\
			\hline
			\multicolumn{8}{c}{PAH1 filter ($8.59\,\mu$m)} \\
			\object{FM~Aql} 		      &  	57548.2671  &  0.81 & 1.23 & $2.04\pm0.04\pm0.10$ &  $5.036\pm0.022\pm0.054$ & \object{HD~186791}  & x \\
			\object{FN~Aql} 		  &  	57548.2763  &  1.06 & 1.14 & $1.28\pm0.06\pm0.06$ &  $5.541\pm0.051\pm0.054$  & \object{HD~187642}  & x \\
			\object{SZ~Aql} 		  &  	57548.2841  &  1.32 & 1.11 & $2.15\pm0.05\pm0.11$ &  $4.979\pm0.023\pm0.054$  & \object{HD~184406} &  \\
			\object{TT~Aql} 		  &  	57548.2918  &  1.36 & 1.12 & $5.70\pm0.05\pm0.29$ &  $3.921\pm0.010\pm0.054$  & \object{HD~184406}  &  \\
			\object{U~Car} 		      &  	57521.0494  &  0.73 & 1.24 & $10.82\pm0.04\pm0.54$ &  $3.225\pm0.004\pm0.054$  & \object{HD~89388}  &  \\
			\object{V~Car} 		      &  	57521.9799  &  0.91 & 1.31 & $1.90\pm0.03\pm0.10$ &  $5.091\pm0.018\pm0.054$  & \object{HD~65662}  &  \\
			\object{VY~Car} 		 &  	57521.0390  &  1.08 & 1.21 & $2.41\pm0.03\pm0.12$ &  $4.857\pm0.014\pm0.054$  & \object{HD~89682} &  \\
			\object{V~Cen} 		   &  	57521.2004  &  0.59 & 1.21 & $3.85\pm0.03\pm0.19$ &  $4.347\pm0.009\pm0.054$  & \object{HD~119193} &  \\
			\object{XX~Cen} 		&  	57505.1804  &  0.75 & 1.20 & $1.42\pm0.03\pm0.07$ &  $5.429\pm0.023\pm0.054$  & \object{HD~119193} &  \\
			\object{$\beta$~Dor} 		&  	57653.3566  &  0.37 & 1.33 & $38.42\pm0.1\pm1.92$ &  $1.849\pm0.003\pm0.054$  & \object{HD~42540} &  \\
			\object{T~Mon} 		&  	57727.1799  &  0.62 & 1.34 & $8.29\pm0.04\pm0.42$ &  $3.514\pm0.005\pm0.054$  & \object{HD~49293}  &  \\
			\object{RY~Sco} 		&  	57520.3837  &  1.52 & 1.13 & $7.03\pm0.04\pm0.35$ &  $3.693\pm0.006\pm0.054$  & \object{HD~168592}  &  \\
			\object{BB~Sgr} 			&  	57548.2579  &  0.83 & 1.00 & $3.66\pm0.04\pm0.18$ &  $4.401\pm0.012\pm0.054$  & \object{HD~175775}  &  \\
			\object{U~Sgr} 				&  	57548.2369  &  1.00 & 1.01 & $6.18\pm0.05\pm0.31$ &  $3.833\pm0.008\pm0.054$  & \object{HD~175775}  &  \\
			\object{WZ~Sgr} 		&  	57548.2474  &  1.23 & 1.01 & $3.53\pm0.05\pm0.18$ &  $4.441\pm0.015\pm0.054$  & \object{HD~169916}  &  \\
			\object{ST~Tau} 		    &  	57704.3532  &  0.95 & 1.39 & $1.39\pm0.07\pm0.07$ &  $5.454\pm0.025\pm0.054$  & \object{HD~37160}  &  \\
			\object{SZ~Tau} 		    &  	57629.3579  &  0.70 & 1.69 & $4.91\pm0.05\pm0.25$ &  $4.083\pm0.023\pm0.054$  & \object{HD~28305}  &  \\
			\object{RY~Vel} 		&  	57521.0286  &  0.79 & 1.19 & $2.76\pm0.03\pm0.14$ &  $4.708\pm0.012\pm0.054$  & \object{HD~89682}  &  \\
			\object{RZ~Vel} 		&  	57521.0163  &  0.60 & 1.23 & $4.91\pm0.04\pm0.25$ &  $4.084\pm0.008\pm0.054$  & \object{HD~67582}  &  \\
			\object{T~Vel} 		    &  	57521.9906  &  0.59 & 1.21 & $1.26\pm0.03\pm0.06$ &  $5.561\pm0.027\pm0.054$  & \object{HD~67582}  &  \\			
			\hline
			\multicolumn{8}{c}{PAH2 filter ($11.25\,\mu$m)} \\
			\object{FM~Aql} 		  &  	57570.1765  &  0.94 & 1.27 & $0.94\pm0.02\pm0.05$ &  $4.722\pm0.027\pm0.054$ & \object{HD~186791}  &  \\
			\object{FN~Aql} 		  &  	57570.2058  &  0.83 & 1.14 & $0.86\pm0.02\pm0.04$ &  $4.820\pm0.028\pm0.054$  & \object{HD~187642}  &  \\
			\object{SZ~Aql} 		  &  	57570.1860  &  1.18 & 1.13 & $0.61\pm0.02\pm0.03$ &  $5.187\pm0.029\pm0.054$  & \object{HD~184406}  & x  \\
			\object{TT~Aql} 		  &  	57570.1957  &  0.76 & 1.12 & $2.69\pm0.02\pm0.13$ &  $3.579\pm0.008\pm0.054$  & \object{HD~184406}  &  x \\
			\object{U~Car} 		      &  	57522.0398  &  0.73 & 1.24 & $3.78\pm0.02\pm0.19$ &  $3.208\pm0.005\pm0.054$  & \object{HD~89388}  &  \\
			\object{V~Car} 		      &  	57729.3141  &  1.36 & 1.24 & $0.60\pm0.01\pm0.03$ &  $5.213\pm0.026\pm0.054$  & \object{HD~65662}  &  \\
			\object{VY~Car} 		  &  	57522.0295  &  0.74 & 1.21 & $0.96\pm0.01\pm0.05$ &  $4.700\pm0.015\pm0.054$  & \object{HD~89682}  &  \\
			\object{V~Cen} 		     &  	57522.0587  &  0.94 & 1.33 & $1.49\pm0.02\pm0.07$ &  $4.22\pm0.011\pm0.054$  & \object{HD~119193}  &  \\
			\object{XX~Cen} 		&  	57522.0493  &  0.96 & 1.27 & $0.64\pm0.01\pm0.03$ &  $5.132\pm0.023\pm0.054$  & \object{HD~119193}  &  \\
			\object{$\beta$~Dor} 		&  	57653.3528  &  0.34 & 1.33 & $13.94\pm0.04\pm0.70$ &  $1.791\pm0.003\pm0.054$  & \object{HD~42540}  &  \\
			\object{T~Mon} 		     &  	57727.1897  &  0.71 & 1.29 & $2.80\pm0.02\pm0.14$ &  $3.534\pm0.006\pm0.054$  & \object{HD~49293}  &  \\
			\object{RY~Sco} 		&  	57520.3936  &  1.61 & 1.17 & $2.38\pm0.02\pm0.12$ &  $3.709\pm0.007\pm0.054$  & \object{HD~168592}  &  \\
			\object{BB~Sgr} 			&  	57570.2162  &  0.86 & 1.00 & $1.33\pm0.01\pm0.07$ &  $4.343\pm0.011\pm0.054$  & \object{HD~175775}  &  \\
			\object{U~Sgr} 				&  	57520.4123  &  1.09 & 1.14 & $2.39\pm0.02\pm0.12$ &  $3.706\pm0.007\pm0.054$  & \object{HD~175775}  &  \\
			\object{WZ~Sgr} 		&  	57520.4029  &  1.44 & 1.14 & $1.58\pm0.02\pm0.08$ &  $4.156\pm0.010\pm0.054$  & \object{HD~169916}  &  \\
			\object{ST~Tau} 		    &  	57704.3424  &  0.96 & 1.35 & $0.27\pm0.02\pm0.01$ &  $6.075\pm0.059\pm0.054$  & \object{HD~37160}  &  \\
			\object{SZ~Tau} 		    &  	57640.3975  &  0.87 & 1.39 & $1.66\pm0.02\pm0.08$ &  $4.101\pm0.010\pm0.054$  & \object{HD~28305}  &  \\
			\object{RY~Vel} 		&  	57522.0213  &  0.79 & 1.19 & $2.76\pm0.01\pm0.14$ &  $4.708\pm0.015\pm0.054$  & \object{HD~89682}  &  \\
			\object{RZ~Vel} 		&  	57522.0019  &  0.62 & 1.19 & $0.95\pm0.02\pm0.05$ &  $4.713\pm0.010\pm0.054$  & \object{HD~67582}  &  \\
			\object{T~Vel} 		    &  	57522.0124  &  0.92 & 1.25 & $0.45\pm0.01\pm0.02$ &  $5.520\pm0.031\pm0.054$  & \object{HD~67582}  &  \\
			\hline
		\end{tabular}
		\label{table__log}
		\tablefoot{Seeing is given at the visible wavelength ($\lambda = 0.5\mu$m). "Cal." denotes the calibrator star we used for the absolute flux calibration. "rej." refers to our photometry we rejected (see Sect.~\ref{section__spips_analysis}).
		}
	\end{table*}
	
	\begin{sidewaystable*}[!ht]
		\centering
		\caption{References for the photometric, spectroscopic, effective temperature and angular diameter measurements used in this study.}
		\begin{tabular}{ccccc|ccccc}
			\hline
			\hline
			Star 		& Ref. phot. &			Ref. RVs  & Ref. $T_\mathrm{eff}$ & Ref. $\theta$ &  Star 		& Ref. phot. &			Ref. RVs  & Ref. $T_\mathrm{eff}$ & Ref. $\theta$ \\
			\hline
			\object{$\eta$~Aql}   & 2,3,4,5,6,7,9,10,12,14  & 14,20,21,37  & 19  & 43,45\tablefootmark{i} &  \object{S~Nor}   &  1,2,3,7,9\tablefootmark{a,b},10\tablefootmark{c,d},11,12,18 & 29,44  & 19  & 44\tablefootmark{i} \\
			\object{FF~Aql}		&	1,2,3,5,7,10\tablefootmark{d},12,14,18\tablefootmark{k},46,47	&14,44,48	& 19  &	49,57	&  \object{TW~Nor}   & 3,7,9\tablefootmark{a,b},10\tablefootmark{c},11,12,15  & 29,38  & 19  & --  \\
			FM~Aql & 1,2,3,4,5,6,7,8,9,10    				&20,21,22 				&19  &  -- &  \object{V340~Nor}   & 1,3,11,12  & 29,38  & 19  & --  \\
			FN~Aql 		 & 1,2,3,4,5,6,7,9,10 				& 20,22 				&19  &  -- &  \object{Y~Oph}   &  1,2,3,5,7,9,10,11,12,13,54 & 21,28  & 19  & 55\tablefootmark{i},56\tablefootmark{i}  \\
			SZ~Aql 		 & 1,2,3,4,5,6,7,8,9\tablefootmark{a,b},10\tablefootmark{c,d},11,12 & 20 						& 19  &  -- &  RY~Sco 		   & 1,2,3,5,9\tablefootmark{a,b},10,11,13 					& 26		    		& 19    &  -- \\
			TT~Aql 		   & 1,2,3,4,5,6,7,8,9\tablefootmark{e},10,13,14 & 20,22,23 			& 19   &  -- &  \object{RU~Sct}   & 1,2,3,5,8,11,12  & 22,38  & 19,39  & --  \\
			\object{U~Aql}   & 2,3,5,7,8,9\tablefootmark{a,b},10,12,14,18\tablefootmark{j}  & 44  & 19  & 44\tablefootmark{i}  &  \object{GY~Sge}  &  3,7,11,12\tablefootmark{l}  & 22,38  &  19  & --   \\
			\object{RT~Aur}   & 1,2,3,4,5,6,8,9\tablefootmark{a,b},10\tablefootmark{c,d},12,14,18  & 14,21,22 &  19 &  53\tablefootmark{i} &  \object{S~Sge}  & 1,2,3,4,5,6,7,10\tablefootmark{c,d},12,14  &  20,21,22 & 19  & --  \\
			\object{$\ell$~Car}  & 2,3,9,10,11,12,15,18,58   & 21,24,50,51,52   & 19   &  43,51  &  BB~Sgr 			& 1,2,3,5,7,9\tablefootmark{a,b},10\tablefootmark{d},16,18					& 22,23					& 19    &  -- \\
			U~Car		      & 1,2,3,9\tablefootmark{a,b},10\tablefootmark{c,d},11,12,13,15,18,36 			& 13,21,24 			& 19,30   & 44\tablefootmark{i}  &  U~Sgr 				& 1,2,3,5,8,9\tablefootmark{a,b},11,12,16,18\tablefootmark{k} 				& 23,24,29 			& 19    &  -- \\
			V~Car 		       & 1,2,3,9\tablefootmark{a,b},10,11 						& 25 						& 19    & --  &  \object{W~Sgr}   & 1,2,3,5,7,10,12,18\tablefootmark{k}  & 28,29,44  &  19 &  43,44\tablefootmark{i} \\
			VY~Car 	      & 1,2,3,9\tablefootmark{a,b},11,13,15,18				 & 24,26 					& 19   & --  &  WZ~Sgr 		&  1,2,3,5,7,8,11,12,13	  		& 22,24,26  		& 32   & --  \\
			\object{CF~Cas}	&  1,2,3,8,9\tablefootmark{h},10\tablefootmark{c},12  & 22,29   & 19   &  -- &  \object{X~Sgr}   & 1,2,3,5,9,10,12,36,40  & 21  & 19  & 43,44\tablefootmark{i}  \\
			\object{DL~Cas}  & 1,2,3,5,6,8,10\tablefootmark{c},12 & 21,22,29  & 19   & --  &  \object{Y~Sgr}  & 1,2,3,5,7,9\tablefootmark{a,b},10\tablefootmark{c,d},12  & 21  & 19  &   \\
			V~Cen		     & 1,2,3,9\tablefootmark{a,b},10\tablefootmark{d},11,12,16,18 				& 24,27 				& 19,31  & --  &  ST~Tau 		    &  1,2,3,4,5,6,8,9\tablefootmark{n},10			  	& 22,29				 & 19    & --  \\
			XX~Cen 	       & 1,2,3,9,10,11,13 					& 24\tablefootmark{f} 						& 19  &  -- &  SZ~Tau 		    &  	1,2,3,4,5,6,9\tablefootmark{a,e},10,11,14,18\tablefootmark{j}  	& 22,29				 & 19    &  -- \\
			\object{$\delta$~Cep}  & 2,3,4,5,6,9\tablefootmark{a,b},10,12,18,40  & 41  & 19    &  42\tablefootmark{i} &  RY~Vel 		&  	1,2,3,9,11,15  				& 13,33  				& 19    &  -- \\
			\object{CD~Cyg}      & 1,2,3,5,6,7,8,9\tablefootmark{a,b},10,12 					& 26\tablefootmark{f} 						& 19  &  -- &  RZ~Vel 		& 1,2,3,9\tablefootmark{a,b},10,11,15 	  				& 21,24,34			   & 19     & --  \\
			\object{X~Cyg}   & 1,2,34,5,6,7,10\tablefootmark{c,d},12,14,18\tablefootmark{k}  &  20,21,22,29,37 & 19  & --  &  T~Vel 		    &    1,2,3,9,11,35						& 	34						& 19    & --  \\		
			$\beta$~Dor 	  & 1,2,3,9\tablefootmark{a,b},10,11,12,17,18 			& 21,28,29			 & 19   & 43  &  \object{S~Vul}   & 3,8,9\tablefootmark{a,b},10\tablefootmark{d},11,12  & 21,22,26  & 19  &  -- \\
			\object{$\zeta$~Gem}   & 2,3,5,6,9,1010\tablefootmark{d},12,14,18\tablefootmark{m},40  & 21,22,23,29  &  19 &   &  \object{SV~Vul}   & 1,2,3,4,5,6,7,8,9\tablefootmark{a,b},10,11,12,14,  & 20,22,29,37  & 19  &   \\ 
			\object{CV~Mon}    & 1,2,3,5,6,8,9,10\tablefootmark{c},11, & 37,38 & 19,39   &  -- &  \object{T~Vul}   & 1,2,3,4,7,9\tablefootmark{a,b},10,12,46  & 14,29  & 19  & 53\tablefootmark{i},57\tablefootmark{i}  \\
			T~Mon 	 		& 1,2,3,5,9\tablefootmark{a,b},11,12,13,18,24 			& 23\tablefootmark{g} 					& 32   &  44\tablefootmark{i} &  \object{U~Vul}   & 1,2,3,4,5,6,8,9\tablefootmark{a,b},10,14  & 20,21,22,29  & 19  & --  \\
			\hline
		\end{tabular}
		\label{table__ref}
		\tablefoot{
			1- \citet{Gaia-Collaboration_2018_08_0}.
			2- \citet{van-Leeuwen_1997_07_0}.
			3- \citet{Berdnikov_2008_04_0}.
			4- \citet{Barnes_1997_06_0}.
			5- \citet{Moffett_1984_07_0}.
			6- \citet{Szabados_1980_01_0}.
			7- \citet{Welch_1984_04_0}.
			8- \citet{Monson_2011_03_0}.
			9- \citet{Wright_2010_12_0}.
			10- \citet{Ishihara_2010_05_0}.
			11- \citet{Laney_1992_04_0}.
			12- \citet{Monson_2012_11_0}.
			13- \citet{Coulson_1985__0}.
			14- \citet{Kiss_1998_07_0}.
			15- \citet{Madore_1975_06_0}.
			16- \citet{Gieren_1981_12_0}.
			17- \citet{Shobbrook_1992_04_0}.
			18- \citet{Marengo_2010_01_0}.
			19- \citet{Luck_2018_10_0}.
			20- \citet{Barnes_2005_02_0}.
			21- \citet{Borgniet_2019_11_0}.
			22- \citet{Gorynya_1998_11_0}.
			23- \citet{Storm_2011_10_0}.
			24- \citet{Bersier_2002_06_0}.
			25- \citet{Lloyd-Evans_1980__0}.
			26- \citet{Anderson_2016_10_0}.
			27- \citet{Gieren_1981_12_0}.
			28- \citet{Petterson_2005_10_0}.
			29- \citet{Bersier_1994_11_0}.
			30- \citet{Usenko_2011_10_0}.
			31- \citet{Usenko_2013_07_0}.
			32- \citet{Kovtyukh_2005_01_0}.
			33- \citet{Pont_1994_05_0}.
			34- \citet{Szabados_2013_09_0}.
			35- \citet{Gieren_1985_08_0}.
			36- \citet{Gallenne_2012_02_0}.
			37- \citet{Storm_2004_02_0}.
			38- \citet{Metzger_1992_02_0}.
			39- \citet{Genovali_2014_06_0}.
			40- \citet{Feast_2008_06_0}.
			41- \citet{Anderson_2015_05_0}.
			42- \citet{Merand_2005_07_0}.
			43- \citet{Kervella_2004_03_0}.
			44- \citet{Gallenne_2019_02_0}.
			45- \citet{Lane_2002_07_0}.
			46- \citet{Szabados_1977_01_0}.
			47- \citet{Szabados_1991_01_0}.
			48- \citet{Evans_1990_05_1}.
			49- \citet{Gallenne_2012_03_0}.
			50- \citet{Anderson_2014_06_0}.
			51- \citet{Anderson_2016_02_0}.
			52- \citet{Anderson_2016_12_0}.
			53- \citet{Gallenne_2015_07_0}.
			54- \citet{Fernie_1995_09_0}.
			55- \citet{Merand_2007_08_0}.
			56- \citet{Gallenne_2011_12_1}.
			57- \citet{Gallenne_2012_03_0}.
			58- \citet{Kervella_2009_05_0}.
			\tablefoottext{a}{Band $W_1$ excluded due to very large error bar.}
			\tablefoottext{b}{Band $W_2$ excluded due to very large error bar.}
			\tablefoottext{c}{Band $S_9$ excluded due to very large error bar.}
			\tablefoottext{d}{Band $L_{18}$ excluded due to very large error bar.}
			\tablefoottext{e}{Band $W_2$ excluded as not consistent with other photometric values, too high value.}
			\tablefoottext{f}{Orbital motion was substracted, see text.}				
			\tablefoottext{g}{We did not succeed in removing the orbital motion, so we only use recent measurements with the good phase coverage.}
			\tablefoottext{h}{Band $W_4$ excluded due to very large error bar.}
			\tablefoottext{i}{Converted to limb-darkened diameter using the linear law approximation from \citet{Hanbury-Brown_1974_06_0} and the $u_\lambda$ coefficient from \citet{Claret_2011_05_0}. Stellar parameters used for choosing $u_\lambda$ are \citep{Luck_2018_10_0}: $v_\mathrm{turb}=4\,\mathrm{m~s^{-1}}$, [Fe/H] = 0, $T_\mathrm{eff}=6000\,$K and $\log{g}=2.0$ for RT~Aur, T~Vul, FF~Aql, $\delta$~Cep and U~Aql, $T_\mathrm{eff}=5500\,$K and $\log{g}=2.0$ for S~Nor, $T_\mathrm{eff}=5250\,$K and $\log{g}=1.0$ for T~Mon and $\ell$~Car, $T_\mathrm{eff}=5000\,$K and $\log{g}=1.0$ for U~Car, $T_\mathrm{eff}=6000\,$K and $\log{g}=1.5$ for Y~Oph and $\eta$~Aql,}
			\tablefoottext{j}{IRAC 5.8$\mu$m, 8$\mu$m and MIPS 70$\mu$m were rejected as they are not consistent with other photometric values ($\sim -0.05$\,mag offset).}
			\tablefoottext{k}{IRAC 5.8$\mu$m and 8$\mu$m were rejected as they are not consistent with other photometric values (more than $-0.05$\,mag offset).}
			\tablefoottext{l}{Not used as not consistent with other photometric values (more than $-0.05$\,mag offset)}
			\tablefoottext{m}{MIPS 70$\mu$m rejected as too large errorbar}
			\tablefoottext{n}{Band $W_4$ excluded due to very large error bar.}
		}
	\end{sidewaystable*}
	
	\subsection{Literature data}
	
	The SPIPS algorithm performs a time-dependent global fit; therefore from the optical to near-IR bands we only retrieved photometric light curves. This includes ground-based observations performed by many astronomers and also data from the \emph{Gaia} data release 2 \citep{Gaia-Collaboration_2016_11_0,Gaia-Collaboration_2018_08_0,Riello_2018_08_0,Evans_2018_08_0} and the Hipparcos mission \citep{Perryman_1997_07_0,van-Leeuwen_1997_07_0}. All references are listed in Table~\ref{table__ref}. $J, H$, and $K$ photometry from \citet{Laney_1992_04_0} and \citet{Feast_2008_06_0} have been transformed from the Carter to CTIO photometric system following the transformation formulae from \citet{Carter_1990_01_0}. Cepheids are known to have nearby hot companions \citep[see e.g.][]{Szabados_1995__0,Szabados_2011_02_0,Kervella_2019_03_0,Kervella_2019_03_1}, so to avoid any flux contamination, $B$-band light curves were not used.
	
	Light curves in the \emph{Spitzer} bands $I_1\sim3.6\,\mu$m and $I_2\sim4.5\,\mu$m were also retrieved from \citet{Monson_2012_11_0}. Unfortunately no other light curves are available in the literature for $\lambda > 3\,\mu$m. In addition to our VISIR mid-IR photometry, we also gathered multi-epoch photometric measurements from WISE \citep[$W_1\sim3.4\,\mu$m, $W_2\sim4.6\,\mu$m, $W_3\sim11.6\,\mu$m and $W_4\sim22.1\,\mu$m,][]{Wright_2010_12_0} and AKARI \citep[$S_9\sim9\,\mu$m and $L_{18}\sim 18\,\mu$m,][]{Ishihara_2010_05_0,Cutri_2012_03_0}. Other measurements from MSX \citep{Egan_2003__0} and IRAS \citep{Helou_1988__0} were not used due to large uncertainties that do not provide any additional constraint ($> 0.1$\,mag). AKARI data do not provide observing dates, so in order to avoid phase mismatch we quadratically added a 0.05\,mag uncertainty. We also rejected all WISE measurements that are in the AllWISE reject table and have a quality image factor < 0.5, which are data that do not meet pre-defined criteria\footnote{That is, if the keyword \emph{cat=0} and \emph{qi\_fact<0.5,} as defined here: \url{https://wise2.ipac.caltech.edu/docs/release/allwise/expsup/sec2_1.html}.} (saturation, S/N < 5, etc.), as suggested by the AllWISE data release products.
	
	We also gathered radial velocities derived from the cross-correlation method if possible. When several data sets exist, we only used the ones with a precision $< 1.5\,\mathrm{km~s^{-1}}$, which are better constraints to the global fit.
	
	Spectroscopic stellar temperatures were retrieved mostly from \citet{Luck_2018_10_0}, who used several previously published Cepheid spectra. Some Cepheids have a good phase coverage in temperature (typically the brightest), while others only have one measurement.
	
	Additionally, when available, we completed our observables with published spectroscopic effective temperatures and interferometric angular diameters. This allowed us to remove the degeneracy with the reddening during the fitting process. For some Cepheids, measurements were made at various pulsation phases, providing better constraints.
	
	\section{SPIPS analysis}
	\label{section__spips_analysis}
	
	Our dataset contains photometry of 45 Cepheids from 0.5\,$\mu$m to 70\,$\mu$m, radial velocities estimated using the cross-correlation method, stellar effective temperatures derived from line-depth ratios, and in certain cases interferometric angular diameters. For some stars, we rejected a few photometric data points from the literature, as noted in Table~\ref{table__ref}. Some of our VISIR measurements were also rejected due to negative excess or a very high value ($>3\sigma$) compared to the trend in excess. For those observations, we checked the weather conditions for which thermal IR observations are strongly sensitive. We noticed that these measurements were executed with some intermittent clouds passing by, which may have altered the photometric transmission, either for the calibrator (resulting in a too strong an IR excess) or the Cepheid (negative IR excess).
	
	Some of the Cepheids are known or suspected binaries. The presence of a companion may affect both the radial velocities and photometric measurements. As previously explained, most of the companions are hot main-sequence stars and are outshone by the Cepheid brightness from the visible wavelength. This is why we decided to not use the $B$ band so that the contamination on the Cepheid photometry would be negligible for wavelengths longer than 0.5$\mu$m. Radial velocities are also affected by the presence of a companion, such that measurements are the combination of the pulsation and orbital motion. Therefore, when the companion is massive enough to impact the velocities (at least at the precision level of the measurements), they need to be corrected before applying the SPIPS modelling. Following the formalism of \citet{Gallenne_2018_11_0,Gallenne_2019_02_0}, we corrected the orbital motion for the Cepheids FF~Aql, U~Aql, DL~Cas, XX~Cen, Y~Oph, S~Sge, W~Sgr and U~Vul in order to only have the pulsation effect. For T~Mon, CD~Cyg, $\delta$~Cep, X~Cyg which are known binaries from long-term trend of the systemic velocity, we did not succeed in fitting the orbital motion, so we only used the more recent RVs with more observations and a good orbital coverage. Others are suspected binaries, but no clear offset is detected between different measurements, so no corrections were applied.
	
	We used the SPIPS modelling tool to perform a global simultaneous fit of the previously listed observations. A full description of the code can be found in \citet{Merand_2015_12_0}\footnote{\url{http://github.com/amerand/SPIPS}}. Briefly, this tool is based on the parallax-of-pulsation method (also called the Baade-Wesselink method), meaning it compares the linear and angular variations of the Cepheid diameters to retrieve physical parameters of the stars, such as the ratio $p$-factor-distance, effective temperature, colour excess, etc., following this equation:
	\begin{displaymath}
		\theta(P_\mathrm{puls}) - \theta(0) \propto \frac{1}{d} \int_0^{P_\mathrm{puls}} v_\mathrm{puls}(t)dt,
	\end{displaymath}
	with the pulsation period $P_\mathrm{puls}$, the distance $d$, the angular diameter $\theta$, and the pulsation velocity $v_\mathrm{puls}$. The radial velocity $v_\mathrm{r}$ is related to the pulsation velocity via the $p$-factor such as $v_\mathrm{puls} = p\,v_\mathrm{r}$ \citep[for a short review, see e.g.][]{Nardetto_2017_01_0}.
	
	The SPIPS code can take several types of data and observables, such as optical and IR magnitudes and colours, radial velocities, effective temperatures, and interferometric angular diameters. The resulting redundancy in the observables ensures a high level of robustness. We already proved the efficiency of SPIPS for both Galactic and Magellanic Cloud Cepheids \citep{Breitfelder_2016_03_0,Kervella_2017_04_0,Gallenne_2017_11_0}. The angular diameter is an important additional observable to constrain the fitted parameters as it is independent of interstellar extinction and allows us to decorrelate the colour excess and the effective temperature, parametrised with the following surface brightness relation \citep{Merand_2015_12_0}:	
	\begin{displaymath}
		m_\lambda = m_0 - C_\lambda \log{T_\mathrm{eff}} - 5 \log{\theta} + A_\lambda E(B-V),
	\end{displaymath}
	with $A_\lambda$ being the bandpass-dependent reddening coefficient, and $m_0, C_\lambda$ a set of parameters describing the relation.
	
	Synthetic photometry is created using the ATLAS9 models \citep{Castelli_2003__0}, interpolated in time over the pulsation cycle with Fourier series or periodic splines functions, and integrated in wavelength using the bandpass of each observing filter\footnote{Using the SVO databases: \url{http://svo2.cab.inta-csic.es/theory/fps/}.}. A Galactic metallicity of 0.03\,dex is used for the models \citep{Luck_1998_02_0}, which is kept fixed during the fitting process. The reddening law and coefficient implemented in SPIPS are Galactic, taken from \citet{Fitzpatrick_1999_01_0} with $R_\mathrm{v}$ =3.1, which is a fixed parameter. Reddening for each photometric bandpass is parametrised with a global colour excess $E(B-V)$ and is computed using a chromatic reddening law, the SED of the star and the bandpass of the filter. It should be noted that this method is much more accurate than usual computations which use only $E(B-V)$ and a generic reddening factor defined for a stellar temperature of 10~000\,K. Cepheids are significantly cooler than 10~000\,K, resulting in reddenings that are not accurately computed with usual assumptions. This also means that, strictly speaking, the $E(B-V)$ we estimate are not directly comparable with values found in the literature. 
	
	Another important aspect is the temperature scale between the temperatures estimated from the SPIPS models and the spectroscopic temperatures estimated from line depth ratios \citep[LDR,][]{Kovtyukh_2000_06_0}. Comparing with temperatures determined from near-IR surface brightness relations, \citet{Lemasle_2020_09_2} estimated that the uncertainties remain below 150\,K at all phases. An ongoing work comparing SPIPS temperatures of stars containing angular diameters with LDR temperatures (Borgniet at al. 2021, private communication) shows that the uncertainties are better ($\sim50$\,K) for short-period Cepheids and can be as high as 150\,K for long periods. This remains of the order of magnitude of the uncertainties on the spectroscopic temperatures ($\sim 50-200$\,K).
	
	In our fitting process, the free parameters are the mean angular diameter $\theta_\mathrm{UD}$, the colour excess $E(B-V)$, the epoch at maximum light MJD$_0$, the pulsation period $P_\mathrm{puls}$, and the period change \citep[if necessary via a polynomial fit, see e.g][]{Kervella_2017_04_0}. The distance and the $p$-factor are fully degenerate, so we have to fix one to obtain the other, although this has no impact on our IR excess study. We kept $p$ fixed to a given value depending on the pulsation period, following the period-$p$-factor relation given by \citet{Gallenne_2017_09_0}. Time-dependent radial velocities and temperatures are described with cubic spline functions with $n$ optimised node positions, $V_\mathrm{n}$ and $T_\mathrm{n}$. The number $n$ depends on the shape of the pulsation curve. Node positions are defined using a comb for the phase axis, and parametrised with an exponent $\alpha$ and a given phase $\phi_0$, meaning that $\alpha$ represents the over-density of nodes at $\phi_0$ ($\alpha = 1$ is a uniform phase coverage, and larger values lead to higher densities). Non-uniform phase coverage is well handled by SPIPS thanks to the spline functions.
	
	Another aspect of the algorithm is the possibility to fit an IR excess using analytical formulae. We adopted a two-parameter power-law model defined as follows:
	\begin{eqnarray}
		\label{equation_power_law_model}
		\Delta m_\lambda = m_\mathrm{obs} - m_\mathrm{model}  = \left\{
		\begin{array}{ll}
			0, & \mathrm{for\,\lambda < 1.2\,\mu m }\\
			a_1 (\lambda - 1.20)^{a_2}, & \mathrm{for\, \lambda \geqslant 1.2\,\mu m}
		\end{array}
		\right.,
	\end{eqnarray}
	where $a_1$ and $a_2$ are additional fitted parameters. Here, we assume that the IR excess is produced by a circumstellar environment and that there is no excess below $1.2\,\mu$m. With the current amount and precision of the IR photometric data, more complicated models would not provide better constraints, this is why we used a simple model.
	
	To check the statistical significance of a possible IR excess, we performed the SPIPS analysis with and without infrared excess (i.e. including or not including the analytical formulae). We quantified the significance in a number of sigma using the following formula for the probability:
	\begin{displaymath}
		P = 1- \mathrm{CDF}_\nu \left( \dfrac{\nu~\chi^2_\mathrm{r,no~cse}}{\chi^2_\mathrm{r,cse}} \right),
	\end{displaymath}
	where $\chi^2_\mathrm{r,no~cse}$ and $\chi^2_\mathrm{r,cse}$ are the minimum chi-square for the model without and with a CSE, and CDF denotes the $\chi^2$ cumulative probability distribution function with $\nu$ degrees of freedom. We then convert the probability into the number of sigmas, $n\sigma$ (e.g. 99.73\,\% = $3\sigma$, 99.99\,\% = $4\sigma$, etc.), which are listed in Table~\ref{table__parameters}. This demonstrates how the model with CSE is significant compared to the model without it. For the following, we chose the following criteria: detection of a CSE if $n\sigma \geqslant 3$, and no detection if $n\sigma < 3$.

	Our fitting procedure was done step by step to ensure a fast convergence. We first fit the node position of the splines functions fixing the stellar parameters to approximate expected values and without including any IR excess. Once the fit converged with a satisfying reduced $\chi^2$, the stellar parameters were included. For the second analysis, we kept the previous fitted parameters, but this time we included the IR excess in the fitting process, starting with $a_1 = 0.1$ and $a_2 = 0.2$.
	
	We found that 13 Cepheids have a significant detection of a CSE. Most of our previously detected CSE are below this threshold, but this is likely due to the low precision in photometric measurements that prevents us from strongly constraining small IR excesses ($\lesssim 0.05$\,mag). We also noticed that most of the Cepheid fits with measured angular diameters have an improved $\chi^2$ when a CSE is included (no significant improvement for FF~Aql). This demonstrates the usefulness of this observable to constrain the presence of a CSE. However, as we only have a few interferometric data points compared to the photometry, it does not strongly contribute to improving the detection level. The reduced $\chi^2$ for the two models are listed in Table~\ref{table__parameters}, together with our derived stellar parameters for our best final model with or without IR excess. In Figs.~\ref{image__spips_etaaql}, \ref{image__spips_ucar}, and \ref{image__spips_betador}, we show the SPIPS model for three Cepheids.

	\begin{sidewaystable*}[]
		\centering
		\caption{Best-fit output parameters given by the SPIPS algorithm.}
		\begin{tabular}{cccccccccc|cc}
			\hline
			\hline
			Star & $P_\mathrm{puls}$   & MJD$_0$ & $E(B-V)$ & $p$-factor & $d$ & $<L>$  & $<T_\mathrm{eff}>$  & $<R>$ & $<\theta_\mathrm{LD}>$ & $\chi^2_r$ & $n\sigma$ \\
			&  (day)						& (day)    & 				&					 & (pc)  &  ($L_\odot$) & $K$ & ($R_\odot$)  & (mas)  & no cse -- cse & \\
			\hline
$\eta$~Aql    &  $7.1768323(81)$  &    $48069.4(1)$  &  $0.166(3)$  &  $1.266(24)$  &  $310(10)$  &  $3368(195)$  &  $5747(19)$  &  $59(2)$  &  $1.760(4)$ &  2.44 -- 1.88 & 5.0  \\
FF~Aql       &  $4.4709129(46)$  &    $47406(2)$  &  $0.272(1)$  &  $1.282(29)$  &  $482(32)$  &  $2704(339)$  &  $6176(21)$  &  $45(3)$  &  $0.8785(9)$ &  1.71 -- 1.63 & 1.4   \\
FM~Aql       &  $6.1145633(74)$  &    $35150.8(1)$  &  $0.677(3)$  &  $1.271(25)$  &  $1180(59)$  &  $3600(360)$  &  $5771(57)$  &  $60(3)$  &  $0.4738(7)$ &  3.53 -- 3.44 & 1.0   \\
FN~Aql       &  $9.482091(39)$  &    $36804.1(3)$  &  $0.544(3)$  &  $1.256(22)$  &  $1096(45)$  &  $2019(156)$  &  $5527(25)$  &  $49(2)$  &  $0.4165(8)$ &  3.60 -- 3.53 & 1.0   \\
SZ~Aql       &  $17.139823(40)$  &    $48314.69(7)$  &  $0.685(3)$  &  $1.236(20)$  &  $2081(48)$  &  $8539(299)$  &  $5442(17)$  &  $104(2)$  &  $0.4655(6)$ &  2.85 -- 2.85 & 0.7   \\
TT~Aql       &  $13.754952(26)$  &    $48308.45(6)$  &  $0.564(2)$  &  $1.243(20)$  &  $1035(24)$  &  $5921(208)$  &  $5504(17)$  &  $85(1)$  &  $0.7617(8)$ &  2.13 -- 2.13 & 0.7   \\
U~Aql        &  $7.024126(12)$  &    $51000.1(1)$  &  $0.424(2)$  &  $1.267(24)$  &  $617(18)$  &  $2642(125)$  &  $5733(21)$  &  $52(1)$  &  $0.7862(6)$ &  1.80 -- 1.79 & 0.7   \\
RT~Aur       &  $3.7283141(50)$  &    $47956.9(3)$  &  $0.083(4)$  &  $1.289(32)$  &  $474(21)$  &  $1451(124)$  &  $6034(66)$  &  $35(1)$  &  $0.685(2)$ &  4.47 -- 3.73 & 3.3   \\
$\ell$~Car   &  $35.55516(14)$  &    $50583.37(6)$  &  $0.217(1)$  &  $1.210(25)$  &  $502(12)$  &  $14052(916)$  &  $4962(75)$  &  $161(2)$  &  $2.976(2)$ &  4.01 -- 3.85 & 1.8   \\
U~Car        &  $38.85711(97)$  &    $48338.4(1)$  &  $0.382(7)$  &  $1.207(26)$  &  $1898(67)$  &  $26642(1584)$  &  $5397(24)$  &  $187(5)$  &  $0.917(3)$ &  3.65 -- 3.18 & 3.3   \\
V~Car        &  $6.6971691(72)$  &    $43781.6(3)$  &  $0.196(4)$  &  $1.268(25)$  &  $877(52)$  &  $1244(146)$  &  $5592(45)$  &  $38(2)$  &  $0.3991(8)$ &  2.49 -- 2.33 & 1.7   \\
VY~Car       &  $18.901635(41)$  &    $48339.30(6)$  &  $0.335(3)$  &  $1.232(20)$  &  $1960(48)$  &  $8709(327)$  &  $5308(13)$  &  $111(2)$  &  $0.5245(7)$ &  1.51 -- 1.50 & 0.7   \\
CF~Cas       &  $4.8750922(93)$  &    $56882.9(9)$  &  $0.595(8)$  &  $1.279(28)$  &  $3075(188)$  &  $1411(168)$  &  $5715(39)$  &  $38(2)$  &  $0.116(1)$ &  10.3 -- 5.76 & 20.3 \\
DL~Cas       &  $8.0003060(86)$  &    $42779.9(3)$  &  $0.580(2)$  &  $1.262(23)$  &  $1918(68)$  &  $3862(250)$  &  $5708(31)$  &  $64(2)$  &  $0.3088(3)$ &  1.08 -- 0.97 & 2.6   \\
XX~Cen       &  $10.958084(28)$  &    $35346.0(1)$  &  $0.38(1)$  &  $1.251(21)$  &  $1780(94)$  &  $5376(595)$  &  $5748(67)$  &  $74(4)$  &  $0.387(1)$ &  5.49 -- 5.26 & 1.3   \\
V~Cen        &  $5.4940145(91)$  &    $48340.2(8)$  &  $0.31(1)$  &  $1.275(27)$  &  $683(40)$  &  $1641(190)$  &  $5830(44)$  &  $40(2)$  &  $0.542(7)$ &  8.39 -- 3.80 & 33.1  \\
$\delta$~Cep   &  $5.366159(13)$  &    $36075.22(9)$  &  $0.113(2)$  &  $1.276(27)$  &  $263(9)$  &  $1949(107)$  &  $5905(29)$  &  $42(1)$  &  $1.492(2)$ &    3.79 -- 2.78 & 6.0 \\
CD~Cyg       &  $17.074411(43)$  &    $48321.55(7)$  &  $0.605(4)$  &  $1.236(20)$  &  $2636(61)$  &  $8089(300)$  &  $5470(16)$  &  $100(2)$  &  $0.354(2)$ &  2.23 -- 1.93 & 3.2   \\
X~Cyg        &  $16.385862(26)$  &    $48319.59(6)$  &  $0.310(2)$  &  $1.237(20)$  &  $1073(23)$  &  $6510(203)$  &  $5314(19)$  &  $95(1)$  &  $0.8263(8)$ &  2.05 -- 2.00 & 1.1   \\
$\beta$~Dor   &  $9.842698(32)$  &    $50275.0(6)$  &  $0.078(4)$  &  $1.255(21)$  &  $313(13)$  &  $3202(284)$  &  $5564(66)$  &  $61(2)$  &  $1.811(3)$ &   5.32 -- 5.31 & 0.7  \\
$\zeta$~Gem   &  $10.149814(17)$  &    $48708.1(3)$  &  $0.020(9)$  &  $1.254(21)$  &  $426(16)$  &  $4413(332)$  &  $5536(34)$  &  $72(3)$  &  $1.58(2)$ &   2.34 -- 2.00 & 3.6  \\
CV~Mon       &  $5.378669(15)$  &    $42772.6(2)$  &  $0.869(4)$  &  $1.276(27)$  &  $1547(87)$  &  $1658(180)$  &  $5829(39)$  &  $40(2)$  &  $0.2405(4)$ &  4.83 -- 4.84 & 0.6   \\
T~Mon        &  $27.03041(40)$  &    $43783.7(1)$  &  $0.299(5)$  &  $1.220(22)$  &  $1341(33)$  &  $12706(477)$  &  $5206(22)$  &  $139(2)$  &  $0.963(2)$ &  2.88 -- 2.63 & 1.8   \\
S~Nor        &  $9.754304(14)$  &    $48320.0(2)$  &  $0.250(2)$  &  $1.255(21)$  &  $1141(52)$  &  $5719(502)$  &  $5616(28)$  &  $80(3)$  &  $0.652(2)$ &  1.95 -- 1.54 & 5.3   \\
TW~Nor       &  $10.78612(15)$  &    $48329(1)$  &  $1.40(2)$  &  $1.252(21)$  &  $2579(353)$  &  $5867(1835)$  &  $5664(218)$  &  $80(10)$  &  $0.287(2)$ &  15.2 -- 15.2 & 0.7 \\
V340~Nor     &  $11.28835(18)$  &    $48336.9(8)$  &  $0.414(3)$  &  $1.250(21)$  &  $2541(156)$  &  $7580(959)$  &  $5598(63)$  &  $93(5)$  &  $0.3394(8)$ &   4.55 -- 1.53 & 32.4 \\
RY~Sco       &  $20.32180(20)$  &    $54670.3(2)$  &  $0.892(3)$  &  $1.230(21)$  &  $1245(67)$  &  $9395(1468)$  &  $5689(168)$  &  $100(5)$  &  $0.746(2)$ &  4.89 -- 4.71 & 1.1   \\
RU~Sct       &  $19.70453(13)$  &    $48335.7(4)$  &  $1.046(4)$  &  $1.231(21)$  &  $2249(94)$  &  $12598(1119)$  &  $5502(62)$  &  $124(5)$  &  $0.512(1)$ &  3.52 -- 3.53 & 0.6   \\
GY~Sge       &  $51.4914(56)$  &    $48310(1)$  &  $1.50(1)$  &  $1.197(29)$  &  $2951(134)$  &  $40288(3417)$  &  $5489(50)$  &  $222(9)$  &  $0.701(2)$ &  2.23 -- 2.14 & 1.2   \\
S~Sge        &  $8.382137(10)$  &    $48311.2(2)$  &  $0.161(2)$  &  $1.261(22)$  &  $614(22)$  &  $2669(168)$  &  $5741(17)$  &  $52(2)$  &  $0.793(1)$ &  1.87 -- 1.75 & 1.9   \\
BB~Sgr       &  $6.637384(11)$  &    $36052.8(2)$  &  $0.347(3)$  &  $1.269(25)$  &  $815(42)$  &  $2348(238)$  &  $5676(50)$  &  $50(2)$  &  $0.5729(9)$ &  1.93 -- 1.92 & 0.7   \\
U~Sgr        &  $6.7457236(73)$  &    $30116.81(6)$  &  $0.489(3)$  &  $1.268(24)$  &  $633(17)$  &  $2611(127)$  &  $5749(44)$  &  $52(1)$  &  $0.758(1)$ &  2.07 -- 1.98 & 1.3   \\
W~Sgr        &  $7.5949735(62)$  &    $48690.7(5)$  &  $0.173(4)$  &  $1.264(23)$  &  $415(16)$  &  $3012(250)$  &  $5796(65)$  &  $55(2)$  &  $1.221(2)$ &  1.60 -- 1.48 & 1.8   \\
WZ~Sgr       &  $21.850929(85)$  &    $48310.3(1)$  &  $0.618(2)$  &  $1.227(21)$  &  $1906(42)$  &  $10861(695)$  &  $5250(76)$  &  $126(2)$  &  $0.6159(5)$ &  3.67 -- 3.67 & 0.7   \\
X~Sgr        &  $7.012741(12)$  &    $48707.8(2)$  &  $0.305(2)$  &  $1.267(24)$  &  $360(14)$  &  $3510(271)$  &  $6080(60)$  &  $53(2)$  &  $1.382(1)$ &  1.77 -- 1.74 & 0.9   \\
Y~Sgr        &  $5.7734082(74)$  &    $48700.3(1)$  &  $0.277(2)$  &  $1.273(26)$  &  $524(25)$  &  $2380(205)$  &  $5811(16)$  &  $48(2)$  &  $0.856(1)$ &  1.56 -- 1.44 & 1.9   \\
ST~Tau       &  $4.0342009(47)$  &    $41762(3)$  &  $0.425(8)$  &  $1.286(30)$  &  $1239(151)$  &  $1988(481)$  &  $6098(42)$  &  $40(5)$  &  $0.300(3)$ &  4.59 -- 3.01 & 8.0   \\
SZ~Tau       &  $3.1488519(48)$  &    $48347(3)$  &  $0.344(2)$  &  $1.295(34)$  &  $713(109)$  &  $2566(872)$  &  $6010(239)$  &  $47(7)$  &  $0.611(1)$ &  2.02 -- 1.81 & 2.2   \\
RY~Vel       &  $28.13729(20)$  &    $53524.9(2)$  &  $0.726(5)$  &  $1.218(23)$  &  $2333(115)$  &  $14904(1619)$  &  $5634(82)$  &  $128(6)$  &  $0.512(1)$ &  1.97 -- 1.86 & 1.5   \\
RZ~Vel       &  $20.390098(57)$  &    $34848.06(9)$  &  $0.42(2)$  &  $1.230(21)$  &  $1595(48)$  &  $9787(615)$  &  $5366(50)$  &  $115(3)$  &  $0.669(3)$ &  3.28 -- 3.06 & 1.7   \\
T~Vel        &  $4.639804(11)$  &    $40713(1)$  &  $0.31(1)$  &  $1.281(29)$  &  $1141(124)$  &  $1553(339)$  &  $5712(66)$  &  $40(4)$  &  $0.329(3)$ &  14.0 -- 6.83 & 25.3 \\
S~Vul        &  $68.5266(47)$  &    $48335(2)$  &  $1.039(4)$  &  $1.188(33)$  &  $4181(152)$  &  $67799(3898)$  &  $5512(45)$  &  $286(7)$  &  $0.6363(8)$ &  2.31 -- 2.32 & 0.6   \\
SV~Vul       &  $44.93754(63)$  &    $48309.2(3)$  &  $0.602(2)$  &  $1.202(28)$  &  $2359(68)$  &  $32781(1261)$  &  $5360(21)$  &  $210(4)$  &  $0.829(1)$ &  2.30 -- 2.30 & 0.7   \\
T~Vul        &  $4.4356696(48)$  &    $41686.8(4)$  &  $0.087(3)$  &  $1.283(29)$  &  $810(45)$  &  $3262(342)$  &  $5943(42)$  &  $54(3)$  &  $0.620(2)$ &  3.99 -- 2.39 & 17.6  \\
U~Vul        &  $7.990765(14)$  &    $48311.2(2)$  &  $0.691(2)$  &  $1.262(23)$  &  $594(27)$  &  $2688(236)$  &  $5869(34)$  &  $50(2)$  &  $0.787(1)$ &  2.20 -- 2.20 & 0.7   \\
			\hline
		\end{tabular}
		\label{table__parameters}
	\end{sidewaystable*}
	
	\begin{table*}[!ht]
		\centering
		\caption{Best-fit output parameters for the power-law IR excess model $a_1 (\lambda - 1.20)^{a_2}$ for Cepheids with a CSE  detected at $> 3\sigma$.}
		\begin{tabular}{ccc|cccccccc}
			\hline
			\hline
			Name  &  $a_1$  &  $a_2$  & $2.2\mu$m  & $3.5\mu$m &  $5.0\mu$m  &  $8.0\mu$m &  $10\mu$m &  $15\mu$m &  $18\mu$m &  $25\mu$m  \\
			&  (mag)  &  &  (mag)   &  (mag)   &  (mag)   &  (mag)   &  (mag)  &  (mag)  &  (mag) &  (mag)     \\
			\hline
$\eta$~Aql   &  $0.054(5)$  &  $0.35(5)$  &  $0.054(5)$  &  $0.072(7)$  &  $0.086(8)$  &  $0.11(1)$  &  $0.12(1)$  &  $0.14(1)$  &  $0.15(1)$  &  $0.17(2)$   \\
RT~Aur      &  $0.051(7)$  &  $0.25(4)$  &  $0.051(7)$  &  $0.063(9)$  &  $0.07(1)$  &  $0.08(1)$  &  $0.09(1)$  &  $0.10(1)$  &  $0.11(1)$  &  $0.11(2)$   \\
U~Car       &  $0.078(7)$  &  $0.18(2)$  &  $0.078(7)$  &  $0.091(8)$  &  $0.099(9)$  &  $0.11(1)$  &  $0.12(1)$  &  $0.13(1)$  &  $0.13(1)$  &  $0.14(1)$   \\
CF~Cas      &  $0.15(2)$  &  $0.21(2)$  &  $0.15(2)$  &  $0.18(2)$  &  $0.20(2)$  &  $0.22(3)$  &  $0.24(3)$  &  $0.26(3)$  &  $0.27(3)$  &  $0.29(4)$   \\
V~Cen       &  $0.10(3)$  &  $0.14(3)$  &  $0.10(3)$  &  $0.11(3)$  &  $0.12(3)$  &  $0.13(3)$  &  $0.14(3)$  &  $0.14(4)$  &  $0.15(4)$  &  $0.16(4)$   \\
$\delta$~Cep  &  $0.012(3)$  &  $0.7(1)$  &  $0.012(3)$  &  $0.021(6)$  &  $0.030(8)$  &  $0.05(1)$  &  $0.05(2)$  &  $0.07(2)$  &  $0.09(2)$  &  $0.11(3)$   \\
CD~Cyg      &  $0.04(1)$  &  $0.27(6)$  &  $0.04(1)$  &  $0.05(1)$  &  $0.06(2)$  &  $0.07(2)$  &  $0.08(2)$  &  $0.09(2)$  &  $0.09(2)$  &  $0.10(3)$   \\
$\zeta$~Gem  &  $0.12(2)$  &  $0.14(2)$  &  $0.12(2)$  &  $0.14(3)$  &  $0.15(3)$  &  $0.16(3)$  &  $0.17(3)$  &  $0.18(3)$  &  $0.18(4)$  &  $0.19(4)$   \\
S~Nor       &  $0.056(5)$  &  $0.23(3)$  &  $0.056(5)$  &  $0.068(7)$  &  $0.076(7)$  &  $0.087(8)$  &  $0.092(9)$  &  $0.10(1)$  &  $0.11(1)$  &  $0.12(1)$   \\
V340~Nor    &  $0.110(5)$  &  $0.08(2)$  &  $0.110(5)$  &  $0.118(5)$  &  $0.123(5)$  &  $0.129(5)$  &  $0.132(6)$  &  $0.137(6)$  &  $0.140(6)$  &  $0.144(6)$   \\
ST~Tau      &  $0.15(2)$  &  $0.28(4)$  &  $0.15(2)$  &  $0.19(3)$  &  $0.22(3)$  &  $0.26(3)$  &  $0.28(4)$  &  $0.31(4)$  &  $0.33(4)$  &  $0.36(5)$   \\
T~Vel       &  $0.09(2)$  &  $0.33(6)$  &  $0.09(2)$  &  $0.11(3)$  &  $0.14(3)$  &  $0.16(4)$  &  $0.18(4)$  &  $0.21(5)$  &  $0.22(5)$  &  $0.25(6)$   \\
T~Vul       &  $0.028(7)$  &  $0.40(7)$  &  $0.028(7)$  &  $0.038(9)$  &  $0.05(1)$  &  $0.06(1)$  &  $0.07(2)$  &  $0.08(2)$  &  $0.09(2)$  &  $0.10(2)$   \\
			\hline
Agv:  & -- & --  &  $0.08(4)$  &  $0.10(5)$  &$0.11(5)$  &$0.13(6)$  &$0.13(6)$  &$0.15(7)$  &$0.16(7)$  &$0.17(8)$ \\

			\hline
		\end{tabular}
		\label{table__excess}
		\end{table*}

		\begin{table}[!ht]
		\centering
		\caption{Upper-limit magnitudes for the presence of a CSE at 10\,$\mu$m.}
		\begin{tabular}{cc}
			\hline
			\hline
			Name  &  10\,$\mu$m upper limit  \\
						&  (mag)    \\
			\hline
			FF~Aql  &  0.047  \\		
			FM~Aql  &  0.087   \\		
			FN~Aql  &  0.103    \\		
			SZ~Aql  &  0.061    \\
			TT~Aql  &  0.026  \\
			U~Aql  &  0.007  \\
			$\ell$~Car  &  0.043  \\		
			V~Car  &  0.032  \\
			VY~Car  &  0.022  \\
			DL~Cas  &  0.017  \\		
			XX~Cen  &  0.076  \\		
			X~Cyg  &  0.021  \\
			$\beta$~Dor &  0.053  \\		
			CV~Mon  &  0.065  \\
			T~Mon  &  0.027  \\		
			TW~Nor  &  0.131  \\
			RY~Sco  &  0.030  \\
			RU~Sct  &  0.045  \\
			GY~Sge  &  0.006  \\
			S~Sge  &  0.015  \\
			BB~Sgr  &  0.034  \\
			U~Sgr  &  0.026  \\
			W~Sgr  &  0.034  \\		
			WZ~Sgr  &  0.017  \\
			X~Sgr  &  0.046  \\
			Y~Sgr  &  0.036  \\
			SZ~Tau & 0.045  \\
			RY~Vel  &  0.022  \\
			RZ~Vel  &  0.042  \\
			S~Vul &  0.026  \\
			SV~Vul  &  0.057  \\
			U~Vul  &  0.050  \\
			\hline
		\end{tabular}
		\label{table__upper_limit_mag}
	\end{table}

	\begin{table*}[!ht]
	\centering
	\caption{Same as Table~\ref{table__excess}, but for CSE detection $< 3\sigma$ threshold.}
	\begin{tabular}{ccc|cccccccc}
		\hline
		\hline
		Name  &  $a_1$  &  $a_2$  & $2.2\mu$m  & $3.5\mu$m &  $5.0\mu$m  &  $8.0\mu$m &  $10\mu$m &  $15\mu$m &  $18\mu$m &  $25\mu$m  \\
		&  (mag)  &  &  (mag)   &  (mag)   &  (mag)   &  (mag)   &  (mag)  &  (mag)  &  (mag)  & (mag)   \\
		\hline
FF~Aql      &  $0.033(8)$  &  $0.4(1)$  &  $0.033(8)$  &  $0.05(1)$  &  $0.06(1)$  &  $0.07(2)$  &  $0.08(2)$  &  $0.09(2)$  &  $0.10(2)$  &  $0.12(3)$   \\
FM~Aql      &  $0.03(2)$  &  $0.19(8)$  &  $0.03(2)$  &  $0.04(2)$  &  $0.04(2)$  &  $0.05(2)$  &  $0.05(3)$  &  $0.06(3)$  &  $0.06(3)$  &  $0.06(3)$   \\
FN~Aql      &  $0.03(1)$  &  $0.2(1)$  &  $0.03(1)$  &  $0.04(2)$  &  $0.04(2)$  &  $0.05(2)$  &  $0.05(2)$  &  $0.06(3)$  &  $0.06(3)$  &  $0.07(3)$   \\
SZ~Aql      &  $0.01(2)$  &  $0.1(2)$  &  $0.01(2)$  &  $0.01(2)$  &  $0.01(2)$  &  $0.01(2)$  &  $0.01(2)$  &  $0.01(2)$  &  $0.01(2)$  &  $0.01(2)$   \\
TT~Aql      &  $0.01(2)$  &  $0.2(3)$  &  $0.01(2)$  &  $0.01(3)$  &  $0.01(3)$  &  $0.02(3)$  &  $0.02(4)$  &  $0.02(4)$  &  $0.02(4)$  &  $0.02(4)$   \\
U~Aql       &  $0.02(1)$  &  $0.1(1)$  &  $0.02(1)$  &  $0.02(1)$  &  $0.02(1)$  &  $0.02(2)$  &  $0.02(2)$  &  $0.02(2)$  &  $0.02(2)$  &  $0.02(2)$   \\
$\ell$~Car  &  $0.03(1)$  &  $0.3(1)$  &  $0.03(1)$  &  $0.04(1)$  &  $0.05(1)$  &  $0.05(1)$  &  $0.06(1)$  &  $0.06(1)$  &  $0.07(1)$  &  $0.07(1)$   \\
V~Car       &  $0.005(7)$  &  $0.7(5)$  &  $0.01(1)$  &  $0.01(1)$  &  $0.01(2)$  &  $0.02(2)$  &  $0.02(3)$  &  $0.03(4)$  &  $0.03(4)$  &  $0.04(6)$   \\
DL~Cas      &  $0.05(1)$  &  $0.2(1)$  &  $0.05(1)$  &  $0.06(1)$  &  $0.07(1)$  &  $0.07(1)$  &  $0.08(2)$  &  $0.09(2)$  &  $0.09(2)$  &  $0.10(2)$   \\
XX~Cen      &  $0.05(1)$  &  $0.21(5)$  &  $0.05(1)$  &  $0.06(1)$  &  $0.06(1)$  &  $0.07(2)$  &  $0.08(2)$  &  $0.08(2)$  &  $0.09(2)$  &  $0.09(2)$   \\
X~Cyg       &  $0.03(1)$  &  $0.32(8)$  &  $0.03(1)$  &  $0.04(1)$  &  $0.05(1)$  &  $0.06(2)$  &  $0.06(2)$  &  $0.07(2)$  &  $0.08(2)$  &  $0.09(2)$   \\
$\beta$~Dor  &  $0.06(2)$  &  $0.05(2)$  &  $0.06(2)$  &  $0.06(2)$  &  $0.07(2)$  &  $0.07(2)$  &  $0.07(2)$  &  $0.07(2)$  &  $0.07(2)$  &  $0.07(2)$   \\
CV~Mon      &  $0.17(3)$  &  $0.11(2)$  &  $0.17(3)$  &  $0.19(4)$  &  $0.20(4)$  &  $0.21(4)$  &  $0.21(4)$  &  $0.23(4)$  &  $0.23(4)$  &  $0.24(5)$   \\
T~Mon       &  $0.05(1)$  &  $0.18(3)$  &  $0.05(1)$  &  $0.06(1)$  &  $0.07(1)$  &  $0.08(1)$  &  $0.08(1)$  &  $0.09(1)$  &  $0.09(1)$  &  $0.10(1)$   \\
RY~Sco      &  $0.05(2)$  &  $0.01(5)$  &  $0.05(2)$  &  $0.05(2)$  &  $0.05(2)$  &  $0.05(2)$  &  $0.05(2)$  &  $0.05(2)$  &  $0.05(2)$  &  $0.05(2)$   \\
S~Sge       &  $0.03(1)$  &  $0.5(2)$  &  $0.03(1)$  &  $0.05(1)$  &  $0.06(2)$  &  $0.08(2)$  &  $0.09(3)$  &  $0.12(3)$  &  $0.13(4)$  &  $0.16(4)$   \\
BB~Sgr      &  $0.06(2)$  &  $0.11(3)$  &  $0.06(2)$  &  $0.07(2)$  &  $0.07(2)$  &  $0.08(3)$  &  $0.08(3)$  &  $0.08(3)$  &  $0.09(3)$  &  $0.09(3)$   \\
U~Sgr       &  $0.02(1)$  &  $0.12(8)$  &  $0.02(1)$  &  $0.02(2)$  &  $0.02(2)$  &  $0.02(2)$  &  $0.02(2)$  &  $0.02(2)$  &  $0.02(2)$  &  $0.02(2)$   \\
W~Sgr       &  $0.22(1)$  &  $0.08(1)$  &  $0.22(1)$  &  $0.23(1)$  &  $0.24(1)$  &  $0.25(1)$  &  $0.26(1)$  &  $0.27(1)$  &  $0.27(1)$  &  $0.28(1)$   \\
X~Sgr       &  $0.11(1)$  &  $0.06(1)$  &  $0.11(1)$  &  $0.11(1)$  &  $0.12(1)$  &  $0.12(1)$  &  $0.12(1)$  &  $0.12(1)$  &  $0.13(1)$  &  $0.13(1)$   \\
Y~Sgr       &  $0.01(2)$  &  $0.4(3)$  &  $0.01(2)$  &  $0.02(2)$  &  $0.02(3)$  &  $0.03(4)$  &  $0.03(4)$  &  $0.04(5)$  &  $0.04(5)$  &  $0.05(6)$   \\
SZ~Tau      &  $0.06(1)$  &  $0.17(4)$  &  $0.06(1)$  &  $0.07(2)$  &  $0.08(2)$  &  $0.09(2)$  &  $0.09(2)$  &  $0.10(2)$  &  $0.10(2)$  &  $0.11(2)$   \\
RY~Vel      &  $0.13(2)$  &  $0.16(4)$  &  $0.13(2)$  &  $0.15(2)$  &  $0.16(3)$  &  $0.18(3)$  &  $0.18(3)$  &  $0.20(3)$  &  $0.20(3)$  &  $0.21(4)$   \\
RZ~Vel      &  $0.08(2)$  &  $0.17(3)$  &  $0.08(2)$  &  $0.09(2)$  &  $0.10(2)$  &  $0.11(2)$  &  $0.12(2)$  &  $0.13(3)$  &  $0.13(3)$  &  $0.14(3)$   \\

		\hline
	\end{tabular}
	\label{table__excess_rejected}
	\end{table*}

\begin{figure}[!h]
		\resizebox{\hsize}{!}{\includegraphics{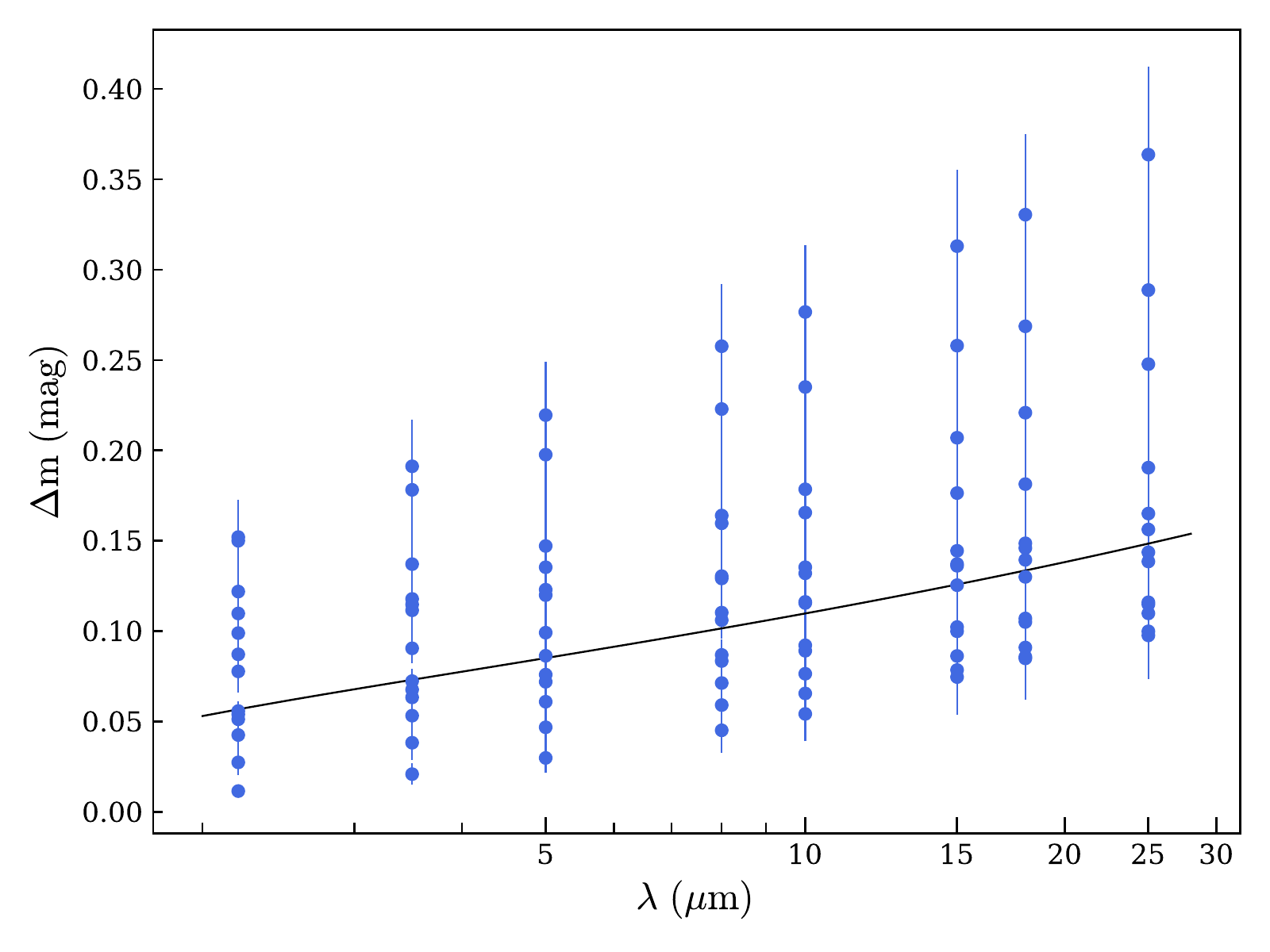}}
		\caption{Relative IR excess with respect to the observing wavelength using the values listed in Table~\ref{table__excess}.}
		\label{figure__table_excess}
	\end{figure}

	We listed the derived CSE parameters for the Cepheids for which we have a detection $> 3\sigma$ in Table~\ref{table__excess}, together with interpolated excess at 2.2, 3.5, 5, 8, 10, 15, 18, and $25\mu$m, corresponding to wavelengths available at the James Webb Space Telescope (JWST) instruments. From this sample, we also estimated the average excess in these bands, using the standard deviation at a given wavelength as uncertainty. The IR excess models for these Cepheids are displayed in Figs.~\ref{image__spips_excess1}, \ref{image__spips_excess2}, and \ref{image__spips_excess3}. We also represent the values of Table~\ref{table__excess} in Fig.~\ref{figure__table_excess}, and we fitted the same power-law model defined in Eq.~\ref{equation_power_law_model}. In order to interpolate or extrapolate at any wavelength, we can use the following fitted relation for $\lambda > 1.2\,\mu$m:
	
	\begin{displaymath}
		\Delta m = 0.057_{\pm0.001}\ (\lambda - 1.20)^{0.303_{\pm0.011}},\ \sigma = 0.062\,mag.
	\end{displaymath}
	
	The Cepheid \object{CE~Cas} was rejected from our SPIPS analysis -as this is a particular case of two orbiting Cepheids- as well as \object{V367~Sct} which is a double-mode Cepheid. We also excluded Y~Oph from the following analysis because we noticed strange behaviour among the data, meaning that while all interferometric measurements are in agreement with each other (from various instruments), they are not very consistent with the photometry. The cause is still under investigation and is possibly related to a physical phenomenon we do not understand yet. We note that \citet{Abt_1954_04_0} already highlighted anomalies for this Cepheid, with its light, colour, and RVs being different to those of other Cepheids with the same period.
	
	
	\section{Discussion}
	\label{section__discussion}
	
	\defcitealias{Groenewegen_2020_03_0}{Gr20}
	\defcitealias{Hocde_2020_01_0}{Ho20}
	\defcitealias{Barmby_2011_11_0}{Ba11}
	\defcitealias{Breitfelder_2016_03_0}{Br16}
	\defcitealias{Kervella_2009_05_0}{Ke09}
	
	Our SPIPS analysis shows that at least 29\,\% (13/45) of our Cepheids have a moderate to large IR excess. The average values are $\Delta m_{K} \sim 0.08$\,mag, $\Delta m_{5\mu \mathrm{m}} \sim 0.11$\,mag, $\Delta m_{10\mu \mathrm{m}} \sim 0.13$\,mag and $\Delta m_{20\mu \mathrm{m}} \sim 0.16$\,mag. Although the physical nature is still not well understood, the main explanation of these IR excesses is the presence of circumstellar envelopes. This supports our previous works suggesting that some Cepheids might harbour a circumstellar environment. Observations already confirmed their existence around a few Cepheids \citep[see e.g.][]{Kervella_2006_03_0,Kervella_2009_05_0,Marengo_2009_01_0,Gallenne_2012_02_0,Nardetto_2016_09_0,Hocde_2020_01_0}. \citet{Kervella_2006_03_0} suggested that CSEs may be created via a mass-loss process enhanced by the pulsation and convection mechanism. This scenario is supported by theoretical models including pulsation and shocks in the atmosphere of the Cepheids. \citet{Neilson_2008_09_0} provided a consistent match between an analytical model of mass loss and the observations of IR excess. Although the knowledge on how these CSEs form is important for our understanding of the Cepheid evolution, we only focused here on the impact of the IR excess on the Cepheid photometry. Our SPIPS analysis using broad-band photometry does not allow us to constrain the formation of these CSEs or the dust composition. For the Cepheids that are below our chosen CSE detection level (i.e. $3\sigma$), upper limit magnitudes at 10\,$\mu$m were estimated for the presence of CSEs from the standard deviation of the residuals, and they are listed in Table~\ref{table__upper_limit_mag}. We cannot exclude IR excess at longer wavelengths as no precise photometry is available.

	\begin{figure}[!h]
	\resizebox{\hsize}{!}{\includegraphics{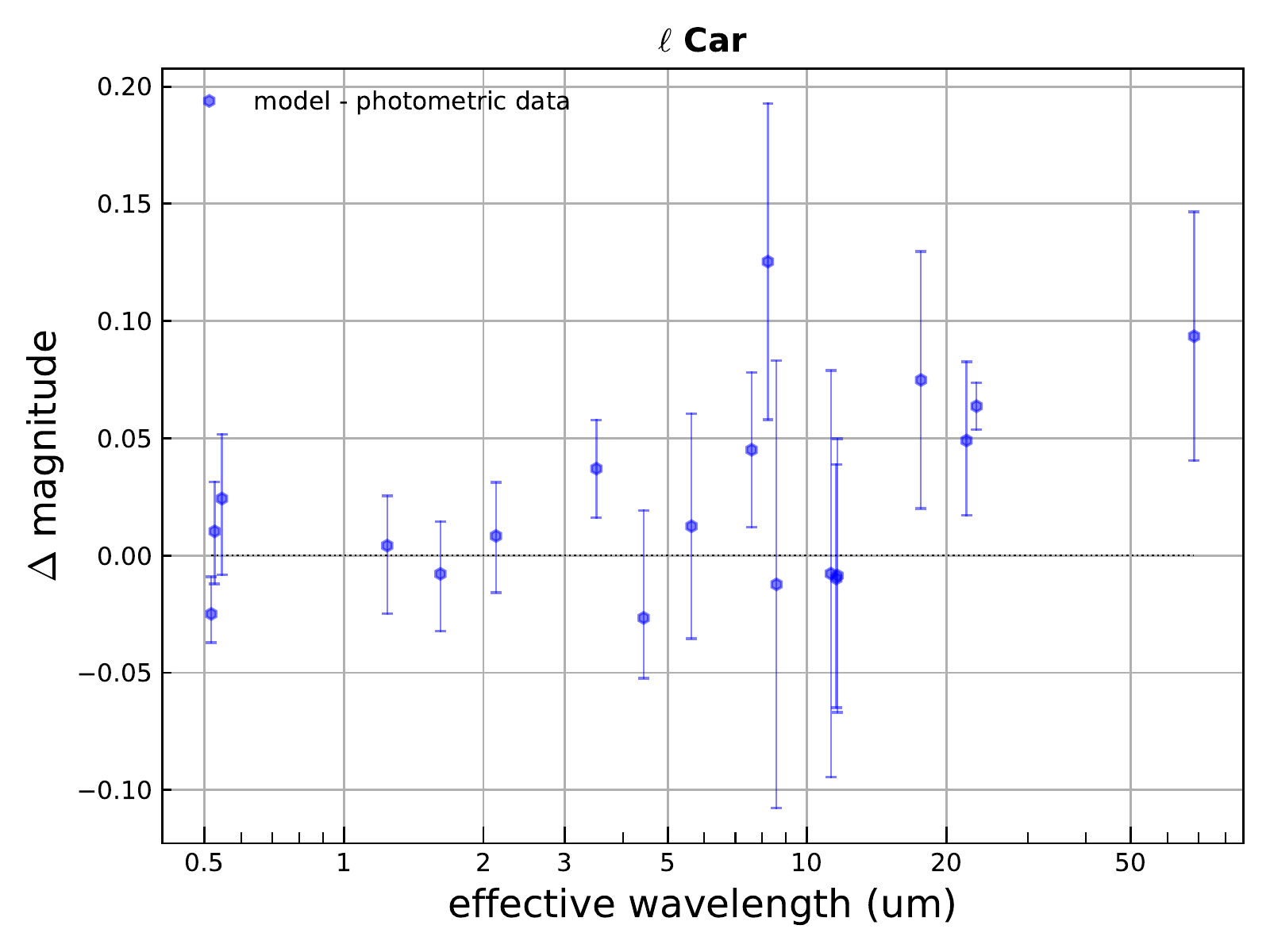}}
	\resizebox{\hsize}{!}{\includegraphics{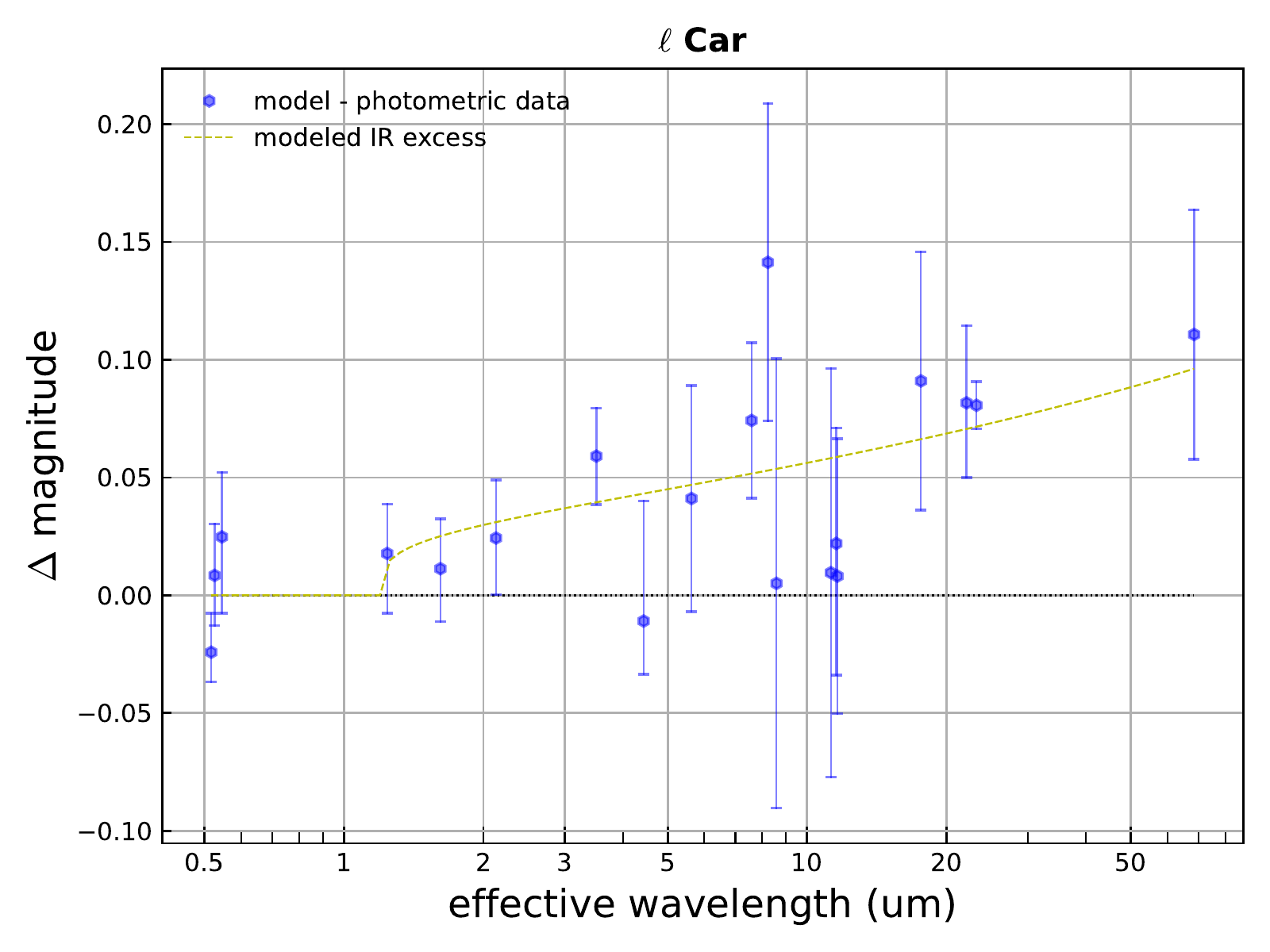}}
	\caption{SPIPS model of $\ell$~Car without (top) and with (bottom) modelling of the IR excess. The r.m.s. of the residuals is 0.043\,mag and 0.035\,mag, respectively, for the model without and with IR excess modelling.}
	\label{image__no_excess_lcar}
	\end{figure}

	\subsection{Comparison with previous works}

	\begin{figure*}[!h]
	\resizebox{\hsize}{!}{\includegraphics{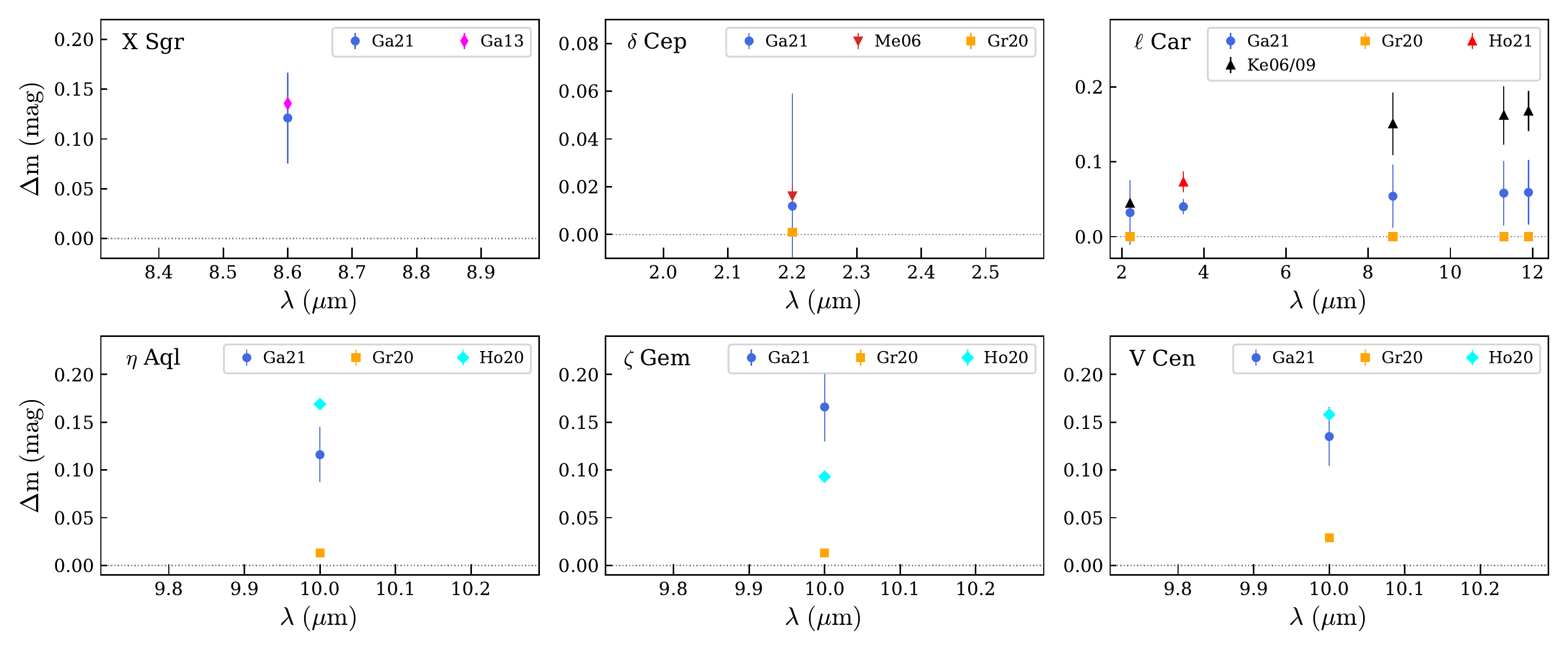}}
	\caption{Comparison of our estimated IR excess with the literature. References are: Ga21: this work, Ga13: \citet{Gallenne_2013_10_0}, Me06: \citet{Merand_2006_07_0}, Gr20: \citet{Groenewegen_2020_03_0}, Ho20: \citet{Hocde_2020_01_0}, Ho21: \citet{Hocde_2021_03_6}. All points from this work are for the model including IR excess. Estimates from \citet{Hocde_2020_01_0} are corrected from their offset (see text). No error bars are listed for the values from \citet{Groenewegen_2020_03_0} as no error estimates are provided in the paper.}
	\label{image__comparison_literature}
	\end{figure*}

	Some of the Cepheids for which a CSE or IR excess was previously reported are below our chosen detection level threshold. As explained previously and according to our statistical criteria defined in Sect.~\ref{section__spips_analysis}, the IR photometric precision is not high enough to strongly constrain faint CSEs. Therefore, although a CSE model might be more adequate for a given Cepheid, low precision photometry prevents a high detection level compared to a model without an envelope. An example is displayed in Fig.~\ref{image__no_excess_lcar} for $\ell$~Car (CSE model significant at $1.8\sigma$ only), where we compare the residuals with or without fitting the IR excess.
	
	Although we rejected the CSE model in some cases, we can still compare them with previous works to check if they are really inconsistent. Interestingly, we found that the CSE flux contributions given by our models provides similar results. The IR excess models for the Cepheids which do not comply with our detection threshold (i.e. $< 3\sigma$) are listed in Table~\ref{table__excess_rejected}. For comparison, we interpolated our IR excess law to the given wavelength (integrating over a filter bandpass does not change the results at the precision level we have), and we took the average standard deviation of the residuals as uncertainties to be conservative. \citet{Gallenne_2012_02_0,Gallenne_2013_10_0} reported a relative CSE contribution of $f = 13.3\pm0.5$\,\% at 8.6\,$\mu$m for the Cepheid X~Sgr, which would be in good agreement with our calculated value of $11.8\pm4.7$\,\%. \citet{Merand_2006_07_0} measured $f = 1.5\pm0.4$\,\% for $\delta$~Cep from $K$-band interferometric measurements, which ia within $1\sigma$ of our estimate $f = 1.1\pm4.4$\,\%. Extended emission was also reported around this Cepheid by \citet{Marengo_2010_12_0} at 5.8, 8.0, 24, and 70\,$\mu$m, possibly originating from the wind of the star and/or its companion, but no excess is stated by the authors. \citet{Kervella_2006_03_0} reported $f = 4.2\pm0.2$\,\% for $\ell$~Car, also from $K$-band interferometry, and would be in agreement with our value of $3.0\pm4.1$\,\%. The flux contributions of $14.9\pm4.4$\,\%, $16.1\pm4.2$\,\%, and $16.7\pm2.9$\,\% estimated from the SED in the PAH1, PAH2, and SiC filters of VISIR by \citet[][hereafter \citetalias{Kervella_2009_05_0}]{Kervella_2009_05_0} are, however, larger than our estimates of $5.1\pm4.1$\,\%, $5.5\pm4.2$\,\% and $5.6\pm4.2$\,\%, respectively. The difference in the model used and a fixed $E(B-V) = 0.147$ in  \citetalias{Kervella_2009_05_0} could explain the deviation. \citet[][hereafter \citetalias{Barmby_2011_11_0}]{Barmby_2011_11_0} discussed tentative evidence of extended emission around $\ell$~Car, which is consistent with the IR excess trend we find. However, we estimated an excess of $\sim 5$\,\% at 10\,$\mu$m, which would mean that the CSE should be more compact. \citet{Hocde_2021_03_6} recently detected an extended emission in the $L$ band for this Cepheid with the new interferometric instrument MATISSE \citep{Lopez_2014_09_6,Matter_2016_08_0}. They estimated a size of $\sim 1.8\,R_\star$  and an IR flux contribution of $7.0\pm1.4$\,\% from a Gaussian envelope model. This would be similar to our CSE model giving $3.8\pm1.0$\,\%.
	
	\citetalias{Barmby_2011_11_0} examined extended IR emission from \emph{Spitzer} IRAC and MIPS observations (for $\lambda > 8$\,$\mu$m). They reported a possible detection around the Cepheid T~Mon and X~Cyg. We did not detect strong IR excess for these stars. The model without CSE provides upper limits of 0.027\,mag and 0.021\,mag (see Table~\ref{table__upper_limit_mag})), respectively, for T~Mon and X~Cyg, so we cannot exclude the presence of a faint CSE for these Cepheids. \citetalias{Barmby_2011_11_0} also reported an unusual elongated extended object at 7\arcsec\ north-east of the Cepheid SZ~Tau that could contaminate their mid-IR photometry. This object is outside our VISIR field of view, and no significant excess is detected from our analysis. \citetalias{Barmby_2011_11_0} also reported signs of some cirrus clouds from the 8\,$\mu$m images for the Cepheids FF~Aql, S~Nor, T~Mon, U~Car, U~Sgr, V~Cen, VY~Car, and X~Cyg, possibly related to the Galactic latitude. We only report a CSE detection for U~Car and V~Cen; however, such cirrus  would have rather cold material compared to compact CSEs, typically $< 100$\,K. Therefore, the impact on our estimated IR excess should be negligible as our analysis is based on wavelengths $< 20$\,$\mu$m. \citetalias{Barmby_2011_11_0} found no evidence of extended IR emission (nor cirrus) for BB Sgr, U~Aql, V350~Sgr, W~Sgr, $\beta$~Dor, $\eta$~Aql, and $\zeta$~Gem, while we detected IR excess for two of them. This seems to support the hypothesis that most of the CSEs are compact.
	
	\citet[][hereafter \citetalias{Groenewegen_2020_03_0}]{Groenewegen_2020_03_0} studied the average SED of a large sample of Cepheids (477) by modelling the IR excess with a dust radiative transfer code. He reported IR excess for only 21 Cepheids (4\,\%). For some Cepheids in common, we also detected an IR excess, but the values reported are significantly different to ours and those of previous works. For instance, for $\delta$~Cep, \citetalias{Groenewegen_2020_03_0} estimated a 1\,mmag excess in the $K$ band, while \citet{Merand_2006_07_0} measured $\sim 16$\,mmag, and our model gives $\sim 12$\,mmag. \citetalias{Groenewegen_2020_03_0} did not succeed in fitting an IR excess for $\ell$~Car, while \citet{Kervella_2006_03_0} measured $\sim 45$\,mmag excess in the $K$ band. We do not report significant IR excess for this Cepheid, but the CSE model would be in agreement with the measurement of \citet{Kervella_2006_03_0}. \citetalias{Groenewegen_2020_03_0} estimated IR excesses $\Delta N = 0.013$, 0.013, and 0.029\,mag respectively for $\eta$~Aql, $\zeta$~Gem and V~Cen, while we report a more significant excess of $\Delta N = 0.116\pm0.029, 0.166\pm0.036$ and $0.135\pm0.031$\,mag, respectively. The possible  explanations for the disagreements between \citetalias{Groenewegen_2020_03_0}  and our results are the differences in the modelling, with a different set of parameters and observables. First, in our approach we fit the colour excess and the angular diameter, which is not the case in \citetalias{Groenewegen_2020_03_0}. Secondly, we have effective temperatures and angular diameters (for some stars) as additional observables, which add more constraints. In addition, \citetalias{Groenewegen_2020_03_0} used pre-calculated averaged photometry and single-epoch NIR data, which may be biased by a phase mismatch, while we performed a time-dependent analysis. Furthermore, the fit quality of \citetalias{Groenewegen_2020_03_0} seems poor with large reduced $\chi^2$ (best is $\chi^2_r  \sim 13$, $\sim 39$ for $\zeta$~Gem and $\sim 31$ for $\ell$~Car). We suspect the SED fit to only be constrained by a few data points with very small error bars (see SED plots in his paper), and it may lead to biased estimates of the IR excesses and would explain the large $\chi^2_r$.
	
	For five Cepheids (RS~Pup, $\zeta$~Gem, $\eta$~Aql, V~Cen, and SU~Cyg), we also noted differences between the CSE analyses of \citet[][hereafter \citetalias{Hocde_2020_01_0}]{Hocde_2020_01_0} and \citetalias{Groenewegen_2020_03_0} who used the same mid-IR \emph{Spitzer} spectra. \citetalias{Hocde_2020_01_0} performed three different analyses: a SPIPS analysis similar to our work with an ad hoc CSE model without the \emph{Spitzer} spectra, a SPIPS analysis including the radiative transfer code \emph{DUSTY} \citep{Ivezic_1999_11_0} considering only \emph{Spitzer} spectra for $\lambda > 5\,\mu$m, and a third analysis with a CSE modelled as a thin shell of ionised gas. The five Cepheids they studied all present an IR excess, with an average value of $\sim 0.08$\,mag, as \citetalias{Groenewegen_2020_03_0} but with smaller excess values. The main differences between the works are the CSE models used and a fitted constant offset for $\lambda <1.2\,\mu$m allowed by  \citetalias{Hocde_2020_01_0} for the thin ionised shell model (i.e. a deficit in the visible due to ionised shell absorption). \citetalias{Groenewegen_2020_03_0} proposed a mixture of metallic iron, warm silicates, and compact aluminium oxide, while \citetalias{Hocde_2020_01_0} suggested a thin shell of ionised gas. We also noticed that the magnitude of the excesses we estimated here is also larger than the values given by the ionised shell model of \citetalias{Hocde_2020_01_0}. At 10\,$\mu$m for $\eta$~Aql, \citetalias{Hocde_2020_01_0} estimated an excess of 0.037\,mag, while here we find $0.099\pm0.029$\,mag. However, the ionised shell model allowed an offset for the visible of $-0.132$\,mag, which shifted the average curve down. The same difference is observed for V~Cen and $\zeta$~Gem; at 10\,$\mu$m, we estimated an IR excess of $0.135\pm0.031$\,mag and $0.166\pm0.036$\,mag, respectively, while \citetalias{Hocde_2020_01_0} evaluated  0.101\,mag with $-0.057$\.mag offset and 0.034\,mag with $-0.059$\,mag offset, respectively. The simple SPIPS model used by \citetalias{Hocde_2020_01_0} using the same analytical formulae to model CSEs gives, nevertheless, results similar to ours. \citet[][hereafter \citetalias{Breitfelder_2016_03_0}]{Breitfelder_2016_03_0} performed a SPIPS analysis of nine Cepheids to derive the $p$-factor with a fixed distance. \citetalias{Breitfelder_2016_03_0} also allowed some IR excess in $H$ and $K$ bands, however they are in general lower than our results here for some Cepheids. There are several explanations for this. First, \citetalias{Breitfelder_2016_03_0} only focused on the $p$-factor determination, and the $H$ and $K$ band excess were simply parametrised with an offset (with the excess in $H$ being half the excess in $K$). Second, \citetalias{Breitfelder_2016_03_0} used photometry from $B$ to $K$ bands only, and the stellar effective temperature as observable was not implemented in SPIPS at that time. In the present work, we added the stellar temperature, which provides more constraints on the photosphere and IR photometry at longer wavelengths, which better constrains the excess. 
	
	In Fig.~\ref{image__comparison_literature}, we graphically summarised the comparison made with the literature for the Cepheids X~Sgr, $\delta$~Cep, $\ell$~Car, $\eta$~Aql, $\zeta$~Gem, and V~Cen.
	
	\begin{figure*}[!h]
		\resizebox{\hsize}{!}{\includegraphics{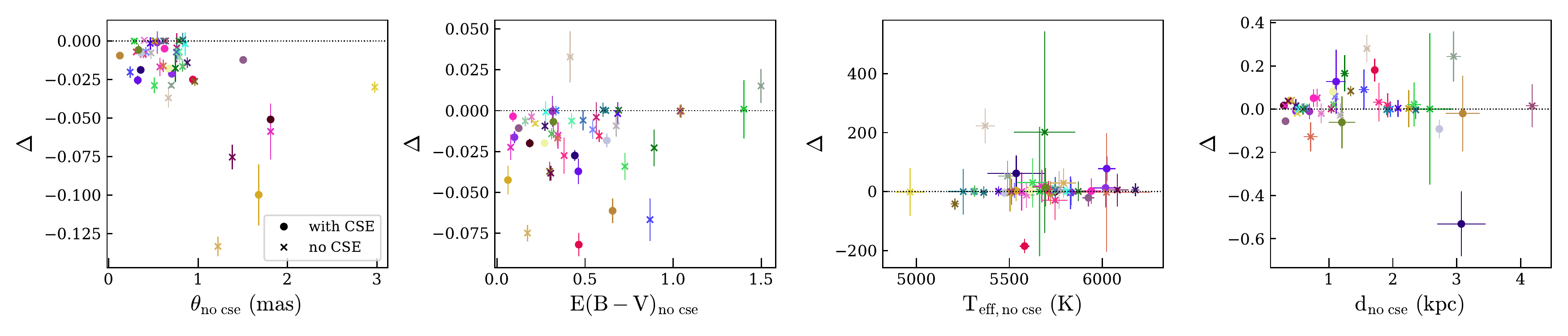}}
		\caption{Comparison of some parameters derived by SPIPS between the models with and without the IR excess law for all the sample. $\Delta$ denotes the difference between the model with CSE and the model without CSE (e.g. for the first panel $\Delta = \theta_\mathrm{with\ cse} - \theta_\mathrm{no\ cse}$). The Cepheids for which we detected a CSE (Table~\ref{table__excess}) are marked with a dot, while others are marked with a cross sign.}
		\label{image__model_comparison}
	\end{figure*}
	
	\begin{figure*}[!h]
		\resizebox{\hsize}{!}{\includegraphics{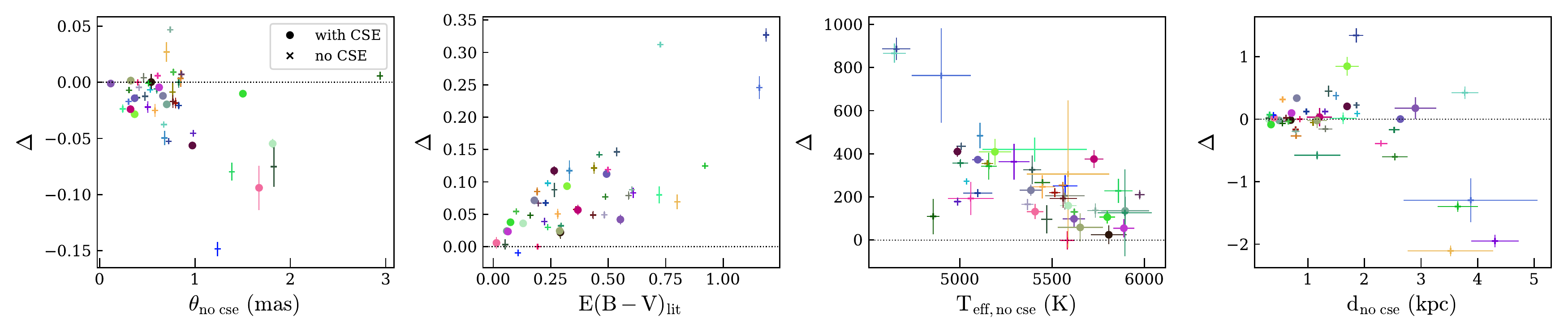}}
		\caption{Same as Fig.~\ref{image__model_comparison}, except that the colour excess is taken from the literature.}
		\label{image__model_comparison_fixed_reddening}
	\end{figure*}
	
	\subsection{Impact of the CSE on the distance estimates from a multi-wavelength PoP analysis}
	
	The presence of IR excess impacts the measured apparent magnitudes, meaning a Cepheid will appear brighter than it actually is. Because such measurements are used to build and calibrate the P-L relations, studying and quantifying the IR excess seems necessary. From our sample  with detected CSE (13 stars), we interpolated a mean IR excess at wavelengths $2.2, 3, 5, 8, 10, 15, 18$, and 25.0\,$\mu$m (wavelengths where the JWST will be operating; see Table~\ref{table__excess}). The given error at each wavelength corresponds to the standard deviation for this sample. We found an average value from 0.08\,mag in $K$ band to 0.13\,mag at 8\,$\mu$m. It is even larger at longer wavelengths with 0.13\,mag  and  0.17\,mag, respectively at 10\,$\mu$m and 25\,$\mu$m. IR wavelengths are usually preferred to build the P-L relations as the scatter decreases with increasing wavelengths \citep[see e.g.][]{Freedman_2010_09_0}, mainly due to the reduced impact of the interstellar extinction and the decreasing effect caused by the temperature (the temperature variation during the pulsation cycle decreases with increasing wavelength). This means that, with the use of a period-luminosity-colour relations, the width in temperature of the instability strip can be neglected at these wavelengths. However, the presence of CSEs and their flux contribution might be a new source of bias. An excess of $\sim 0.08$\,mag in the $K$-band photometry would produce a bias of $\sim 4$\,\% on the distance, while we would have $\sim 7$\,\% at 10-20\,$\mu$m according to our ad hoc analytical law, where the JWST will be operating. To check the effect of the presence and modelling of IR excess, in Fig.~\ref{image__model_comparison} we compare the angular diameters, colour excesses, effective temperatures, and distances estimated by SPIPS with and without modelling of the IR excess for the whole sample. We see that the angular diameter and reddening are the most impacted parameters if no CSE is included, while there seems to be no impact on the temperatures and distances on average. Without modelling the CSE, the IR excess is 'compensated' by increasing the size of the star and its reddening. The most interesting point here is that there is no bias on the distance estimate (mean offset of $16\pm111$\,pc). We can therefore conclude that for a large sample of Cepheids in a given galaxy, the presence of CSEs would not bias the average distance, but it would likely contribute to the scatter.
	
	As an additional test to check for possible biases on the distance estimates, we performed a third SPIPS analysis by fixing $E(B-V)$ and not including a CSE, which we then compared to the CSE model (with fitted colour excess). We retrieved colour excess values mostly from \citet{Turner_2016_10_0}, or otherwise from \citet[][for V~Car and XX~Cen]{Fouque_2007_12_0}. In the literature, all the PoP analyses to determine Cepheid distances use only two photometric bands, fix the colour excess, and ignore possible IR excess (probably less important at $\lambda < 3\,\mu$m). Here we can check the efficiency of SPIPS, which includes a lot more observables to constrain more parameters. In Fig.~\ref{image__model_comparison_fixed_reddening}, we plot the same observables as in Fig.~\ref{image__model_comparison}, but for a fixed colour excess. We clearly see the impact of fixing $E(B-V)$ on the temperatures and distances. As expected, the closest Cepheids are less impacted as they are less affected by interstellar extinction, while the scatter on the distance increases with more distant Cepheids. It is worth mentioning that the fits for most of the Cepheids is poor, particularly for the temperatures and diameters, which cannot be properly constrained. Examples are displayed in Appendix~\ref{appendix__spips_model_fixed_reddening} for three Cepheids. This impact is only problematic for MW Cepheids that are more affected by reddening variations according to the distance and position of the Cepheid. This will be the main limitation of an accurate P-L relation from the next Gaia parallaxes, and a multi-wavelength analysis seems to be necessary to estimate accurate luminosities.
	
	Taking the Magellanic Clouds (MC) as examples of extragalactic cases, \citet{Wielgorski_2017_06_0} already demonstrated that their distances estimated from multi-band photometry ($V, J, H$ and $K_\mathrm{s}$) do not depend on the reddening (assuming the reddening law is correct), although SMC Cepheids are on average 0.045\,mag less reddened than LMC Cepheids. \citet{Inno_2016_12_5} showed, however, that the use of optical wavelengths leads to more systematics than near-IR wavelengths due to the uncertainty of the adopted reddening law. This is why a multi-wavelength approach including the fit of the colour excess as implemented in SPIPS offers a promising route for a precise distance estimate. \citet{Gallenne_2017_11_0} also showed that MC Cepheids have a mean magnitude offset of $\sim 0.06$\,mag at $\sim4\,\mu$m, which could bias the distance estimate by $\sim3$\,\% . However, as previously demonstrated for the Galactic case, the impact on the distance will likely be mitigated by using a multi-band fitting of a large sample of Cepheids, or by restricting the study to wavelengths $< 3\,\mu$m to avoid non-negligible contamination from IR excess. On the other hand, if we assume that all Cepheids have an IR excess emission, the distances estimated from the P-L relations would still be valid as they would be calibrated for a Cepheid plus a CSE, and the presence of IR excess would probably contribute to the dispersion of the relations.
	
	\subsection{Correlation with IR excess}
	
\begin{figure}[!h]
		\resizebox{\hsize}{!}{\includegraphics{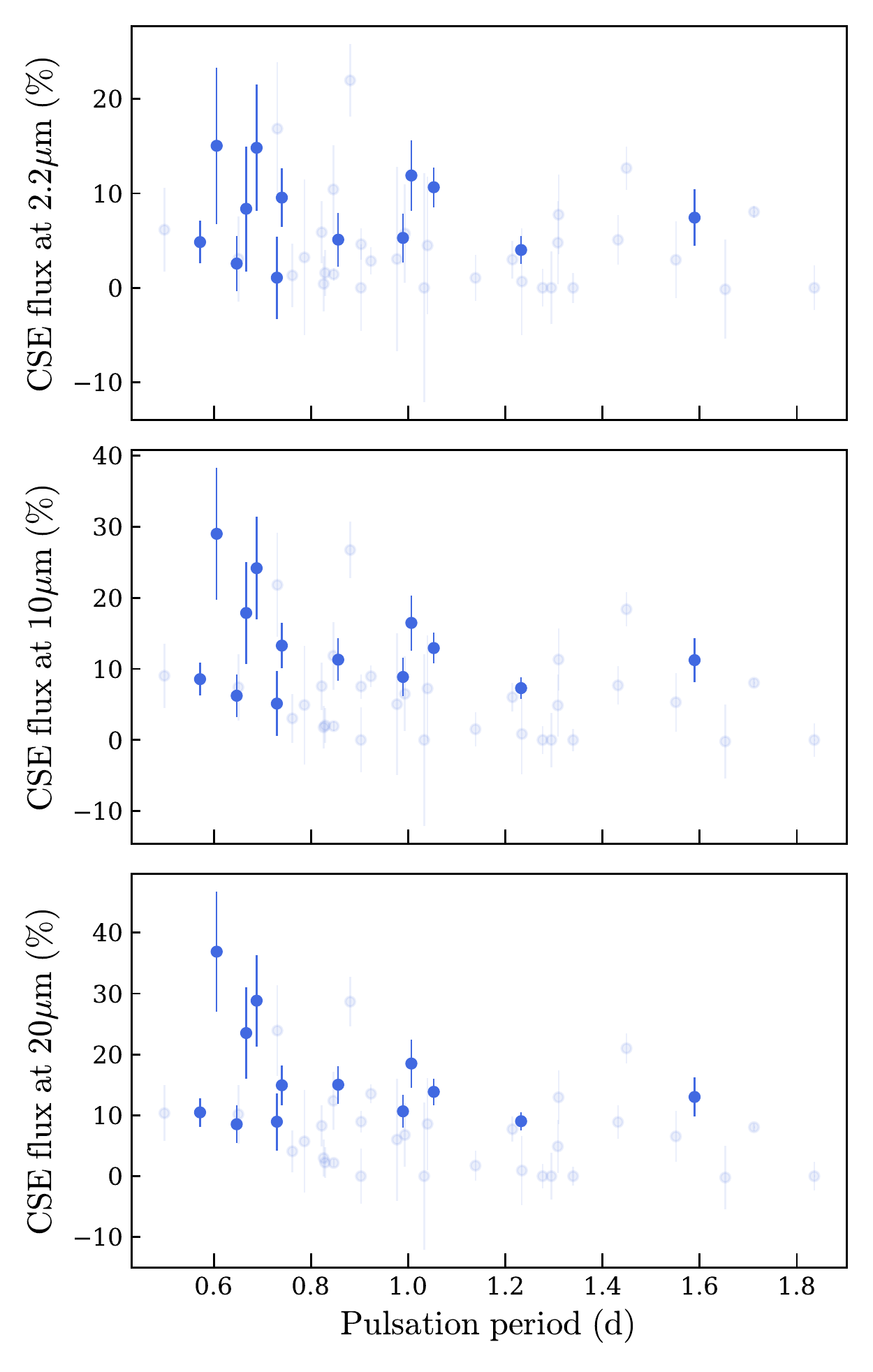}}
		\caption{Estimated relative IR flux at 2.2$\mu$m, 10$\mu$m, and 20$\mu$m with respect to the pulsation period of the Cepheids. The blue dots are the Cepheids for which the IR excess is detected at more than $3\sigma$, and shaded points are detections $<3\sigma$.}
		\label{image__period_excess}
	\end{figure}

	In our previous works on MW Cepheids \citep{Merand_2007_08_0,Gallenne_2012_02_0,Gallenne_2013_10_0}, we noticed a possible correlation between the pulsation period and the relative IR flux ($F_\mathrm{cse}/F_\mathrm{Cepheid}$) at wavelengths 2.2\,$\mu$m and 8.6\,$\mu$m. However, our sample was limited to only eight stars. In \citet{Gallenne_2017_11_0}, we applied a SPIPS analysis for a sample of 29 Magellanic Cloud Cepheids and found a constant offset at 3.6\,$\mu$m and 4.5\,$\mu$m. In Fig.~\ref{image__period_excess}, we display the estimated relative CSE flux at 2.2\,$\mu$m, 10\,$\mu$m and 20\,$\mu$m with respect to the pulsation period for the Cepheids with IR excess detected at more than $3\sigma$ (blue dots). We do not see any specific correlation with the period as we previously suspected for MW Cepheids, but we do see average offsets of $7.7\pm4.3$\,\%, $13.3\pm6.8$\,\%, and $16.3\pm8.3$\,\%, respectively, at 2.2\,$\mu$m, 10\,$\mu$m, and 20\,$\mu$m (standard deviation taken as uncertainties), as found for MC Cepheids \citep{Gallenne_2017_11_0}. This constant trend was also reported in the $K$ band by \citet{Trahin_2019_11_0} from a SPIPS analysis of 72 MW Cepheids ($K$ band was the longest wavelength used for his study). Even when assuming all Cepheids with an IR excess model, there is no correlation (shaded dots in Fig.~\ref{image__period_excess}). We previously formulated the hypothesis that longer period Cepheids (heavier star) may experience higher mass-loss rates, but it does not seem to be the case. Longer period Cepheids with a larger amplitude of radius variation and lower gravity would  have an enhanced mass loss, and therefore more IR excess. This is not what we observe here, although our sample of long periods is limited. This is in disagreement with the theoretical work of \citet{Neilson_2008_09_0} predicting that the mass-loss rate should increase with period, either using only radiation  as the main driving mechanism or a combination of radiative driving and pulsation. The same trend is predicted when including shock effects near the stellar surface in the models to generate the wind, but with a larger scatter than the models without including shocks. However, their models are based on an IR excess produced by cold dusty winds. Other mechanisms such as hot ionised winds might be the origin, as tested by \citetalias{Hocde_2020_01_0}. On the other hand, short-period Cepheids have a longer lifetime inside the instability strip (IS) than long periods, therefore, in the case of a pulsation-driven mass-loss mechanism, short-period Cepheids may accumulate more IR excess compared to long-period ones. However, according to the current prescription from evolutionary models, the mass-loss rate also impacts the lifetime inside the IS by reducing the extent in effective temperature of the blue loop \citep{Matthews_2012_01_0,Bono_2000_11_4}. A high mass-loss rate would significantly shorten the Cepheid's lifetime.
	
	We also checked for possible correlations between the IR excess and the stellar temperature, IR colours, and metallicity, and there is also no clear correlation. The diagram $\mathrm{T_{eff}}$ - IR excess, colour - IR excess and [Fe/H] - IR excess are displayed in Appendix~\ref{image__ir_correllation}.
	
\begin{figure*}[!h]
		\resizebox{\hsize}{!}{\includegraphics{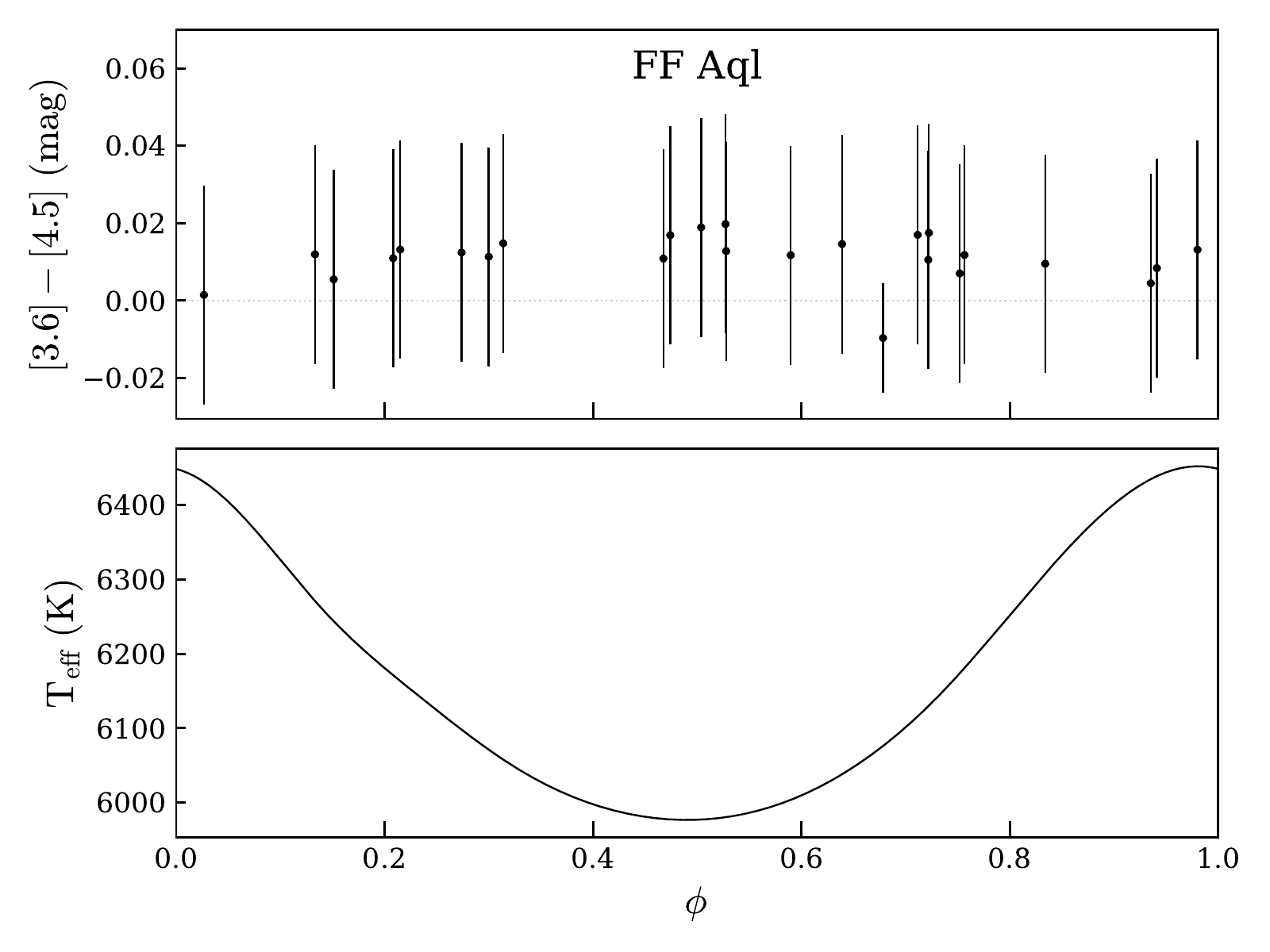}\includegraphics{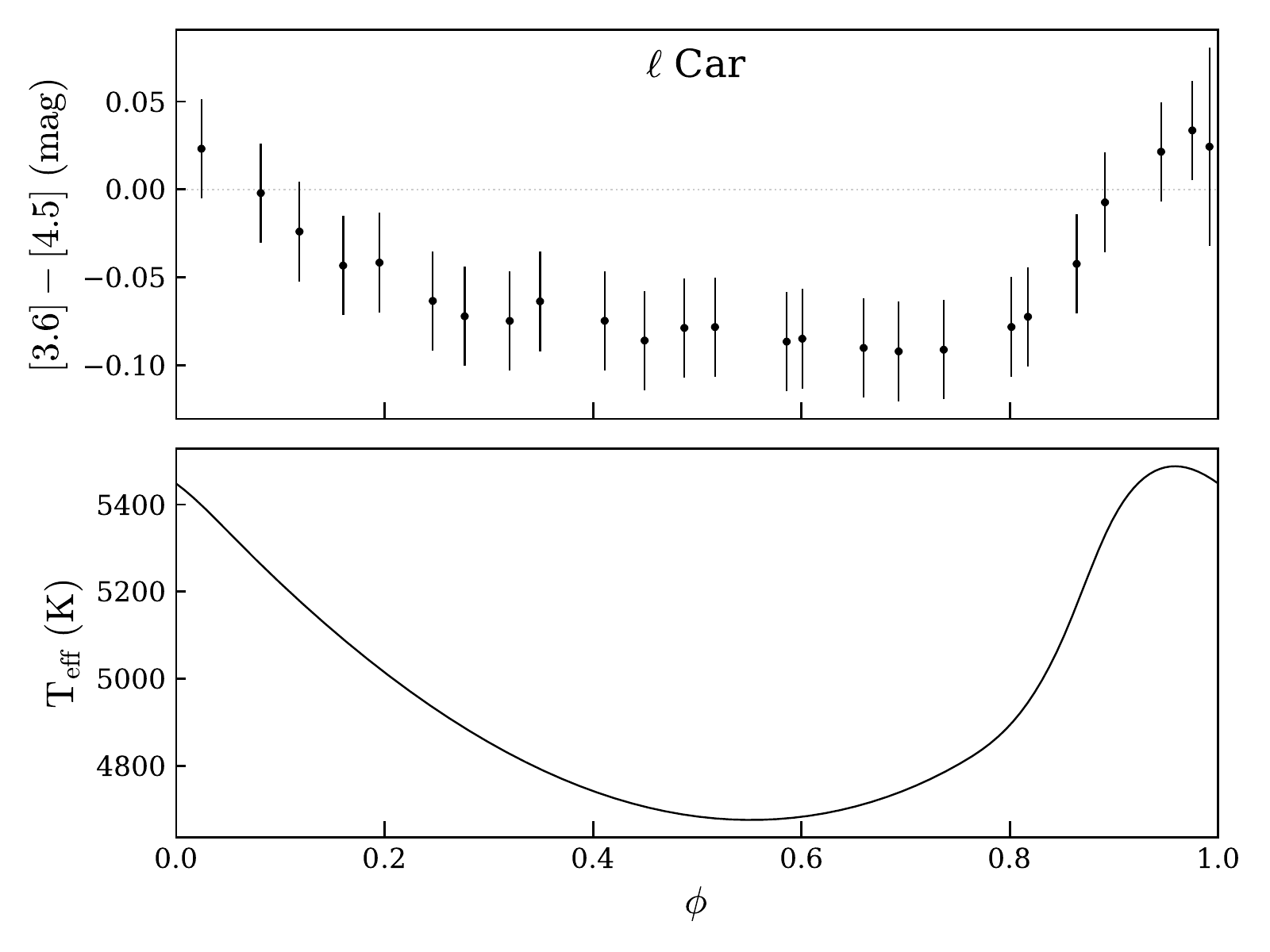}}
		\caption{\emph{Spitzer} colour and stellar effective temperature versus the pulsation phase for the Cepheid FF~Aql and $\ell$~Car.}
		\label{image__CO_cepheids}
	\end{figure*}

\begin{figure}[!h]
		\resizebox{\hsize}{!}{\includegraphics{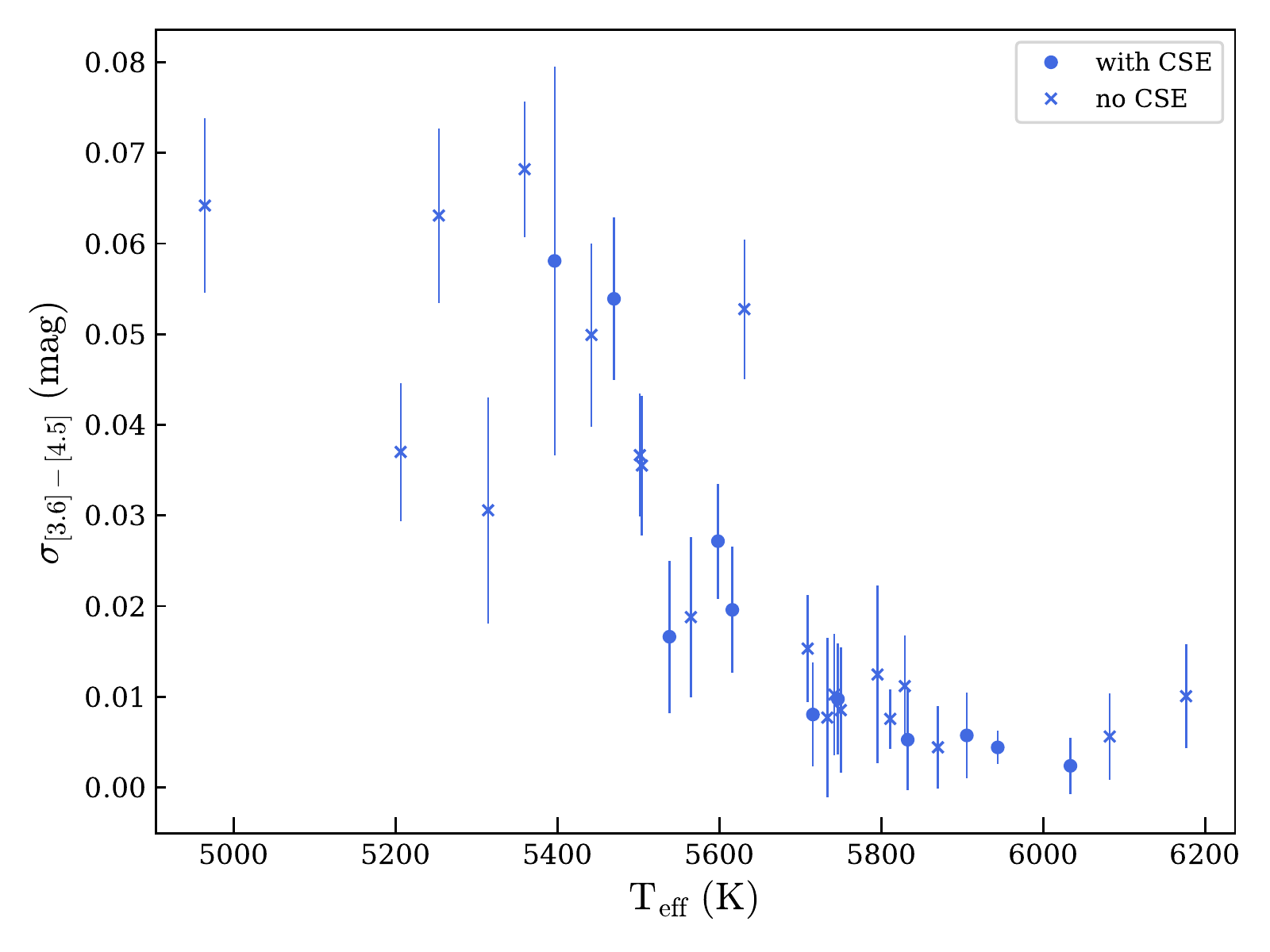}}
		\caption{Colour dispersion-temperature diagram showing that long-period Cepheids (cooler) are more affected by CO absorption than short periods (hotter). Error bars estimated as the standard deviation of the residuals between the observed colour and a Fourier fits.}
		\label{image__CO_correlation}
	\end{figure}

	Finally, we also investigated a possible correlation between these CSEs and the presence of carbon monoxide (CO) reported in previous studies \citep{Marengo_2010_01_0,Scowcroft_2011_12_0,Monson_2012_11_0}. There are cyclic variations in the $[3.6]-[4.5]$ colour induced by the CO absorption. This is because the 4.5\,$\mu$m band is affected by CO molecular absorption which mostly impacts long-period (cooler) Cepheids. This behaviour is pulsation-cycle dependent as the Cepheid becomes hotter during the contraction phase, where the CO dissociates, and cooler during its expansion, where the CO recombines and therefore induces more absorption. This feature is clearly seen in Fig.~\ref{image__CO_cepheids} for the short-period Cepheid FF~Aql (4.4\,d) and the long-period Cepheid $\ell$~Car (35.6\,d). We see that the short period Cepheid is not affected by the CO absorption, while the impact is clearly seen for the long period, and correlated with the photospheric temperature. In Fig.~\ref{image__CO_correlation}, we display the average standard deviation during the pulsation cycle of the colour $[3.6]-[4.5]$ versus the stellar effective temperature for all our sample. As expected, there is a correlation; cooler Cepheids (longer pulsation periods) have the largest scatter. This was also reported by \citet{Monson_2012_11_0} with respect to the pulsation period. However, we do not see any correlation with the presence of CSE. This lead us to the conclusion that the CO molecules are likely to be mostly from the photosphere.
	
	\subsection{Rate of period change, instability strip crossing, and correlation with IR excess}

\begin{figure}[!h]
	\resizebox{\hsize}{!}{\includegraphics{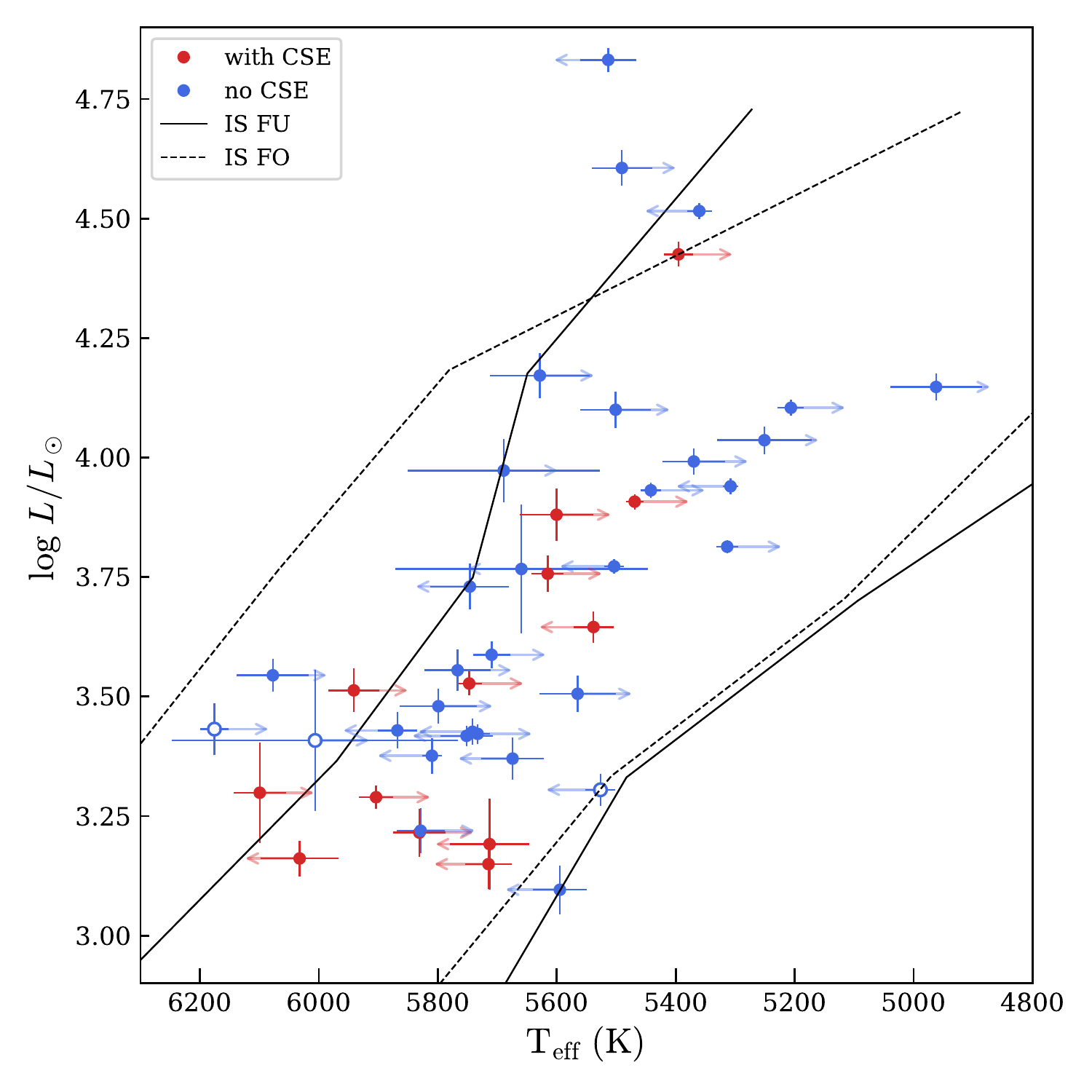}}
	\caption{HR diagram for our sample of Cepheids. The arrow represents the direction the Cepheid evolves through the IS. Fundamental mode Cepheids (FU) are marked with filled circles, while first overtone Cepheids (FO) are marked with open circles.}
	\label{image__hrdiagram}
	\end{figure}

	Our SPIPS analysis also provides an estimate of the rate of period change $\dot{P}$. The algorithm allows us to fit the period as a constant, or including a period change with a polynomial variation with respect to time (as a function of $t - T_0$). This is necessary, especially for long-period Cepheids which often display subtle yet complex period changes over time. Stellar evolution models of Cepheids predict that the period should change: a linear change in time is expected, to the first order, due to the Cepheid slowly changing radius. Interestingly, the rate is different enough for a Cepheid crossing the IS for the first, second, or third time, that the evolution status of an individual star can be determined from the change of period. In order to compare the predicted linear change of period to the observed one, SPIPS performs a linear fit to the fitted polynomial period change in order to estimate an overall linear trend for the observed time range. The rates of period change are listed in Table~\ref{table__period_change}, together with the IS crossing number derived from the model prediction of \citet{Fadeyev_2014_05_0}. We note that the models were calculated for $M_\mathrm{ZAMS} > 6\,M_\odot$, hence we extrapolated for smaller masses so that we could compare with literature values. To be conservative, we estimated $\dot{P}$ for each Cepheid as the mean value between the model with and without IR excess, and used the standard deviation as uncertainty. In most cases, the rate of period change is similar between models for a given Cepheid, except a few cases. This shows that the presence (or lack thereof) of a CSE has no significant impact on the period change in the SPIPS framework. In Fig.~\ref{image__hrdiagram}, we display the HR diagram for our sample, distinguishing the Cepheids with a detected IR excess. We do not see any correlation between the IR excess and the IS crossing number or the position of the Cepheid in the IS. All Cepheids in our sample seem to be in the second or third crossing, except RY~Vel, which would be in the first crossing. This is not surprising as the shell hydrogen burning phase (first crossing) has a shorter lifetime than the core helium burning phase (second and third crossing). We also have three first overtone (FO) Cepheids in our sample (FF~Aql, FN~Aql, and SZ~Tau, open circles in Fig.~\ref{image__hrdiagram}), and no specific correlation is detected for them either.
		
	Previous estimates of the rate of period change are also listed in Table~\ref{table__period_change} for comparison. They are mostly in agreement, particularly for the increasing or decreasing trend. The ones in disagreement might be due to the data used, spanning a different time baseline. Most of the literature data are based on observations roughly from 1900-2000, while we only used data from 1980 due to the lack of precision prior to this date. A switch from period decrease (increase) to period increase (decrease) is also possible, as previously reported for the Cepheid \object(WZ~Car), for example \citep{Turner_2003_01_0}. The case of RT~Aur is also intriguing as we found the same decreasing rate of $-0.14\,\mathrm{s\,yr^{-1}}$ as \citet{Turner_1998_01_0}, while \citet{Turner_2007_11_0} reported an increasing rate of 0.082\,$\mathrm{s\,yr^{-1}}$. Another interesting case is U~Sgr where literature values differ from each other and our estimate. \citet{Turner_1998_01_0} reported $\dot{P} = 0.33\,\mathrm{s\,yr^{-1}}$ corresponding to the star evolving in the third crossing of the IS, while we report a decreasing trend and the Cepheid as being in its second crossing. \citet{Majaess_2013_12_0} found this Cepheid to be in the third crossing, but with a different period evolution of 0.073$\,\mathrm{s\,yr^{-1}}$. However, we see from their O-C diagram (see their Fig.~3), spanning $\sim 140$\,yr of observations, that the residuals with the polynomial fit still exhibits sinusoidal-like variations. This may indicate that the period evolution may occur on a shorter time scale.

	\begin{table*}[!ht]
		\centering
		\caption{Derived rate of period change and IS crossing number. Other measured rates of period change from the literature are also listed in the last column.}
		\begin{tabular}{ccc|c}
			\hline
			\hline
			Name  & $\dot{P}$  & Crossing & Lit. \\
						&  $s~yr^{-1}$   &  & 	$s~yr^{-1}$  \\
			\hline
$\eta$~Aql  &      $0.234\pm0.001$  &  2   & $0.255\pm0.092$\tablefootmark{a}, 0.24\tablefootmark{b} \\
FF~Aql  &         $0.124\pm0.001$  &  3   &  0.08\tablefootmark{b}, $0.072\pm0.011$\tablefootmark{f} \\
FM~Aql  &         $-0.258\pm0.010$  &  3    &  0.01\tablefootmark{b} \\
FN~Aql  &         $-0.447\pm0.083$  &  2   & $-0.37$\tablefootmark{b} \\
SZ~Aql  &         $3.904\pm0.014$  &  3   & 5.72\tablefootmark{b} \\
TT~Aql  &         $-0.752\pm0.006$  &  2   &  0.33\tablefootmark{b}  \\
U~Aql  &         $0.101\pm0.011$  &  3  &  $-0.19$\tablefootmark{b}  \\
RT~Aur  &         $-0.140\pm0.020$  &  2   & $-0.14$\tablefootmark{b}, $0.082\pm0.012$\tablefootmark{g}  \\
$\ell$~Car  &    $16.965\pm0.001$  &  3        & 118.6\tablefootmark{b}  \\
U~Car  &         $78.478\pm1.146$  &  3   &  --   \\
V~Car  &         $-0.687\pm0.013$  &  2    &  --   \\
VY~Car  &         $-35.324\pm0.000$  &  2       & $-75.2$\tablefootmark{c}    \\
CF~Cas  &         $-0.033\pm0.000$  &  2   &  $-0.072$\tablefootmark{d}  \\
DL~Cas  &         $0.569\pm0.006$  &  3   & 0.02\tablefootmark{b}, $0.109\pm0.037$\tablefootmark{i} \\
XX~Cen  &         $-4.878\pm0.004$  &  2   & $-4.47$\tablefootmark{b}  \\
V~Cen  &         $0.152\pm0.007$  &  3  &  --  \\
$\delta$~Cep  &      $0.067\pm0.006$  &  3    & $-0.1006\pm0.0002$\tablefootmark{a}, $-0.08$\tablefootmark{b} \\
CD~Cyg  &         $1.704\pm0.065$  &  3   & $5.223\pm0.458$\tablefootmark{e}  \\
X~Cyg  &         $1.637\pm0.010$  &  3  & $1.460\pm0.267$\tablefootmark{e}  \\
$\beta$~Dor  &      $0.331\pm0.021$  &  3        & $0.468\pm0.016$\tablefootmark{a}, 0.91\tablefootmark{b}  \\
$\zeta$~Gem  &      $-1.311\pm0.000$  &  2   & $-3.10\pm0.17$\tablefootmark{a}, $-3.40$\tablefootmark{b}  \\
CV~Mon  &         $-0.006\pm0.052$  &  3    & 0.42\tablefootmark{b} \\
T~Mon  &         $4.273\pm3.455$  &  3  &  12.5\tablefootmark{b} \\
S~Nor  &         $0.784\pm0.081$  &  3  &  -- \\
TW~Nor  &         $-2.377\pm0.018$  &  2    & --   \\
V340~Nor  &         $0.966\pm0.308$  &  3     & --   \\
RY~Sco  &         $1.227\pm0.219$  &  3   &  12.73\tablefootmark{b}  \\
RU~Sct  &         $5.179\pm0.015$  &  3   & 10.1\tablefootmark{b}    \\
GY~Sge  &         $-93.681\pm68.486$  &  3   & 743.0\tablefootmark{b}  \\
S~Sge  &         $-0.034\pm0.001$  &  2   &  0.35\tablefootmark{b}   \\
BB~Sgr  &         $-0.299\pm0.019$  &  2   &  0.68\tablefootmark{b}  \\
U~Sgr  &         $-0.330\pm0.000$  &  2   & 0.33\tablefootmark{b}, 0.073\tablefootmark{d}  \\
W~Sgr  &         $0.098\pm0.014$  &  3  &  -- \\
WZ~Sgr  &         $4.961\pm0.023$  &  3   & --   \\
X~Sgr  &         $0.091\pm0.003$  &  3  & 1.49\tablefootmark{b}  \\
Y~Sgr  &         $-0.074\pm0.002$  &  2   & --  \\
ST~Tau  &         $0.062\pm0.030$  &  3   & 0.06\tablefootmark{b}  \\
SZ~Tau  &         $-0.305\pm0.002$  &  3    & $-0.353\pm0.156$\tablefootmark{a}, -0.49\tablefootmark{b} \\
RY~Vel  &         $17.213\pm0.200$  &  1     &  -- \\
RZ~Vel  &         $8.875\pm0.011$  &  3   &  --  \\
T~Vel  &         $-0.013\pm0.008$  &  2   & --  \\
S~Vul  &         $-934.298\pm5.731$  &  2    &  561.1\tablefootmark{b}  \\
SV~Vul  &         $-238.527\pm0.086$  &  2   & $-231.2\pm7.9$\tablefootmark{a},  $-214.3\pm5.5$\tablefootmark{j}  \\
T~Vul  &         $-0.315\pm0.000$  &  3   &  $-0.24\pm0.05$\tablefootmark{h} \\
U~Vul  &         $-0.151\pm0.001$  &  2   & 0.04\tablefootmark{b} \\
			\hline
		\end{tabular}
		\label{table__period_change}
		\tablefoot{\tablefoottext{a}{\citet{Engle_2015_02_0}.}
		\tablefoottext{b}{\citet{Turner_1998_01_0}}
		\tablefoottext{c}{\citet{Berdnikov_2004_02_0}}
		\tablefoottext{d}{\citet{Majaess_2013_12_0}}
		\tablefoottext{e}{\citet{Turner_1999_06_0}}
		\tablefoottext{f}{\citet{Berdnikov_2014_04_0}}
		\tablefoottext{g}{\citet{Turner_2007_11_0}}
		\tablefoottext{h}{\citet{Meyer_2006_08_0}}
		\tablefoottext{i}{\citet{Pop_2004_10_0}}
		\tablefoottext{j}{\citet{Turner_2004_08_0}}
	}
	\end{table*}

	\section{Conclusions}
	\label{section__conclusion}
	
	We presented a multi-wavelength and time-dependent modelling of 45 Galactic Cepheids using the SPIPS algorithm in order to investigate the IR excess coming from the circumstellar envelops. We used radial velocities, stellar effective temperatures, angular diameters, and photometry from 0.5\,$\mu$m to 70\,$\mu$m  retrieved from the literature. We also reported new mid-IR photometry with VLT/VISIR in the PAH1 (8.6\,$\mu$m) and PAH2 (11.3\,$\mu$m) filters.
	
	We detected IR excess at $> 3\sigma$ for 13 Cepheids, which we attribute to the presence of a CSE, while the detections for other Cepheids are marginal with the current photometric precision level. We estimated an average IR excess in various photometric bands of $\sim 0.08$\,mag, $\sim 0.11$\,mag, $\sim 0.13$\,mag and $\sim 0.17$\,mag, respectively, at 2.2\,$\mu$m, 5\,$\mu$m, 10\,$\mu$m, and 25\,$\mu$m. We also showed that Cepheid distance estimates from the SPIPS modelling are not biased by the presence of IR excess. The most affected parameters are the angular diameters and the colour excesses, which 'compensate' the presence of IR excess. However, we demonstrated that not fitting the colour excess can lead to biased distance estimates. The impact of these CSEs on the extragalactic P-L relations still need to be quantified; however it is likely mitigated as long-period Cepheids do not seem to exhibit more IR excess than short-period ones. We therefore expect short period Cepheids to be more biased due to their lower brightness. The presence of CSE does not seem to be linked to the evolution stage of the Cepheid either (i.e. first, second, or third crossing).
	
	The SPIPS analysis also allowed us to deduce average stellar parameters, such as the pulsation period stellar luminosity, effective temperature, and linear and angular radii (see Table~\ref{table__parameters}).
	
	With this larger sample of Cepheids compared to our previous works, we did not detect any correlation between the CSE relative flux and the pulsation period as previously suspected for MW Cepheids. The IR excess tends to be approximatively constant at a given wavelength. There is also no correlation between the existence of the CSE and the CO molecular feature seen at 4.5\,$\mu$m, which is therefore mostly related to the stellar photosphere. 
	
	We emphasise the need for more precise light curves at near-IR bands and more precise photometry at longer wavelengths to better constrain the properties of these CSEs. The JWST will provide light curves of Cepheids from 1\,$\mu$m to 30\,$\mu$m and allow additional extragalactic studies of the CSEs. For MW Cepheids, JWST light curves in several bands combined with Gaia parallaxes in a SPIPS analysis will provide unbiased and precise luminosity for a precise calibration of the MW P-L relations.

	
	\begin{acknowledgements}
		The authors acknowledge the support of the French Agence Nationale de la Recherche (ANR), under grant ANR-15-CE31-0012-01 (project UnlockCepheids). We acknowledge financial support from "Programme National de Physique Stellaire" (PNPS) of CNRS/INSU, France. The research leading to these results has received funding from the European Research Council (ERC) under the European Union's Horizon 2020 research and innovation programme under grant agreement No 695099 (project CepBin) and from the National Science Center, Poland grants MAESTRO UMO-2017/26/A/ST9/00446 and BEETHOVEN UMO-2018/31/G/ST9/03050. We acknowledge support from the IdP II 2015 0002 64 and DIR/WK/2018/09 grants of the Polish Ministry of Science and Higher Education. Support from the BASAL Centro de Astrof\'isica y Tecnolog\'ias Afines (CATA) AFB-170002, the Millenium Institute of Astrophysics (MAS) of the Iniciativa Cient\'ifica Milenio del Ministerio de Econom\'ia, Fomento y Turismo de Chile, project IC120009. This work has made use of data from the European Space Agency (ESA) mission \emph{Gaia} (\url{https://www.cosmos.esa.int/gaia}), processed by the {\it Gaia} Data Processing and Analysis Consortium (DPAC, \url{https://www.cosmos.esa.int/web/gaia/dpac/consortium}). Funding for the DPAC has been provided by national institutions, in particular the institutions participating in the \emph{Gaia} Multilateral Agreement. This publication makes use of data products from the Wide-field Infrared Survey Explorer, which is a joint project of the University of California, Los Angeles, and the Jet Propulsion Laboratory/California Institute of Technology, funded by the National Aeronautics and Space Administration.
	\end{acknowledgements}
	
	
	\bibliographystyle{aa}   
	\bibliography{/Users/agallenn/Sciences/Articles/bibliographie}

\begin{thebibliography}{130}
\expandafter\ifx\csname natexlab\endcsname\relax\def\natexlab#1{#1}\fi

\bibitem[{{Abt}(1954)}]{Abt_1954_04_0}
{Abt}, H.~A. 1954, \pasp, 66, 65

\bibitem[{{Anderson}(2014)}]{Anderson_2014_06_0}
{Anderson}, R.~I. 2014, \aap, 566, L10

\bibitem[{{Anderson}(2016)}]{Anderson_2016_12_0}
{Anderson}, R.~I. 2016, \mnras, 463, 1707

\bibitem[{{Anderson} {et~al.}(2016{\natexlab{a}}){Anderson}, {Casertano},
  {Riess}, {Melis}, {Holl}, {Semaan}, {Papics}, {Blanco-Cuaresma}, {Eyer},
  {Mowlavi}, {Palaversa}, \& {Roelens}}]{Anderson_2016_10_0}
{Anderson}, R.~I., {Casertano}, S., {Riess}, A.~G., {et~al.}
  2016{\natexlab{a}}, \apjs, 226, 18

\bibitem[{{Anderson} {et~al.}(2016{\natexlab{b}}){Anderson}, {M\'erand},
  {Kervella}, {Breitfelder}, {LeBouquin}, {Eyer}, {Gallenne}, {Palaversa},
  {Semaan}, {Saesen}, \& {Mowlavi}}]{Anderson_2016_02_0}
{Anderson}, R.~I., {M\'erand}, A., {Kervella}, P., {et~al.} 2016{\natexlab{b}},
  \mnras, 455, 4231

\bibitem[{{Anderson} {et~al.}(2015){Anderson}, {Sahlmann}, {Holl}, {Eyer},
  {Palaversa}, {Mowlavi}, {S{\"u}veges}, \& {Roelens}}]{Anderson_2015_05_0}
{Anderson}, R.~I., {Sahlmann}, J., {Holl}, B., {et~al.} 2015, \apj, 804, 144

\bibitem[{{Barmby} {et~al.}(2011){Barmby}, {Marengo}, {Evans}, {Bono},
  {Huelsman}, {Su}, {Welch}, \& {Fazio}}]{Barmby_2011_11_0}
{Barmby}, P., {Marengo}, M., {Evans}, N.~R., {et~al.} 2011, \aj, 141, 42

\bibitem[{{Barnes} {et~al.}(1997){Barnes}, {Fernley}, {Frueh}, {Navas},
  {Moffett}, \& {Skillen}}]{Barnes_1997_06_0}
{Barnes}, III, T.~G., {Fernley}, J.~A., {Frueh}, M.~L., {et~al.} 1997, \pasp,
  109, 645

\bibitem[{{Barnes} {et~al.}(2005){Barnes}, {Jeffery}, {Montemayor}, \&
  {Skillen}}]{Barnes_2005_02_0}
{Barnes}, III, T.~G., {Jeffery}, E.~J., {Montemayor}, T.~J., \& {Skillen}, I.
  2005, \apjs, 156, 227

\bibitem[{{Berdnikov}(2008)}]{Berdnikov_2008_04_0}
{Berdnikov}, L.~N. 2008, VizieR Online Data Catalog: II/285, originally
  published in: Sternberg Astronomical Institute, Moscow, 2285

\bibitem[{{Berdnikov} \& {Turner}(2004)}]{Berdnikov_2004_02_0}
{Berdnikov}, L.~N. \& {Turner}, D.~G. 2004, AAT, 23, 123

\bibitem[{{Berdnikov} {et~al.}(2014){Berdnikov}, {Turner}, \&
  {Henden}}]{Berdnikov_2014_04_0}
{Berdnikov}, L.~N., {Turner}, D.~G., \& {Henden}, A.~A. 2014, Astronomy
  Reports, 58, 240

\bibitem[{{Bersier}(2002)}]{Bersier_2002_06_0}
{Bersier}, D. 2002, \apjs, 140, 465

\bibitem[{{Bersier} {et~al.}(1994){Bersier}, {Burki}, {Mayor}, \&
  {Duquennoy}}]{Bersier_1994_11_0}
{Bersier}, D., {Burki}, G., {Mayor}, M., \& {Duquennoy}, A. 1994, \aaps, 108,
  25

\bibitem[{{Bono} {et~al.}(2000){Bono}, {Caputo}, {Cassisi}, {Marconi},
  {Piersanti}, \& {Tornamb{\`e}}}]{Bono_2000_11_4}
{Bono}, G., {Caputo}, F., {Cassisi}, S., {et~al.} 2000, \apj, 543, 955

\bibitem[{{Borgniet} {et~al.}(2019){Borgniet}, {Kervella}, {Nardetto},
  {Gallenne}, {M{\'e}rand}, {Anderson}, {Aufdenberg}, {Breuval}, {Gieren},
  {Hocd{\'e}}, {Javanmardi}, {Lagadec}, {Pietrzy{\'n}ski}, \&
  {Trahin}}]{Borgniet_2019_11_0}
{Borgniet}, S., {Kervella}, P., {Nardetto}, N., {et~al.} 2019, \aap, 631, A37

\bibitem[{{Breitfelder} {et~al.}(2016){Breitfelder}, {M{\'e}rand}, {Kervella},
  {Gallenne}, {Szabados}, {Anderson}, \& {Le Bouquin}}]{Breitfelder_2016_03_0}
{Breitfelder}, J., {M{\'e}rand}, A., {Kervella}, P., {et~al.} 2016, \aap, 587,
  A117

\bibitem[{{Caputo} {et~al.}(2005){Caputo}, {Bono}, {Fiorentino}, {Marconi}, \&
  {Musella}}]{Caputo_2005_08_0}
{Caputo}, F., {Bono}, G., {Fiorentino}, G., {Marconi}, M., \& {Musella}, I.
  2005, \apj, 629, 1021

\bibitem[{{Carter}(1990)}]{Carter_1990_01_0}
{Carter}, B.~S. 1990, \mnras, 242, 1

\bibitem[{{Castelli} \& {Kurucz}(2003)}]{Castelli_2003__0}
{Castelli}, F. \& {Kurucz}, R.~L. 2003, in IAU Symp., Vol. 210, Modelling of
  Stellar Atmospheres, ed. {N.~Piskunov, W.~W.~Weiss, \& D.~F.~Gray} (ASP), A20

\bibitem[{{Claret} \& {Bloemen}(2011)}]{Claret_2011_05_0}
{Claret}, A. \& {Bloemen}, S. 2011, \aap, 529, A75

\bibitem[{{Cohen} {et~al.}(1999){Cohen}, {Walker}, {Carter}, {Hammersley},
  {Kidger}, \& {Noguchi}}]{Cohen_1999_04_0}
{Cohen}, M., {Walker}, R.~G., {Carter}, B., {et~al.} 1999, \aj, 117, 1864

\bibitem[{{Coulson} \& {Caldwell}(1985)}]{Coulson_1985__0}
{Coulson}, I.~M. \& {Caldwell}, J.~A.~R. 1985, South African Astron. Observ.
  Circ., 9, 5

\bibitem[{{Cutri} {et~al.}(2012){Cutri}, {Wright}, {Conrow}, {Bauer},
  {Benford}, {Brandenburg}, {Dailey}, {Eisenhardt}, {Evans}, {Fajardo-Acosta},
  {Fowler}, {Gelino}, {Grillmair}, {Harbut}, {Hoffman}, {Jarrett},
  {Kirkpatrick}, {Leisawitz}, {Liu}, {Mainzer}, {Marsh}, {Masci}, {McCallon},
  {Padgett}, {Ressler}, {Royer}, {Skrutskie}, {Stanford}, {Wyatt}, {Tholen},
  {Tsai}, {Wachter}, {Wheelock}, {Yan}, {Alles}, {Beck}, {Grav}, {Masiero},
  {McCollum}, {McGehee}, {Papin}, \& {Wittman}}]{Cutri_2012_03_0}
{Cutri}, R.~M., {Wright}, E.~L., {Conrow}, T., {et~al.} 2012, {Explanatory
  Supplement to the WISE All-Sky Data Release Products}, Explanatory Supplement
  to the WISE All-Sky Data Release Products

\bibitem[{{Egan} {et~al.}(2003){Egan}, {Price}, {Kraemer}, {Mizuno}, {Carey},
  {Wright}, {Engelke}, {Cohen}, \& {Gugliotti}}]{Egan_2003__0}
{Egan}, M.~P., {Price}, S.~D., {Kraemer}, K.~E., {et~al.} 2003, VizieR Online
  Data Catalog: V/114. Originally published in: Air Force Research Laboratory
  Technical Report AFRL-VS-TR-2003-1589 (2003), 5114, 0

\bibitem[{{Engle}(2015)}]{Engle_2015_02_0}
{Engle}, S. 2015, PhD thesis, James Cook University

\bibitem[{{Engle} {et~al.}(2014){Engle}, {Guinan}, {Harper}, {Neilson}, \&
  {Remage Evans}}]{Engle_2014_10_0}
{Engle}, S.~G., {Guinan}, E.~F., {Harper}, G.~M., {Neilson}, H.~R., \& {Remage
  Evans}, N. 2014, \apj, 794, 80

\bibitem[{{Evans} {et~al.}(2018){Evans}, {Riello}, {De Angeli}, {Carrasco},
  {Montegriffo}, {Fabricius}, {Jordi}, {Palaversa}, {Diener}, {Busso},
  {Cacciari}, {van Leeuwen}, {Burgess}, {Davidson}, {Harrison}, {Hodgkin},
  {Pancino}, {Richards}, {Altavilla}, {Balaguer-N{\'u}{\~n}ez}, {Barstow},
  {Bellazzini}, {Brown}, {Castellani}, {Cocozza}, {De Luise}, {Delgado},
  {Ducourant}, {Galleti}, {Gilmore}, {Giuffrida}, {Holl}, {Kewley}, {Koposov},
  {Marinoni}, {Marrese}, {Osborne}, {Piersimoni}, {Portell}, {Pulone},
  {Ragaini}, {Sanna}, {Terrett}, {Walton}, {Wevers}, \&
  {Wyrzykowski}}]{Evans_2018_08_0}
{Evans}, D.~W., {Riello}, M., {De Angeli}, F., {et~al.} 2018, \aap, 616, A4

\bibitem[{{Evans} {et~al.}(1990){Evans}, {Welch}, {Scarfe}, \&
  {Teays}}]{Evans_1990_05_1}
{Evans}, N.~R., {Welch}, D.~L., {Scarfe}, C.~D., \& {Teays}, T.~J. 1990, \aj,
  99, 1598

\bibitem[{{Fadeyev}(2014)}]{Fadeyev_2014_05_0}
{Fadeyev}, Y.~A. 2014, Astronomy Letters, 40, 301

\bibitem[{{Feast} {et~al.}(2008){Feast}, {Laney}, {Kinman}, {van Leeuwen}, \&
  {Whitelock}}]{Feast_2008_06_0}
{Feast}, M.~W., {Laney}, C.~D., {Kinman}, T.~D., {van Leeuwen}, F., \&
  {Whitelock}, P.~A. 2008, \mnras, 386, 2115

\bibitem[{{Fernie} {et~al.}(1995){Fernie}, {Khoshnevissan}, \&
  {Seager}}]{Fernie_1995_09_0}
{Fernie}, J.~D., {Khoshnevissan}, M.~H., \& {Seager}, S. 1995, \aj, 110, 1326

\bibitem[{{Fitzpatrick}(1999)}]{Fitzpatrick_1999_01_0}
{Fitzpatrick}, E.~L. 1999, \pasp, 111, 63

\bibitem[{{Fouqu{\'e}} {et~al.}(2007){Fouqu{\'e}}, {Arriagada}, {Storm},
  {Barnes}, {Nardetto}, {M{\'e}rand}, {Kervella}, {Gieren}, {Bersier},
  {Benedict}, \& {McArthur}}]{Fouque_2007_12_0}
{Fouqu{\'e}}, P., {Arriagada}, P., {Storm}, J., {et~al.} 2007, \aap, 476, 73

\bibitem[{{Freedman} \& {Madore}(2010)}]{Freedman_2010_09_0}
{Freedman}, W.~L. \& {Madore}, B.~F. 2010, \araa, 48, 673

\bibitem[{{Gaia Collaboration} {et~al.}(2018){Gaia Collaboration}, {Brown},
  {Vallenari}, {Prusti}, {de Bruijne}, {Babusiaux}, {Bailer-Jones}, {Biermann},
  {Evans}, {Eyer}, \& et~al.}]{Gaia-Collaboration_2018_08_0}
{Gaia Collaboration}, {Brown}, A.~G.~A., {Vallenari}, A., {et~al.} 2018, \aap,
  616, A1

\bibitem[{{Gaia Collaboration} {et~al.}(2016){Gaia Collaboration}, {Prusti},
  {de Bruijne}, {Brown}, {Vallenari}, {Babusiaux}, {Bailer-Jones}, {Bastian},
  {Biermann}, {Evans}, \& et~al.}]{Gaia-Collaboration_2016_11_0}
{Gaia Collaboration}, {Prusti}, T., {de Bruijne}, J.~H.~J., {et~al.} 2016,
  \aap, 595, A1

\bibitem[{{Gallenne}(2011)}]{Gallenne_2011_12_1}
{Gallenne}, A. 2011, PhD thesis, Universit\'e Paris VI

\bibitem[{{Gallenne} {et~al.}(2019){Gallenne}, {Kervella}, {Borgniet},
  {M{\'e}rand}, {Pietrzy{\'n}ski}, {Gieren}, {Monnier}, {Schaefer}, {Evans},
  {Anderson}, {Baron}, {Roettenbacher}, \& {Karczmarek}}]{Gallenne_2019_02_0}
{Gallenne}, A., {Kervella}, P., {Borgniet}, S., {et~al.} 2019, \aap, 622, A164

\bibitem[{{Gallenne} {et~al.}(2018){Gallenne}, {Kervella}, {Evans}, {Proffitt},
  {Monnier}, {M{\'e}rand}, {Nelan}, {Winston}, {Pietrzy{\'n}ski}, {Schaefer},
  {Gieren}, {Anderson}, {Borgniet}, {Kraus}, {Roettenbacher}, {Baron},
  {Pilecki}, {Taormina}, {Graczyk}, {Mowlavi}, \& {Eyer}}]{Gallenne_2018_11_0}
{Gallenne}, A., {Kervella}, P., {Evans}, N.~R., {et~al.} 2018, \apj, 867, 121

\bibitem[{{Gallenne} {et~al.}(2012{\natexlab{a}}){Gallenne}, {Kervella}, \&
  {M{\'e}rand}}]{Gallenne_2012_02_0}
{Gallenne}, A., {Kervella}, P., \& {M{\'e}rand}, A. 2012{\natexlab{a}}, \aap,
  538, A24

\bibitem[{{Gallenne} {et~al.}(2012{\natexlab{b}}){Gallenne}, {Kervella},
  {M{\'e}rand}, {McAlister}, {ten Brummelaar}, {Coud{\'e} du Foresto},
  {Sturmann}, {Sturmann}, {Turner}, {Farrington}, \&
  {Goldfinger}}]{Gallenne_2012_03_0}
{Gallenne}, A., {Kervella}, P., {M{\'e}rand}, A., {et~al.} 2012{\natexlab{b}},
  \aap, 541, A87

\bibitem[{{Gallenne} {et~al.}(2017{\natexlab{a}}){Gallenne}, {Kervella},
  {M{\'e}rand}, {Pietrzy{\'n}ski}, {Gieren}, {Nardetto}, \&
  {Trahin}}]{Gallenne_2017_11_0}
{Gallenne}, A., {Kervella}, P., {M{\'e}rand}, A., {et~al.} 2017{\natexlab{a}},
  \aap, 608, A18

\bibitem[{{Gallenne} {et~al.}(2013){Gallenne}, {M{\'e}rand}, {Kervella},
  {Chesneau}, {Breitfelder}, \& {Gieren}}]{Gallenne_2013_10_0}
{Gallenne}, A., {M{\'e}rand}, A., {Kervella}, P., {et~al.} 2013, \aap, 558,
  A140

\bibitem[{{Gallenne} {et~al.}(2017{\natexlab{b}}){Gallenne}, {M{\'e}rand},
  {Kervella}, {Evans}, {Proffitt}, {Pietrz{\'n}ski}, \&
  {Gieren}}]{Gallenne_2017_09_0}
{Gallenne}, A., {M{\'e}rand}, A., {Kervella}, P., {et~al.} 2017{\natexlab{b}},
  in European Physical Journal Web of Conferences, Vol. 152, European Physical
  Journal Web of Conferences, 03007

\bibitem[{{Gallenne} {et~al.}(2010){Gallenne}, {M{\'e}rand}, {Kervella}, \&
  {Girard}}]{Gallenne_2010_12_0}
{Gallenne}, A., {M{\'e}rand}, A., {Kervella}, P., \& {Girard}, J.~H.~V. 2010,
  \aap, 527, A51

\bibitem[{{Gallenne} {et~al.}(2015){Gallenne}, {M{\'e}rand}, {Kervella},
  {Monnier}, {Schaefer}, {Baron}, {Breitfelder}, {Le Bouquin}, {Roettenbacher},
  {Gieren}, {Pietrzy{\'n}ski}, {McAlister}, {ten Brummelaar}, {Sturmann},
  {Sturmann}, {Turner}, {Ridgway}, \& {Kraus}}]{Gallenne_2015_07_0}
{Gallenne}, A., {M{\'e}rand}, A., {Kervella}, P., {et~al.} 2015, \aap, 579, A68

\bibitem[{{Genovali} {et~al.}(2014){Genovali}, {Lemasle}, {Bono}, {Romaniello},
  {Fabrizio}, {Ferraro}, {Iannicola}, {Laney}, {Nonino}, {Bergemann},
  {Buonanno}, {Fran{\c c}ois}, {Inno}, {Kudritzki}, {Matsunaga}, {Pedicelli},
  {Primas}, \& {Th{\'e}venin}}]{Genovali_2014_06_0}
{Genovali}, K., {Lemasle}, B., {Bono}, G., {et~al.} 2014, \aap, 566, A37

\bibitem[{{Gieren}(1981)}]{Gieren_1981_12_0}
{Gieren}, W. 1981, \apjs, 47, 315

\bibitem[{{Gieren}(1985)}]{Gieren_1985_08_0}
{Gieren}, W.~P. 1985, \apj, 295, 507

\bibitem[{{Gorynya} {et~al.}(1998){Gorynya}, {Samus}, {Sachkov}, {Rastorguev},
  {Glushkova}, \& {Antipin}}]{Gorynya_1998_11_0}
{Gorynya}, N.~A., {Samus}, N.~N., {Sachkov}, M.~E., {et~al.} 1998, Astronomy
  Letters, 24, 815

\bibitem[{{Groenewegen}(2020)}]{Groenewegen_2020_03_0}
{Groenewegen}, M.~A.~T. 2020, \aap, 635, A33

\bibitem[{{Hanbury Brown} {et~al.}(1974){Hanbury Brown}, {Davis}, {Lake}, \&
  {Thompson}}]{Hanbury-Brown_1974_06_0}
{Hanbury Brown}, R., {Davis}, J., {Lake}, R.~J.~W., \& {Thompson}, R.~J. 1974,
  \mnras, 167, 475

\bibitem[{{Helou} \& {Walker}(1988)}]{Helou_1988__0}
{Helou}, G. \& {Walker}, D.~W. 1988, in Infrared astronomical satellite (IRAS)
  catalogs and atlases. Volume 7, p.1-265, Vol.~7 (STI)

\bibitem[{{Hocd{\'e}} {et~al.}(2020{\natexlab{a}}){Hocd{\'e}}, {Nardetto},
  {Borgniet}, {Lagadec}, {Kervella}, {M{\'e}rand}, {Evans}, {Gillet},
  {Mathias}, {Chiavassa}, {Gallenne}, {Breuval}, \&
  {Javanmardi}}]{Hocde_2020_09_0}
{Hocd{\'e}}, V., {Nardetto}, N., {Borgniet}, S., {et~al.} 2020{\natexlab{a}},
  \aap, 641, A74

\bibitem[{{Hocd{\'e}} {et~al.}(2020{\natexlab{b}}){Hocd{\'e}}, {Nardetto},
  {Lagadec}, {Niccolini}, {Domiciano de Souza}, {M{\'e}rand}, {Kervella},
  {Gallenne}, {Marengo}, {Trahin}, {Gieren}, {Pietrzy{\'n}ski}, {Borgniet},
  {Breuval}, \& {Javanmardi}}]{Hocde_2020_01_0}
{Hocd{\'e}}, V., {Nardetto}, N., {Lagadec}, E., {et~al.} 2020{\natexlab{b}},
  \aap, 633, A47

\bibitem[{{Hocd{\'e}} {et~al.}(2021){Hocd{\'e}}, {Nardetto}, {Matter},
  {Lagadec}, {M{\'e}rand}, {Cruzal{\`e}bes}, {Meilland}, {Millour}, {Lopez},
  {Berio}, {Weigelt}, {Petrov}, {Isbell}, {Jaffe}, {Kervella}, {Glindemann},
  {Sch{\"o}ller}, {Allouche}, {Gallenne}, {Domiciano de Souza}, {Niccolini},
  {Kokoulina}, {Varga}, {Lagarde}, {Augereau}, {van Boekel}, {Bristow},
  {Henning}, {Hofmann}, {Zins}, {Danchi}, {Delbo}, {Dominik}, {G{\'a}mez
  Rosas}, {Klarmann}, {Hron}, {Hogerheijde}, {Meisenheimer}, {Pantin},
  {Paladini}, {Robbe-Dubois}, {Schertl}, {Stee}, {Waters}, {Lehmitz},
  {Bettonvil}, {Heininger}, {Bristow}, {Woillez}, {Wolf}, {Yoffe}, {Szabados},
  {Chiavassa}, {Borgniet}, {Breuval}, {Javanmardi}, {{\'A}brah{\'a}m},
  {Abadie}, {Abuter}, {Accardo}, {Adler}, {Ag{\'o}cs}, {Alonso}, {Antonelli},
  {B{\"o}hm}, {Bailet}, {Bazin}, {Beckmann}, {Beltran}, {Boland}, {Bourget},
  {Brast}, {Bresson}, {Burtscher}, {Buter}, {Castillo}, {Chelli}, {Cid},
  {Clausse}, {Connot}, {Conzelmann}, {De Haan}, {Ebert}, {Elswijk}, {Fantei},
  {Frahm}, {G{\'a}mez Rosas}, {Gabasch}, {Garces}, {Girard}, {Glazenborg},
  {Gont{\'e}}, {Gonz{\'a}lez Herrera}, {Graser}, {Guajardo}, {Guitton},
  {Hanenburg}, {Haubois}, {Hubin}, {Huerta}, {Idserda}, {Ives}, {Jakob},
  {Jask{\'o}}, {Jochum}, {Klein}, {Kragt}, {Kroes}, {Kuindersma}, {Labadie},
  {Laun}, {Le Poole}, {Leinert}, {Lizon}, {Lopez}, {Marcotto}, {Mauclert},
  {Maurer}, {Mehrgan}, {Meisner}, {Meixner}, {Mellein}, {Mohr}, {Morel},
  {Mosoni}, {Navarro}, {Neumann}, {Nu{\ss}baum}, {Pallanca}, {Pasquini},
  {Percheron}, {Phan Duc}, {Pott}, {Pozna}, {Ridinger}, {Rigal}, {Riquelme},
  {Rivinius}, {Roelfsema}, {Rohloff}, {Rousseau}, {Schuhler}, {Schuil},
  {Shabun}, {Soulain}, {Stephan}, {ter Horst}, {Tromp}, {Vakili}, {van Duin},
  {Venema}, {Vinther}, {Wittkowski}, \& {Wrhel}}]{Hocde_2021_03_6}
{Hocd{\'e}}, V., {Nardetto}, N., {Matter}, A., {et~al.} 2021, \aap, in press
  [eprint: 2103.17014]

\bibitem[{{Inno} {et~al.}(2016){Inno}, {Bono}, {Matsunaga}, {Fiorentino},
  {Marconi}, {Lemasle}, {da Silva}, {Soszy{\'n}ski}, {Udalski}, {Romaniello},
  \& {Rix}}]{Inno_2016_12_5}
{Inno}, L., {Bono}, G., {Matsunaga}, N., {et~al.} 2016, \apj, 832, 176

\bibitem[{{Ishihara} {et~al.}(2010){Ishihara}, {Onaka}, {Kataza}, {Salama},
  {Alfageme}, {Cassatella}, {Cox}, {Garc{\'{\i}}a-Lario}, {Stephenson},
  {Cohen}, {Fujishiro}, {Fujiwara}, {Hasegawa}, {Ita}, {Kim}, {Matsuhara},
  {Murakami}, {M{\"u}ller}, {Nakagawa}, {Ohyama}, {Oyabu}, {Pyo}, {Sakon},
  {Shibai}, {Takita}, {Tanab{\'e}}, {Uemizu}, {Ueno}, {Usui}, {Wada},
  {Watarai}, {Yamamura}, \& {Yamauchi}}]{Ishihara_2010_05_0}
{Ishihara}, D., {Onaka}, T., {Kataza}, H., {et~al.} 2010, \aap, 514, A1

\bibitem[{{Ivezic} {et~al.}(1999){Ivezic}, {Nenkova}, \&
  {Elitzur}}]{Ivezic_1999_11_0}
{Ivezic}, Z., {Nenkova}, M., \& {Elitzur}, M. 1999, {DUSTY: Radiation transport
  in a dusty environment}, astrophysics Source Code Library

\bibitem[{{Kervella} {et~al.}(2019{\natexlab{a}}){Kervella}, {Gallenne},
  {Evans}, {Szabados}, {Arenou}, {M{\'e}rand}, {Nardetto}, {Gieren}, \&
  {Pietrzynski}}]{Kervella_2019_03_1}
{Kervella}, P., {Gallenne}, A., {Evans}, N.~R., {et~al.} 2019{\natexlab{a}},
  \aap, 623, A117

\bibitem[{{Kervella} {et~al.}(2019{\natexlab{b}}){Kervella}, {Gallenne},
  {Remage Evans}, {Szabados}, {Arenou}, {M{\'e}rand}, {Proto}, {Karczmarek},
  {Nardetto}, {Gieren}, \& {Pietrzynski}}]{Kervella_2019_03_0}
{Kervella}, P., {Gallenne}, A., {Remage Evans}, N., {et~al.}
  2019{\natexlab{b}}, \aap, 623, A116

\bibitem[{{Kervella} {et~al.}(2009){Kervella}, {M{\'e}rand}, \&
  {Gallenne}}]{Kervella_2009_05_0}
{Kervella}, P., {M{\'e}rand}, A., \& {Gallenne}, A. 2009, \aap, 498, 425

\bibitem[{{Kervella} {et~al.}(2006){Kervella}, {M{\'e}rand}, {Perrin}, \&
  {Coud{\'e} Du Foresto}}]{Kervella_2006_03_0}
{Kervella}, P., {M{\'e}rand}, A., {Perrin}, G., \& {Coud{\'e} Du Foresto}, V.
  2006, \aap, 448, 623

\bibitem[{{Kervella} {et~al.}(2004){Kervella}, {Nardetto}, {Bersier},
  {Mourard}, \& {Coud{\'e} du Foresto}}]{Kervella_2004_03_0}
{Kervella}, P., {Nardetto}, N., {Bersier}, D., {Mourard}, D., \& {Coud{\'e} du
  Foresto}, V. 2004, \aap, 416, 941

\bibitem[{{Kervella} {et~al.}(2017){Kervella}, {Trahin}, {Bond}, {Gallenne},
  {Szabados}, {M{\'e}rand}, {Breitfelder}, {Dailloux}, {Anderson},
  {Fouqu{\'e}}, {Gieren}, {Nardetto}, \&
  {Pietrzy{\'n}ski}}]{Kervella_2017_04_0}
{Kervella}, P., {Trahin}, B., {Bond}, H.~E., {et~al.} 2017, \aap, 600, A127

\bibitem[{{Kiss}(1998)}]{Kiss_1998_07_0}
{Kiss}, L.~L. 1998, \mnras, 297, 825

\bibitem[{{Kovtyukh} {et~al.}(2005){Kovtyukh}, {Andrievsky}, {Belik}, \&
  {Luck}}]{Kovtyukh_2005_01_0}
{Kovtyukh}, V.~V., {Andrievsky}, S.~M., {Belik}, S.~I., \& {Luck}, R.~E. 2005,
  \aj, 129, 433

\bibitem[{{Kovtyukh} \& {Gorlova}(2000)}]{Kovtyukh_2000_06_0}
{Kovtyukh}, V.~V. \& {Gorlova}, N.~I. 2000, \aap, 358, 587

\bibitem[{{Lagage} {et~al.}(2004){Lagage}, {Pel}, {Authier}, {Belorgey},
  {Claret}, {Doucet}, {Dubreuil}, {Durand}, {Elswijk}, {Girardot}, {K{\"a}ufl},
  {Kroes}, {Lortholary}, {Lussignol}, {Marchesi}, {Pantin}, {Peletier},
  {Pirard}, {Pragt}, {Rio}, {Schoenmaker}, {Siebenmorgen}, {Silber}, {Smette},
  {Sterzik}, \& {Veyssiere}}]{Lagage_2004_09_0}
{Lagage}, P.~O., {Pel}, J.~W., {Authier}, M., {et~al.} 2004, The Messenger,
  117, 12

\bibitem[{{Laher} {et~al.}(2012){Laher}, {Gorjian}, {Rebull}, {Masci},
  {Fowler}, {Helou}, {Kulkarni}, \& {Law}}]{Laher_2012_07_0}
{Laher}, R.~R., {Gorjian}, V., {Rebull}, L.~M., {et~al.} 2012, \pasp, 124, 737

\bibitem[{{Lane} {et~al.}(2002){Lane}, {Creech-Eakman}, \&
  {Nordgren}}]{Lane_2002_07_0}
{Lane}, B.~F., {Creech-Eakman}, M.~J., \& {Nordgren}, T.~E. 2002, \apj, 573,
  330

\bibitem[{{Laney} \& {Stobie}(1992)}]{Laney_1992_04_0}
{Laney}, C.~D. \& {Stobie}, R.~S. 1992, \aaps, 93, 93

\bibitem[{{Lemasle} {et~al.}(2020){Lemasle}, {Hanke}, {Storm}, {Bono}, \&
  {Grebel}}]{Lemasle_2020_09_2}
{Lemasle}, B., {Hanke}, M., {Storm}, J., {Bono}, G., \& {Grebel}, E.~K. 2020,
  \aap, 641, A71

\bibitem[{{Lloyd Evans}(1980)}]{Lloyd-Evans_1980__0}
{Lloyd Evans}, T. 1980, SAAO Circ., 1, 257

\bibitem[{{Lopez} {et~al.}(2014){Lopez}, {Lagarde}, {Jaffe}, {Petrov},
  {Sch{\"o}ller}, {Antonelli}, {Beckmann}, {Berio}, {Bettonvil}, {Glindemann},
  {Gonzalez}, {Graser}, {Hofmann}, {Millour}, {Robbe-Dubois}, {Venema}, {Wolf},
  {Henning}, {Lanz}, {Weigelt}, {Agocs}, {Bailet}, {Bresson}, {Bristow},
  {Dugu{\'e}}, {Heininger}, {Kroes}, {Laun}, {Lehmitz}, {Neumann}, {Augereau},
  {Avila}, {Behrend}, {van Belle}, {Berger}, {van Boekel}, {Bonhomme},
  {Bourget}, {Brast}, {Clausse}, {Connot}, {Conzelmann}, {Cruzal{\`e}bes},
  {Csepany}, {Danchi}, {Delbo}, {Delplancke}, {Dominik}, {van Duin}, {Elswijk},
  {Fantei}, {Finger}, {Gabasch}, {Gay}, {Girard}, {Girault}, {Gitton},
  {Glazenborg}, {Gont{\'e}}, {Guitton}, {Guniat}, {De Haan}, {Haguenauer},
  {Hanenburg}, {Hogerheijde}, {ter Horst}, {Hron}, {Hugues}, {Hummel},
  {Idserda}, {Ives}, {Jakob}, {Jasko}, {Jolley}, {Kiraly}, {K{\"o}hler},
  {Kragt}, {Kroener}, {Kuindersma}, {Labadie}, {Leinert}, {Le Poole}, {Lizon},
  {Lucuix}, {Marcotto}, {Martinache}, {Martinot-Lagarde}, {Mathar}, {Matter},
  {Mauclert}, {Mehrgan}, {Meilland}, {Meisenheimer}, {Meisner}, {Mellein},
  {Menardi}, {Menut}, {Merand}, {Morel}, {Mosoni}, {Navarro}, {Nussbaum},
  {Ottogalli}, {Palsa}, {Panduro}, {Pantin}, {Parra}, {Percheron}, {Duc},
  {Pott}, {Pozna}, {Przygodda}, {Rabbia}, {Richichi}, {Rigal}, {Roelfsema},
  {Rupprecht}, {Schertl}, {Schmidt}, {Schuhler}, {Schuil}, {Spang},
  {Stegmeier}, {Thiam}, {Tromp}, {Vakili}, {Vannier}, {Wagner}, \&
  {Woillez}}]{Lopez_2014_09_6}
{Lopez}, B., {Lagarde}, S., {Jaffe}, W., {et~al.} 2014, The Messenger, 157, 5

\bibitem[{{Luck}(2018)}]{Luck_2018_10_0}
{Luck}, R.~E. 2018, \aj, 156, 171

\bibitem[{{Luck} {et~al.}(1998){Luck}, {Moffett}, {Barnes}, \&
  {Gieren}}]{Luck_1998_02_0}
{Luck}, R.~E., {Moffett}, T.~J., {Barnes}, III, T.~G., \& {Gieren}, W.~P. 1998,
  \aj, 115, 605

\bibitem[{{Madore}(1975)}]{Madore_1975_06_0}
{Madore}, B.~F. 1975, \apjs, 29, 219

\bibitem[{{Majaess} {et~al.}(2013){Majaess}, {Carraro}, {Moni Bidin},
  {Bonatto}, {Berdnikov}, {Balam}, {Moyano}, {Gallo}, {Turner}, {Lane},
  {Gieren}, {Borissova}, {Kovtyukh}, \& {Beletsky}}]{Majaess_2013_12_0}
{Majaess}, D., {Carraro}, G., {Moni Bidin}, C., {et~al.} 2013, \aap, 560, A22

\bibitem[{{Marengo} {et~al.}(2009){Marengo}, {Evans}, {Barmby}, {Bono}, \&
  {Welch}}]{Marengo_2009_01_0}
{Marengo}, M., {Evans}, N.~R., {Barmby}, P., {Bono}, G., \& {Welch}, D. 2009,
  in The Evolving ISM in the Milky Way and Nearby Galaxies

\bibitem[{{Marengo} {et~al.}(2010{\natexlab{a}}){Marengo}, {Evans}, {Barmby},
  {Bono}, {Welch}, \& {Romaniello}}]{Marengo_2010_01_0}
{Marengo}, M., {Evans}, N.~R., {Barmby}, P., {et~al.} 2010{\natexlab{a}}, \apj,
  709, 120

\bibitem[{{Marengo} {et~al.}(2010{\natexlab{b}}){Marengo}, {Evans}, {Barmby},
  {Matthews}, {Bono}, {Welch}, {Romaniello}, {Huelsman}, {Su}, \&
  {Fazio}}]{Marengo_2010_12_0}
{Marengo}, M., {Evans}, N.~R., {Barmby}, P., {et~al.} 2010{\natexlab{b}}, \apj,
  725, 2392

\bibitem[{{Matter} {et~al.}(2016){Matter}, {Lagarde}, {Petrov}, {Berio},
  {Robbe-Dubois}, {Lopez}, {Antonelli}, {Allouche}, {Cruzalebes}, {Millour},
  {Bazin}, \& {Bourg{\`e}s}}]{Matter_2016_08_0}
{Matter}, A., {Lagarde}, S., {Petrov}, R.~G., {et~al.} 2016, in Society of
  Photo-Optical Instrumentation Engineers (SPIE) Conference Series, Vol. 9907,
  Optical and Infrared Interferometry and Imaging V, ed. F.~{Malbet}, M.~J.
  {Creech-Eakman}, \& P.~G. {Tuthill}, 990728

\bibitem[{{Matthews} {et~al.}(2012){Matthews}, {Marengo}, {Evans}, \&
  {Bono}}]{Matthews_2012_01_0}
{Matthews}, L.~D., {Marengo}, M., {Evans}, N.~R., \& {Bono}, G. 2012, \apj,
  744, 53

\bibitem[{{M{\'e}rand} {et~al.}(2007){M{\'e}rand}, {Aufdenberg}, {Kervella},
  {Foresto}, {ten Brummelaar}, {McAlister}, {Sturmann}, {Sturmann}, \&
  {Turner}}]{Merand_2007_08_0}
{M{\'e}rand}, A., {Aufdenberg}, J.~P., {Kervella}, P., {et~al.} 2007, \apj,
  664, 1093

\bibitem[{{M{\'e}rand} {et~al.}(2015){M{\'e}rand}, {Kervella}, {Breitfelder},
  {Gallenne}, {Coud{\'e} du Foresto}, {ten Brummelaar}, {McAlister}, {Ridgway},
  {Sturmann}, {Sturmann}, \& {Turner}}]{Merand_2015_12_0}
{M{\'e}rand}, A., {Kervella}, P., {Breitfelder}, J., {et~al.} 2015, \aap, 584,
  A80

\bibitem[{{M{\'e}rand} {et~al.}(2006){M{\'e}rand}, {Kervella}, {Coud{\'e} Du
  Foresto}, {Perrin}, {Ridgway}, {Aufdenberg}, {Ten Brummelaar}, {McAlister},
  {Sturmann}, {Sturmann}, {Turner}, \& {Berger}}]{Merand_2006_07_0}
{M{\'e}rand}, A., {Kervella}, P., {Coud{\'e} Du Foresto}, V., {et~al.} 2006,
  \aap, 453, 155

\bibitem[{{M{\'e}rand} {et~al.}(2005){M{\'e}rand}, {Kervella}, {Coud{\'e} du
  Foresto}, {Ridgway}, {Aufdenberg}, {ten Brummelaar}, {Berger}, {Sturmann},
  {Sturmann}, {Turner}, \& {McAlister}}]{Merand_2005_07_0}
{M{\'e}rand}, A., {Kervella}, P., {Coud{\'e} du Foresto}, V., {et~al.} 2005,
  \aap, 438, L9

\bibitem[{{Metzger} {et~al.}(1992){Metzger}, {Caldwell}, \&
  {Schechter}}]{Metzger_1992_02_0}
{Metzger}, M.~R., {Caldwell}, J. A.~R., \& {Schechter}, P.~L. 1992, \aj, 103,
  529

\bibitem[{{Meyer}(2006)}]{Meyer_2006_08_0}
{Meyer}, R. 2006, Open European Journal on Variable Stars, 0046, 1

\bibitem[{{Moffett} \& {Barnes}(1984)}]{Moffett_1984_07_0}
{Moffett}, T.~J. \& {Barnes}, III, T.~G. 1984, \apjs, 55, 389

\bibitem[{{Monson} {et~al.}(2012){Monson}, {Freedman}, {Madore}, {Persson},
  {Scowcroft}, {Seibert}, \& {Rigby}}]{Monson_2012_11_0}
{Monson}, A.~J., {Freedman}, W.~L., {Madore}, B.~F., {et~al.} 2012, \apj, 759,
  146

\bibitem[{{Monson} \& {Pierce}(2011)}]{Monson_2011_03_0}
{Monson}, A.~J. \& {Pierce}, M.~J. 2011, \apjs, 193, 12

\bibitem[{{Nardetto} {et~al.}(2008){Nardetto}, {Groh}, {Kraus}, {Millour}, \&
  {Gillet}}]{Nardetto_2008_10_0}
{Nardetto}, N., {Groh}, J.~H., {Kraus}, S., {Millour}, F., \& {Gillet}, D.
  2008, \aap, 489, 1263

\bibitem[{{Nardetto} {et~al.}(2016){Nardetto}, {M{\'e}rand}, {Mourard},
  {Storm}, {Gieren}, {Fouqu{\'e}}, {Gallenne}, {Graczyk}, {Kervella},
  {Neilson}, {Pietrzynski}, {Pilecki}, {Breitfelder}, {Berio}, {Challouf},
  {Clausse}, {Ligi}, {Mathias}, {Meilland}, {Perraut}, {Poretti}, {Rainer},
  {Spang}, {Stee}, {Tallon-Bosc}, \& {ten Brummelaar}}]{Nardetto_2016_09_0}
{Nardetto}, N., {M{\'e}rand}, A., {Mourard}, D., {et~al.} 2016, \aap, 593, A45

\bibitem[{{Nardetto} {et~al.}(2006){Nardetto}, {Mourard}, {Kervella},
  {Mathias}, {M{\'e}rand}, \& {Bersier}}]{Nardetto_2006_07_0}
{Nardetto}, N., {Mourard}, D., {Kervella}, P., {et~al.} 2006, \aap, 453, 309

\bibitem[{{Nardetto} {et~al.}(2017){Nardetto}, {Poretti}, {Rainer}, {Fokin},
  {Mathias}, {Anderson}, {Gallenne}, {Gieren}, {Graczyk}, {Kervella},
  {M{\'e}rand}, {Mourard}, {Neilson}, {Pietrzynski}, {Pilecki}, \&
  {Storm}}]{Nardetto_2017_01_0}
{Nardetto}, N., {Poretti}, E., {Rainer}, M., {et~al.} 2017, \aap, 597, A73

\bibitem[{{Neilson} \& {Lester}(2008)}]{Neilson_2008_09_0}
{Neilson}, H.~R. \& {Lester}, J.~B. 2008, \apj, 684, 569

\bibitem[{{Neilson} {et~al.}(2010){Neilson}, {Ngeow}, {Kanbur}, \&
  {Lester}}]{Neilson_2010_06_0}
{Neilson}, H.~R., {Ngeow}, C., {Kanbur}, S.~M., \& {Lester}, J.~B. 2010, \apj,
  716, 1136

\bibitem[{{Perryman} {et~al.}(1997){Perryman}, {Lindegren}, {Kovalevsky},
  {Hoeg}, {Bastian}, {Bernacca}, {Cr{\'e}z{\'e}}, {Donati}, {Grenon}, {van
  Leeuwen}, {van der Marel}, {Mignard}, {Murray}, {Le Poole}, {Schrijver},
  {Turon}, {Arenou}, {Froeschl{\'e}}, \& {Petersen}}]{Perryman_1997_07_0}
{Perryman}, M.~A.~C., {Lindegren}, L., {Kovalevsky}, J., {et~al.} 1997, \aap,
  323, L49

\bibitem[{{Petterson} {et~al.}(2005){Petterson}, {Cottrell}, {Albrow}, \&
  {Fokin}}]{Petterson_2005_10_0}
{Petterson}, O.~K.~L., {Cottrell}, P.~L., {Albrow}, M.~D., \& {Fokin}, A. 2005,
  \mnras, 362, 1167

\bibitem[{{Pont} {et~al.}(1994){Pont}, {Burki}, \& {Mayor}}]{Pont_1994_05_0}
{Pont}, F., {Burki}, G., \& {Mayor}, M. 1994, \aaps, 105, 165

\bibitem[{{Pop} {et~al.}(2004){Pop}, {Turcu}, \& {Codreanu}}]{Pop_2004_10_0}
{Pop}, A., {Turcu}, V., \& {Codreanu}, S. 2004, \apss, 293, 393

\bibitem[{{Riello} {et~al.}(2018){Riello}, {De Angeli}, {Evans}, {Busso},
  {Hambly}, {Davidson}, {Burgess}, {Montegriffo}, {Osborne}, {Kewley},
  {Carrasco}, {Fabricius}, {Jordi}, {Cacciari}, {van Leeuwen}, \&
  {Holland}}]{Riello_2018_08_0}
{Riello}, M., {De Angeli}, F., {Evans}, D.~W., {et~al.} 2018, \aap, 616, A3

\bibitem[{{Schmidt}(2015)}]{Schmidt_2015_11_0}
{Schmidt}, E.~G. 2015, \apj, 813, 29

\bibitem[{{Sch{\"u}tz} \& {Sterzik}(2005)}]{Schutz_2005__0}
{Sch{\"u}tz}, O. \& {Sterzik}, M. 2005, in High Resolution Infrared
  Spectroscopy in Astronomy, ed. {H.~U.~K{\"a}ufl, R.~Siebenmorgen, \&
  A.~Moorwood}, Proceedings of an ESO Workshop held at Garching, Germany,
  104--108

\bibitem[{{Scowcroft} {et~al.}(2011){Scowcroft}, {Freedman}, {Madore},
  {Monson}, {Persson}, {Seibert}, {Rigby}, \& {Sturch}}]{Scowcroft_2011_12_0}
{Scowcroft}, V., {Freedman}, W.~L., {Madore}, B.~F., {et~al.} 2011, \apj, 743,
  76

\bibitem[{{Shobbrook}(1992)}]{Shobbrook_1992_04_0}
{Shobbrook}, R.~R. 1992, \mnras, 255, 486

\bibitem[{{Storm} {et~al.}(2004){Storm}, {Carney}, {Gieren}, {Fouqu{\'e}},
  {Latham}, \& {Fry}}]{Storm_2004_02_0}
{Storm}, J., {Carney}, B.~W., {Gieren}, W.~P., {et~al.} 2004, \aap, 415, 531

\bibitem[{{Storm} {et~al.}(2011){Storm}, {Gieren}, {Fouqu{\'e}}, {Barnes},
  {Pietrzy{\'n}ski}, {Nardetto}, {Weber}, {Granzer}, \&
  {Strassmeier}}]{Storm_2011_10_0}
{Storm}, J., {Gieren}, W., {Fouqu{\'e}}, P., {et~al.} 2011, \aap, 534, A94

\bibitem[{{Szabados}(1977)}]{Szabados_1977_01_0}
{Szabados}, L. 1977, Commun. of the Konkoly Observatory Hungary, 70, 1

\bibitem[{{Szabados}(1980)}]{Szabados_1980_01_0}
{Szabados}, L. 1980, Mitt. Sternw. Ungar. Akad. Wiss, 76

\bibitem[{{Szabados}(1991)}]{Szabados_1991_01_0}
{Szabados}, L. 1991, Commun. of the Konkoly Observatory Hungary, 96, 123

\bibitem[{{Szabados}(1995)}]{Szabados_1995__0}
{Szabados}, L. 1995, in ASPC Series, Vol.~83, IAU Colloq. 155: Astrophysical
  Applications of Stellar Pulsation, ed. {R.~S.~Stobie \& P.~A.~Whitelock}, 357

\bibitem[{{Szabados} {et~al.}(2013){Szabados}, {Anderson}, {Derekas}, {Kiss},
  {Szalai}, {Sz{\'e}kely}, \& {Christiansen}}]{Szabados_2013_09_0}
{Szabados}, L., {Anderson}, R.~I., {Derekas}, A., {et~al.} 2013, \mnras, 434,
  870

\bibitem[{{Szabados} {et~al.}(2011){Szabados}, {Kiss}, \&
  {Klagyivik}}]{Szabados_2011_02_0}
{Szabados}, L., {Kiss}, Z.~T., \& {Klagyivik}, P. 2011, in EAS Publications
  Series, Vol.~45, 441--444

\bibitem[{{Trahin}(2019)}]{Trahin_2019_11_0}
{Trahin}, B. 2019, PhD thesis, Universit\'e PSL
  (https://hal.archives-ouvertes.fr/tel-02372923)

\bibitem[{{Turner} {et~al.}(2003){Turner}, {Berdnikov}, \&
  {Abdel-Sabour}}]{Turner_2003_01_0}
{Turner}, D., {Berdnikov}, L.~N., \& {Abdel-Sabour}, M.~A. 2003, \jrasc, 97,
  216

\bibitem[{{Turner}(1998)}]{Turner_1998_01_0}
{Turner}, D.~G. 1998, Journal of the American Association of Variable Star
  Observers (JAAVSO), 26, 101

\bibitem[{{Turner}(2016)}]{Turner_2016_10_0}
{Turner}, D.~G. 2016, \rmxaa, 52, 223

\bibitem[{{Turner} \& {Berdnikov}(2004)}]{Turner_2004_08_0}
{Turner}, D.~G. \& {Berdnikov}, L.~N. 2004, \aap, 423, 335

\bibitem[{{Turner} {et~al.}(2007){Turner}, {Bryukhanov}, {Balyuk}, {Gain},
  {Grabovsky}, {Grigorenko}, {Klochko}, {Kosa-Kiss}, {Kosinsky}, {Kushmar},
  {Mamedov}, {Narkevich}, {Pogosyants}, {Semenyuta}, {Sergey}, {Schukin},
  {Strigelsky}, {Tamello}, {Lane}, \& {Majaess}}]{Turner_2007_11_0}
{Turner}, D.~G., {Bryukhanov}, I.~S., {Balyuk}, I.~I., {et~al.} 2007, \pasp,
  119, 1247

\bibitem[{{Turner} {et~al.}(1999){Turner}, {Horsford}, \&
  {MacMillan}}]{Turner_1999_06_0}
{Turner}, D.~G., {Horsford}, A.~J., \& {MacMillan}, J.~D. 1999, Journal of the
  American Association of Variable Star Observers (JAAVSO), 27, 5

\bibitem[{{Usenko} {et~al.}(2011){Usenko}, {Berdnikov}, {Kravtsov}, {Kniazev},
  {Chini}, {Hoffmeister}, {Stahl}, \& {Drass}}]{Usenko_2011_10_0}
{Usenko}, I.~A., {Berdnikov}, L.~N., {Kravtsov}, V.~V., {et~al.} 2011,
  Astronomy Letters, 37, 718

\bibitem[{{Usenko} {et~al.}(2013){Usenko}, {Kniazev}, {Berdnikov}, {Kravtsov},
  \& {Fokin}}]{Usenko_2013_07_0}
{Usenko}, I.~A., {Kniazev}, A.~Y., {Berdnikov}, L.~N., {Kravtsov}, V.~V., \&
  {Fokin}, A.~B. 2013, Astronomy Letters, 39, 432

\bibitem[{{van Leeuwen} {et~al.}(1997){van Leeuwen}, {Evans}, {Grenon},
  {Grossmann}, {Mignard}, \& {Perryman}}]{van-Leeuwen_1997_07_0}
{van Leeuwen}, F., {Evans}, D.~W., {Grenon}, M., {et~al.} 1997, \aap, 323, L61

\bibitem[{{Welch} {et~al.}(1984){Welch}, {Wieland}, {McAlary}, {McGonegal},
  {Madore}, {McLaren}, \& {Neugebauer}}]{Welch_1984_04_0}
{Welch}, D.~L., {Wieland}, F., {McAlary}, C.~W., {et~al.} 1984, \apjs, 54, 547

\bibitem[{{Wielg{\'o}rski} {et~al.}(2017){Wielg{\'o}rski}, {Pietrzy{\'n}ski},
  {Gieren}, {G{\'o}rski}, {Kudritzki}, {Zgirski}, {Bresolin}, {Storm},
  {Matsunaga}, {Graczyk}, \& {Soszy{\'n}ski}}]{Wielgorski_2017_06_0}
{Wielg{\'o}rski}, P., {Pietrzy{\'n}ski}, G., {Gieren}, W., {et~al.} 2017, \apj,
  842, 116

\bibitem[{{Wright} {et~al.}(2010){Wright}, {Eisenhardt}, {Mainzer}, {Ressler},
  {Cutri}, {Jarrett}, {Kirkpatrick}, {Padgett}, {McMillan}, {Skrutskie},
  {Stanford}, {Cohen}, {Walker}, {Mather}, {Leisawitz}, {Gautier}, {McLean},
  {Benford}, {Lonsdale}, {Blain}, {Mendez}, {Irace}, {Duval}, {Liu}, {Royer},
  {Heinrichsen}, {Howard}, {Shannon}, {Kendall}, {Walsh}, {Larsen}, {Cardon},
  {Schick}, {Schwalm}, {Abid}, {Fabinsky}, {Naes}, \&
  {Tsai}}]{Wright_2010_12_0}
{Wright}, E.~L., {Eisenhardt}, P.~R.~M., {Mainzer}, A.~K., {et~al.} 2010, \aj,
  140, 1868

\end{thebibliography}
	
	
	\begin{appendix} 
		
		\section{SPIPS model for three Cepheids, fitting a CSE with an analytical model.}
		\label{appendix__spips_model}
		
		\begin{figure*}[!h]
			\resizebox{\hsize}{!}{\includegraphics{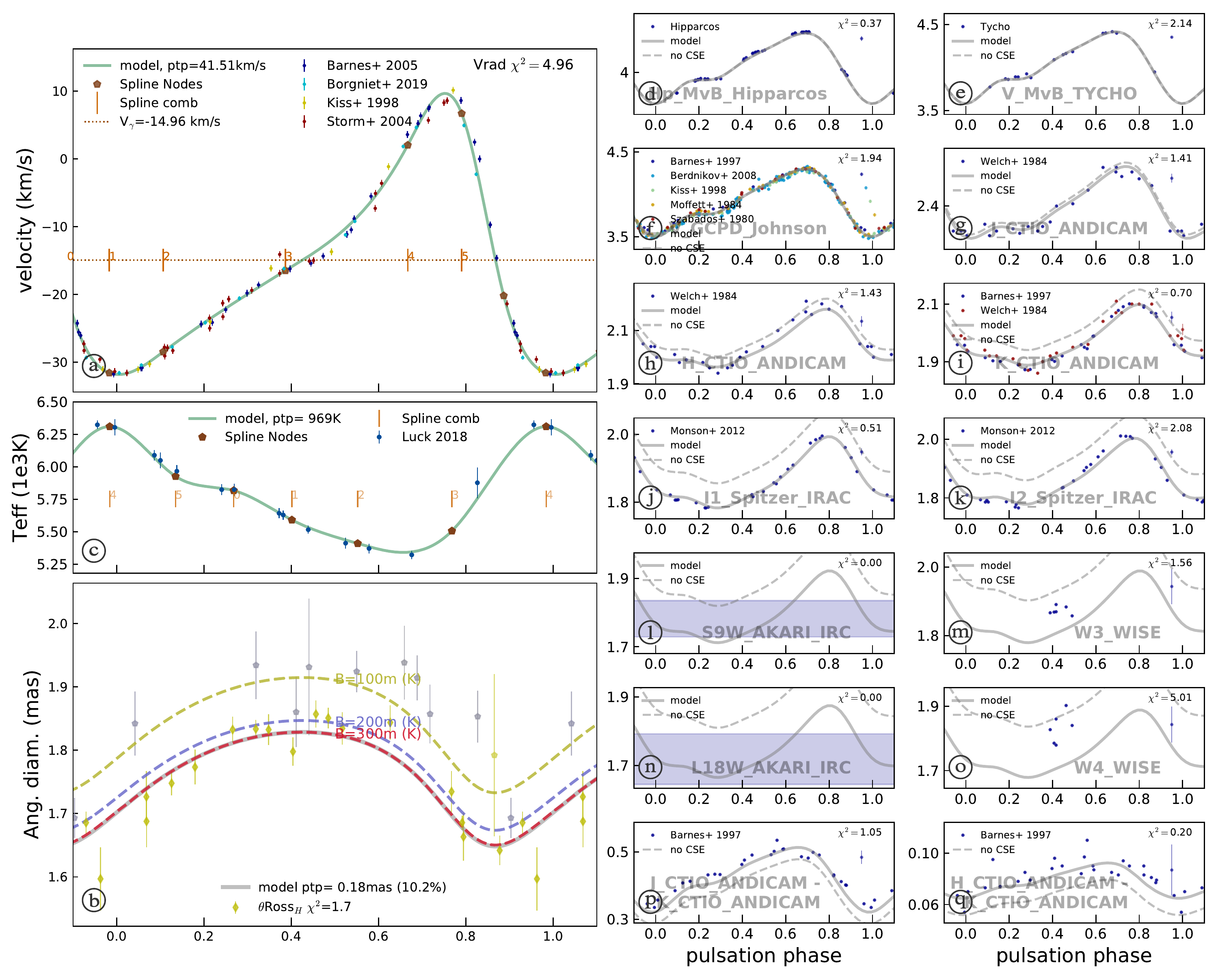}}
			\caption{SPIPS model of $\eta$~Aql with a CSE.}
			\label{image__spips_etaaql}
		\end{figure*}
		
		\begin{figure*}[!h]
			\resizebox{\hsize}{!}{\includegraphics{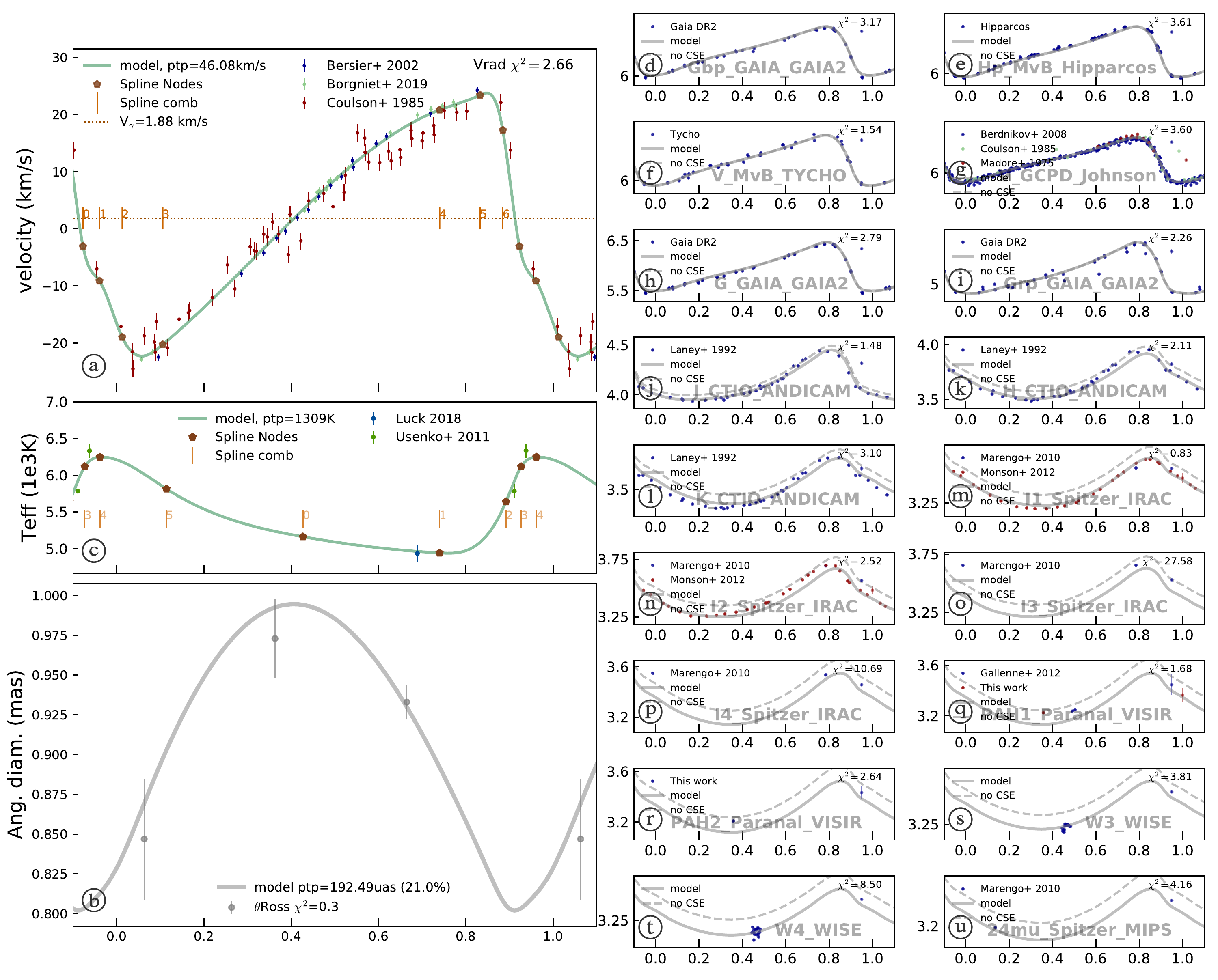}}
			\caption{SPIPS model of U~Car with a CSE.}
			\label{image__spips_ucar}
		\end{figure*}
		
		\begin{figure*}[!h]
			\resizebox{\hsize}{!}{\includegraphics{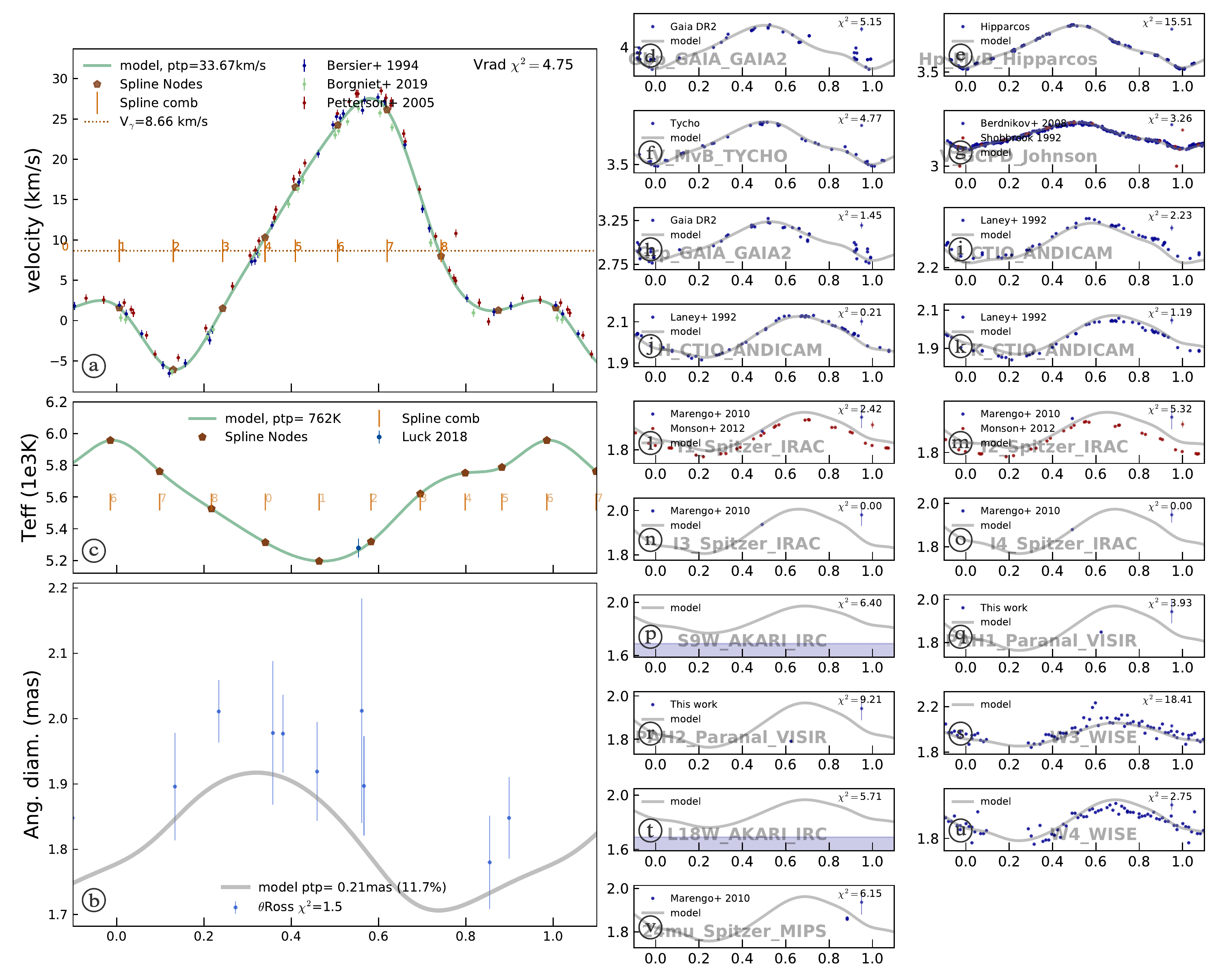}}
			\caption{SPIPS model of $\beta$~Dor without a CSE.}
			\label{image__spips_betador}
		\end{figure*}
		
		\begin{figure*}[!h]
			\resizebox{\hsize}{!}{\includegraphics{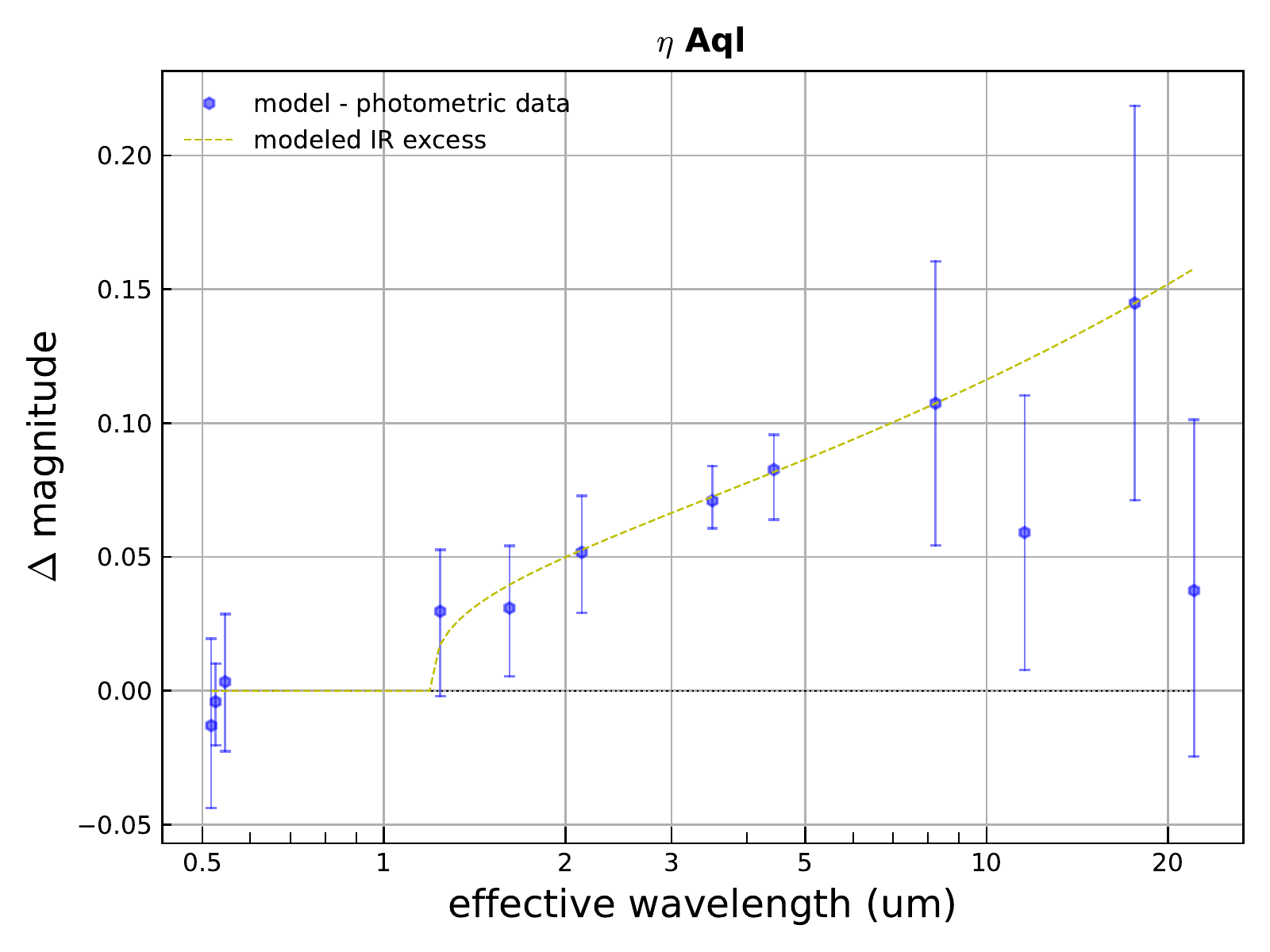}\includegraphics{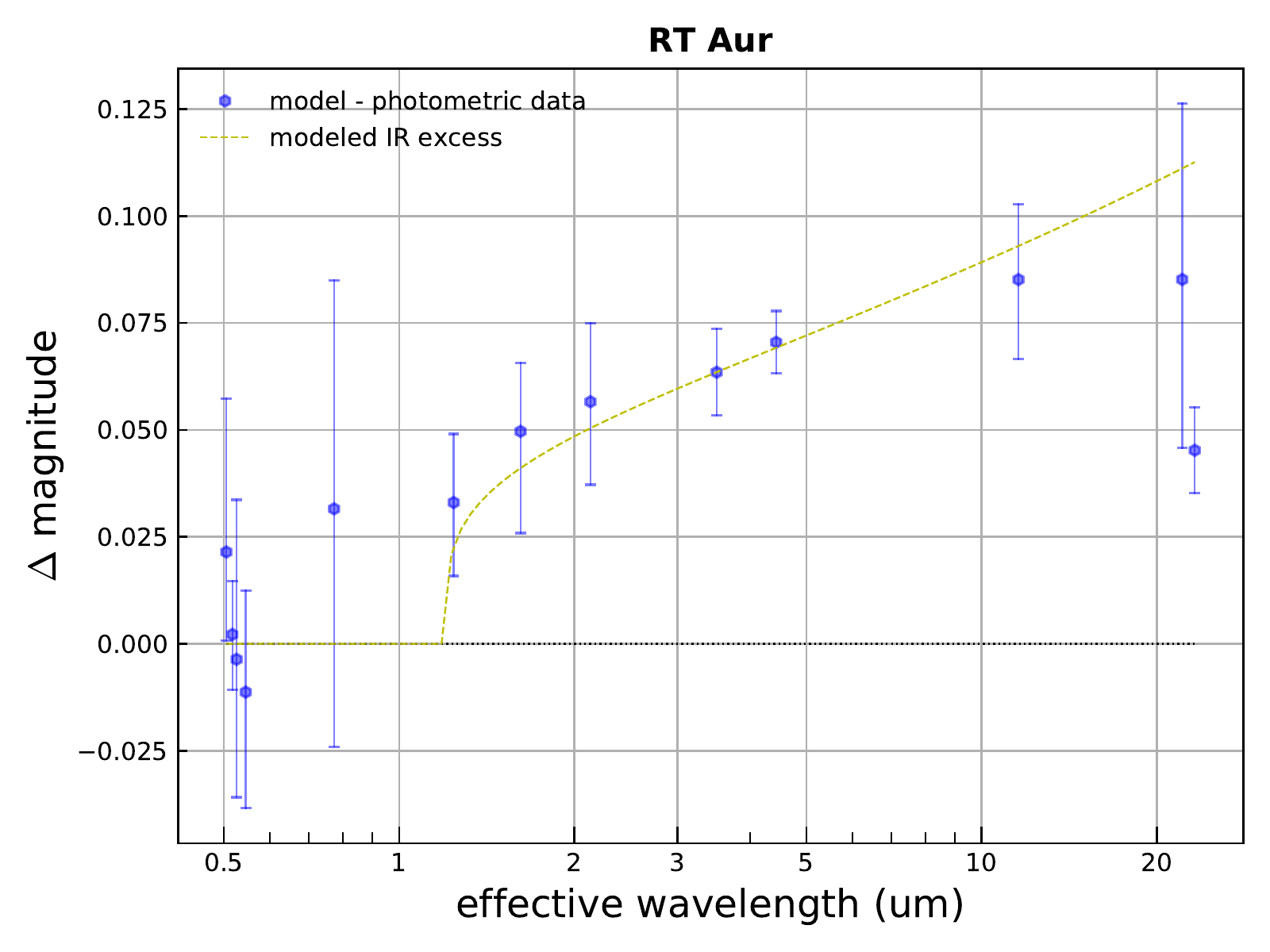}}
			\resizebox{\hsize}{!}{\includegraphics{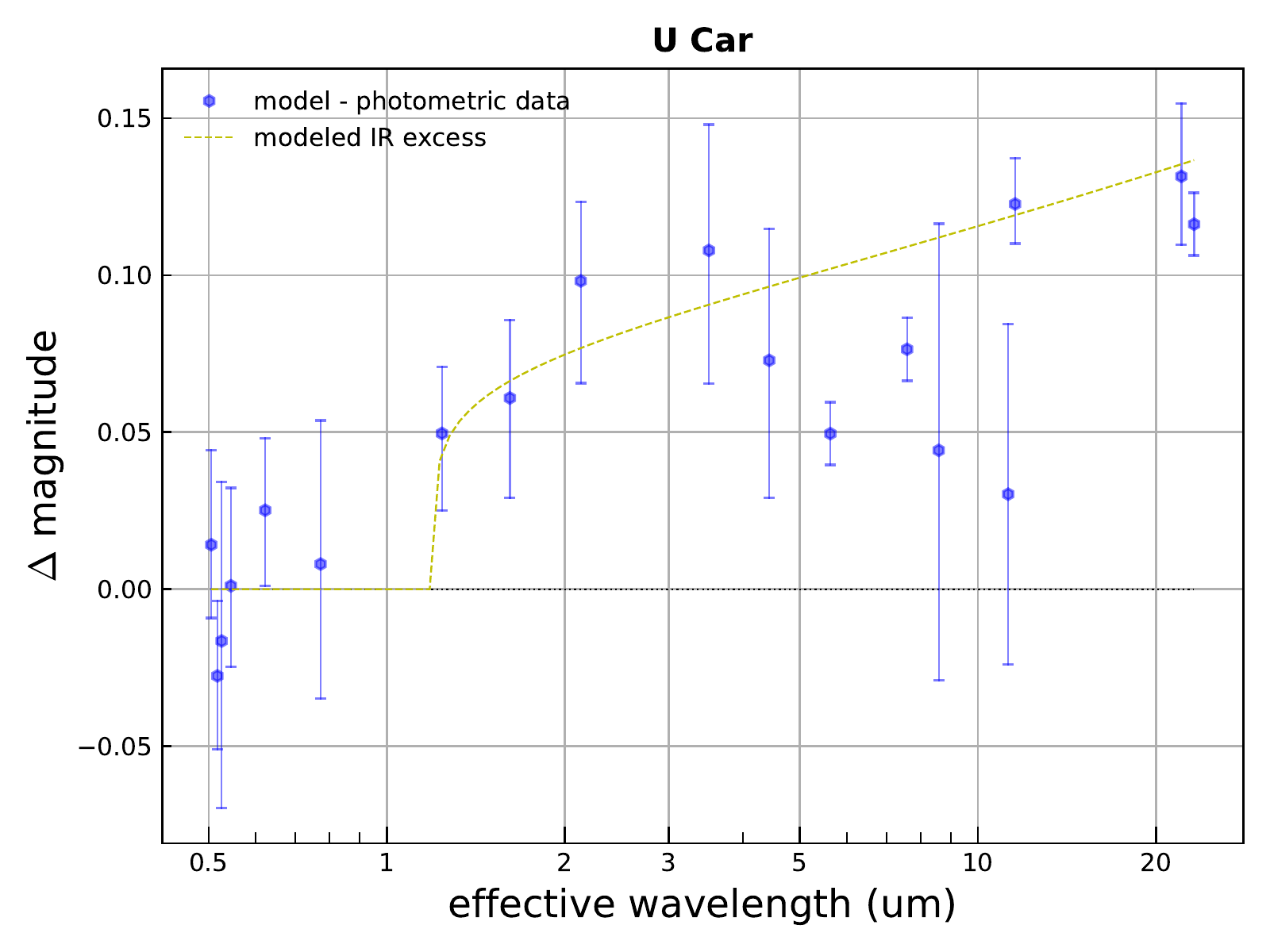}\includegraphics{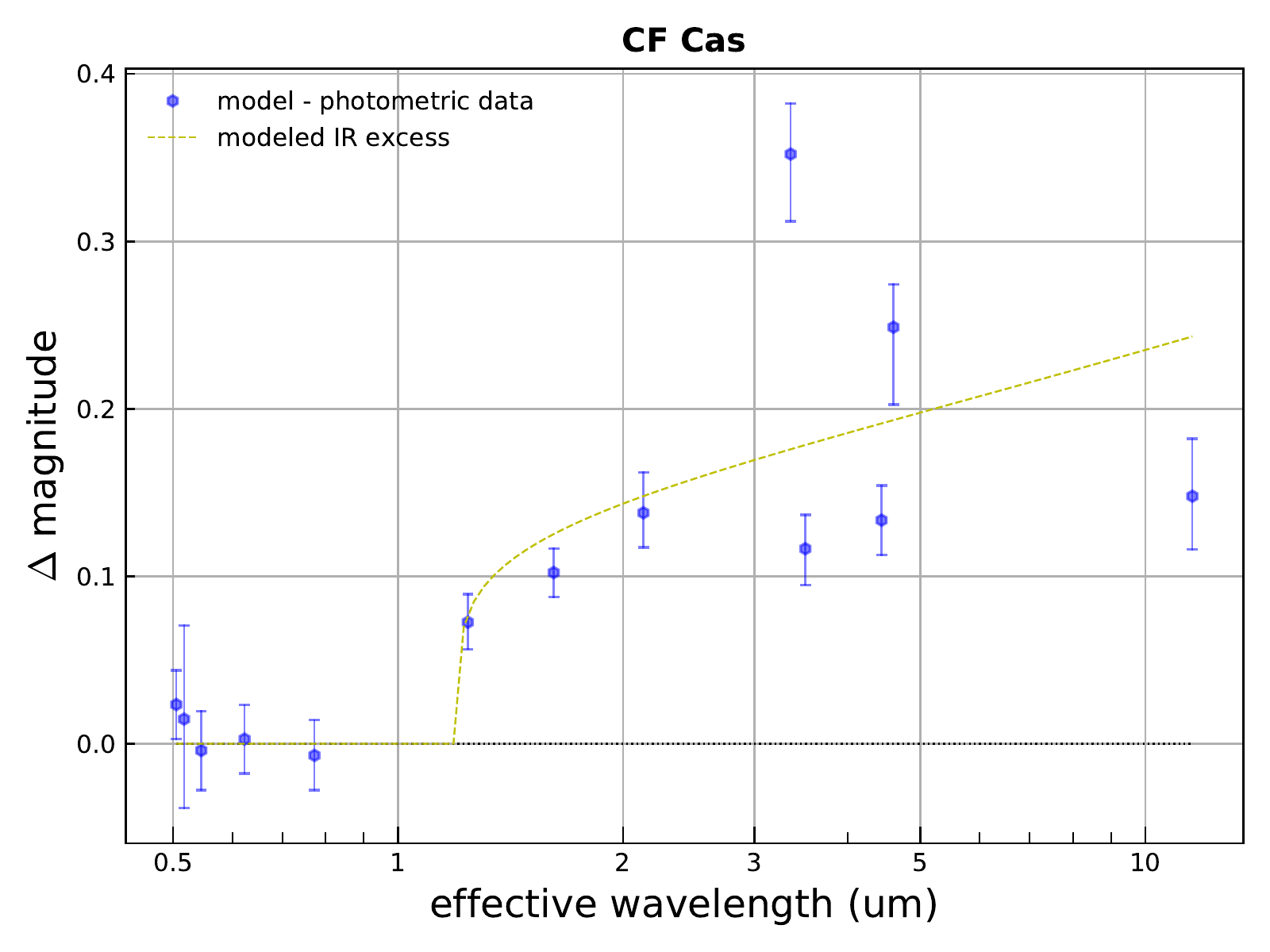}}
			\resizebox{\hsize}{!}{\includegraphics{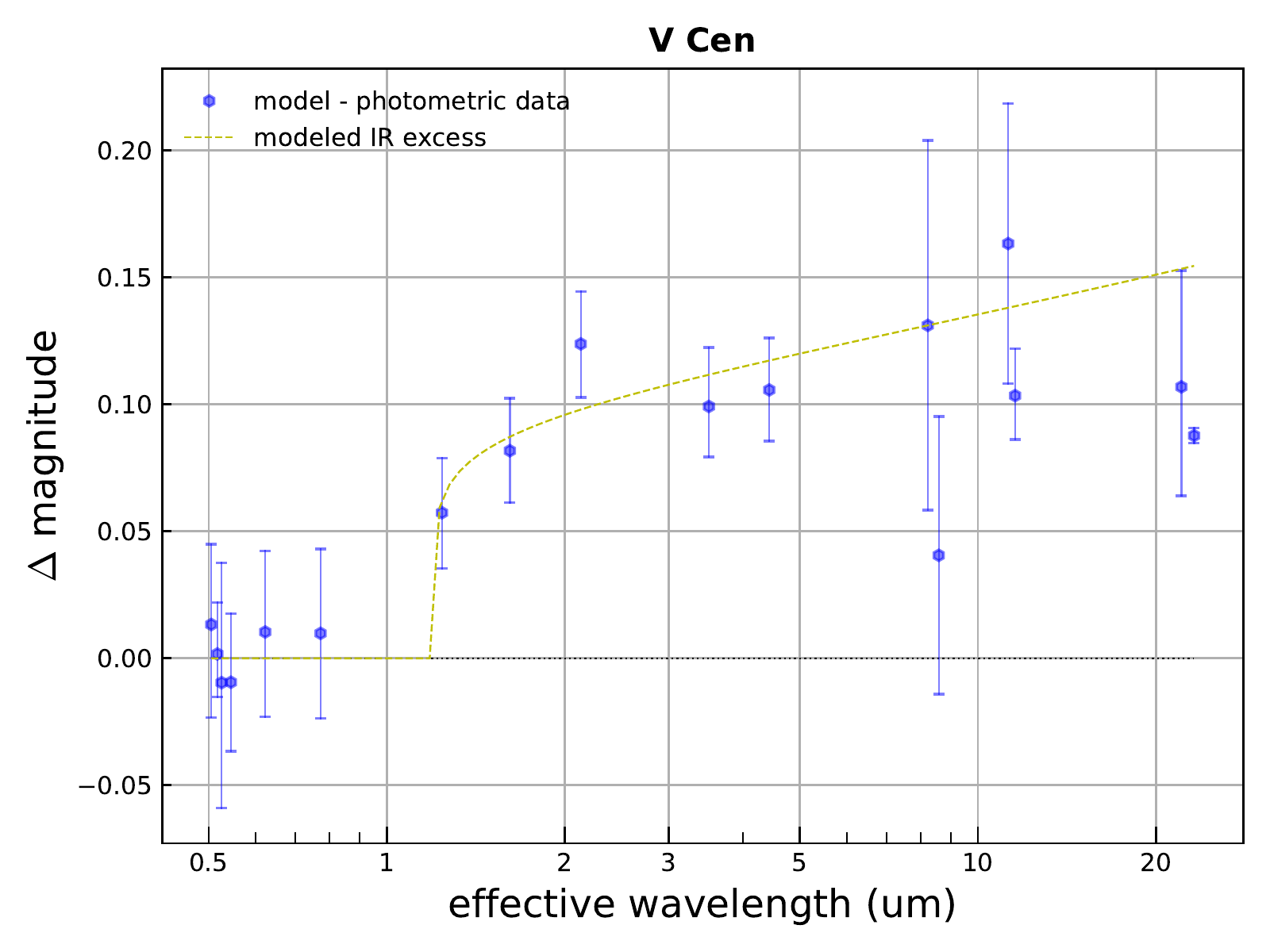}\includegraphics{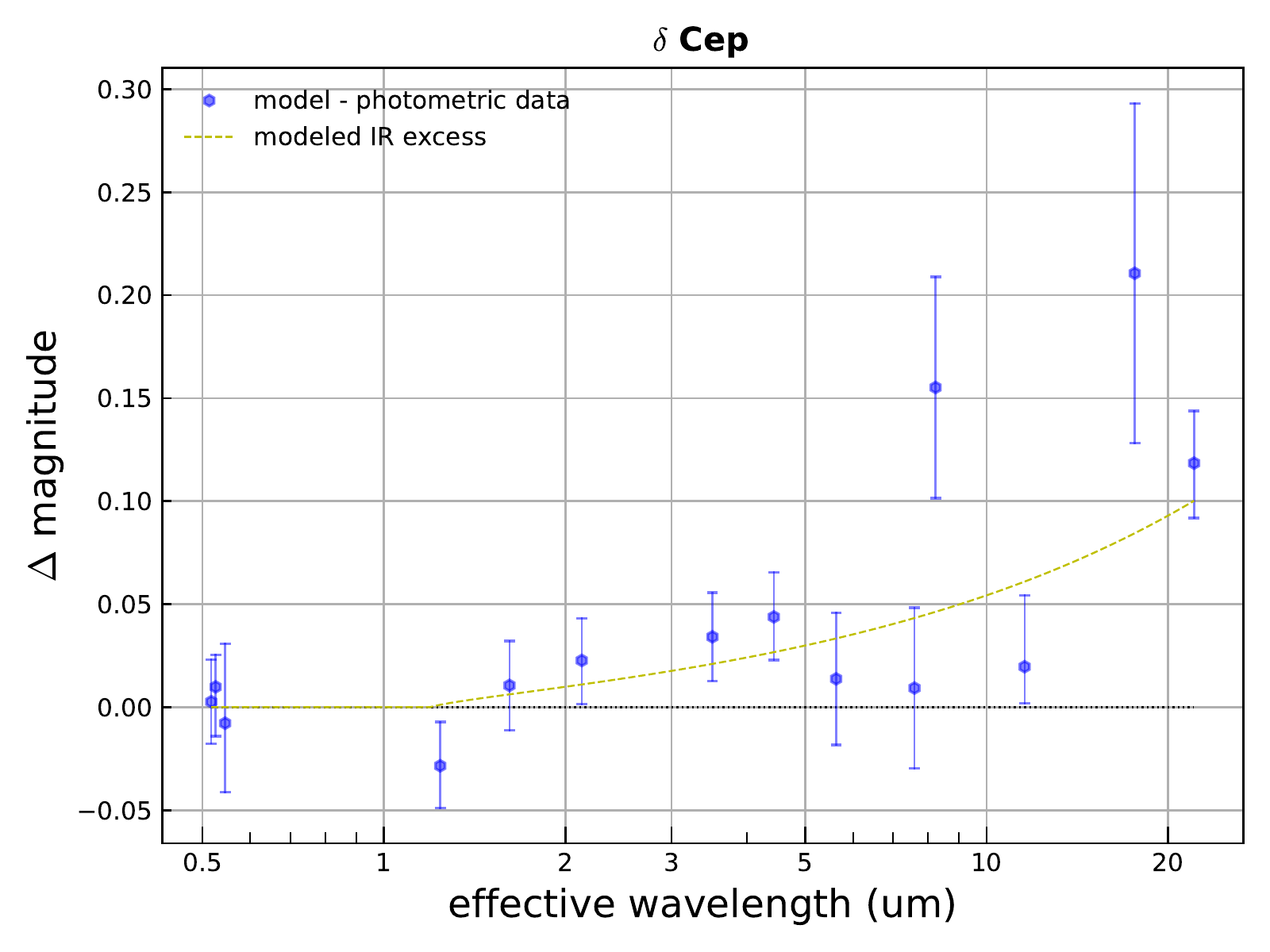}}
			\caption{SPIPS IR-excess model for the Cepheids $\eta$~Aql,, RT~Aur, U~Car, CF~Cas, V~Cen, and $\delta$~Cep, for which we detected a CSE at more than $3\sigma$.}
			\label{image__spips_excess1}
		\end{figure*}
		\begin{figure*}[!h]
			\resizebox{\hsize}{!}{\includegraphics{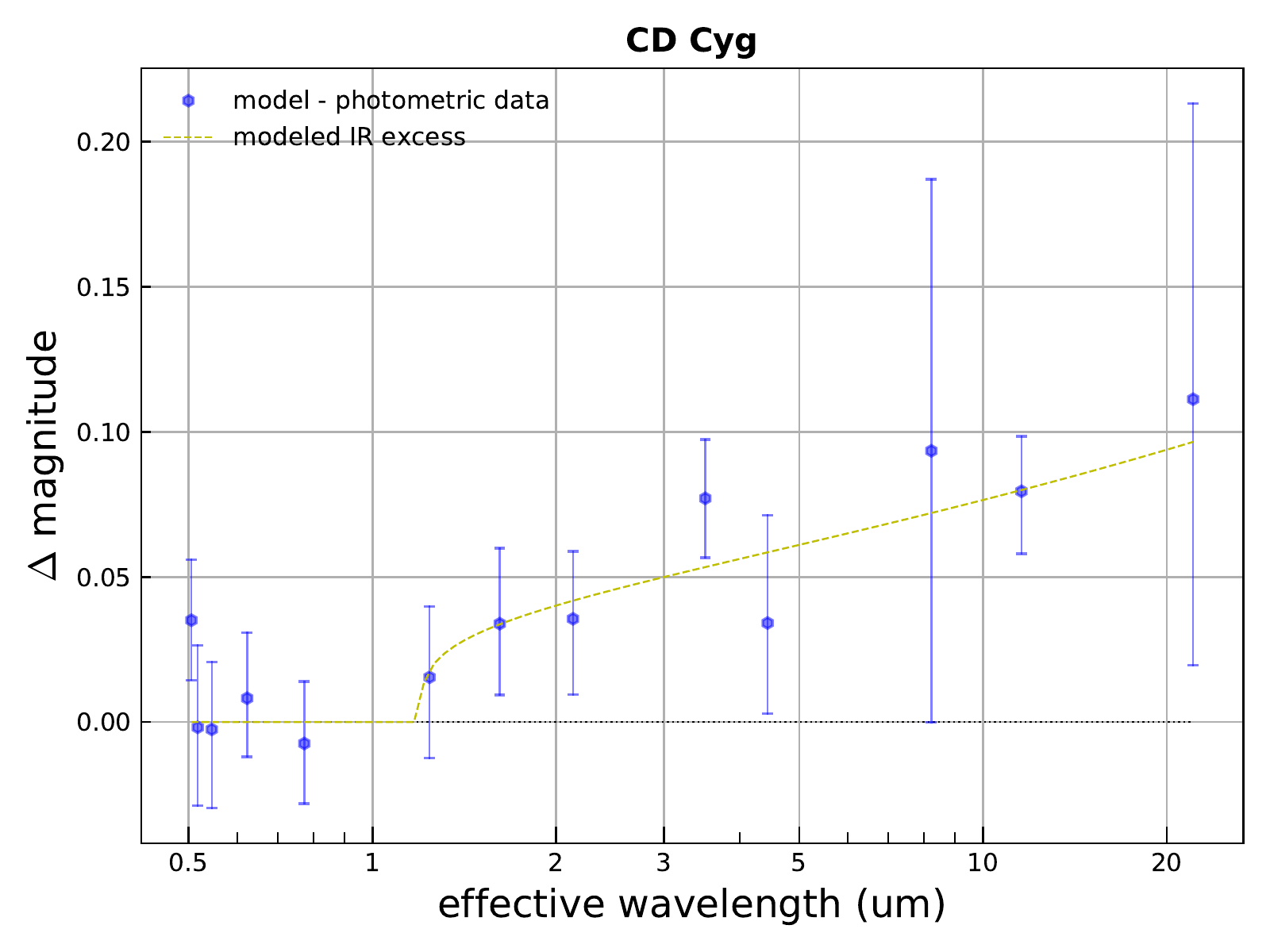}\includegraphics{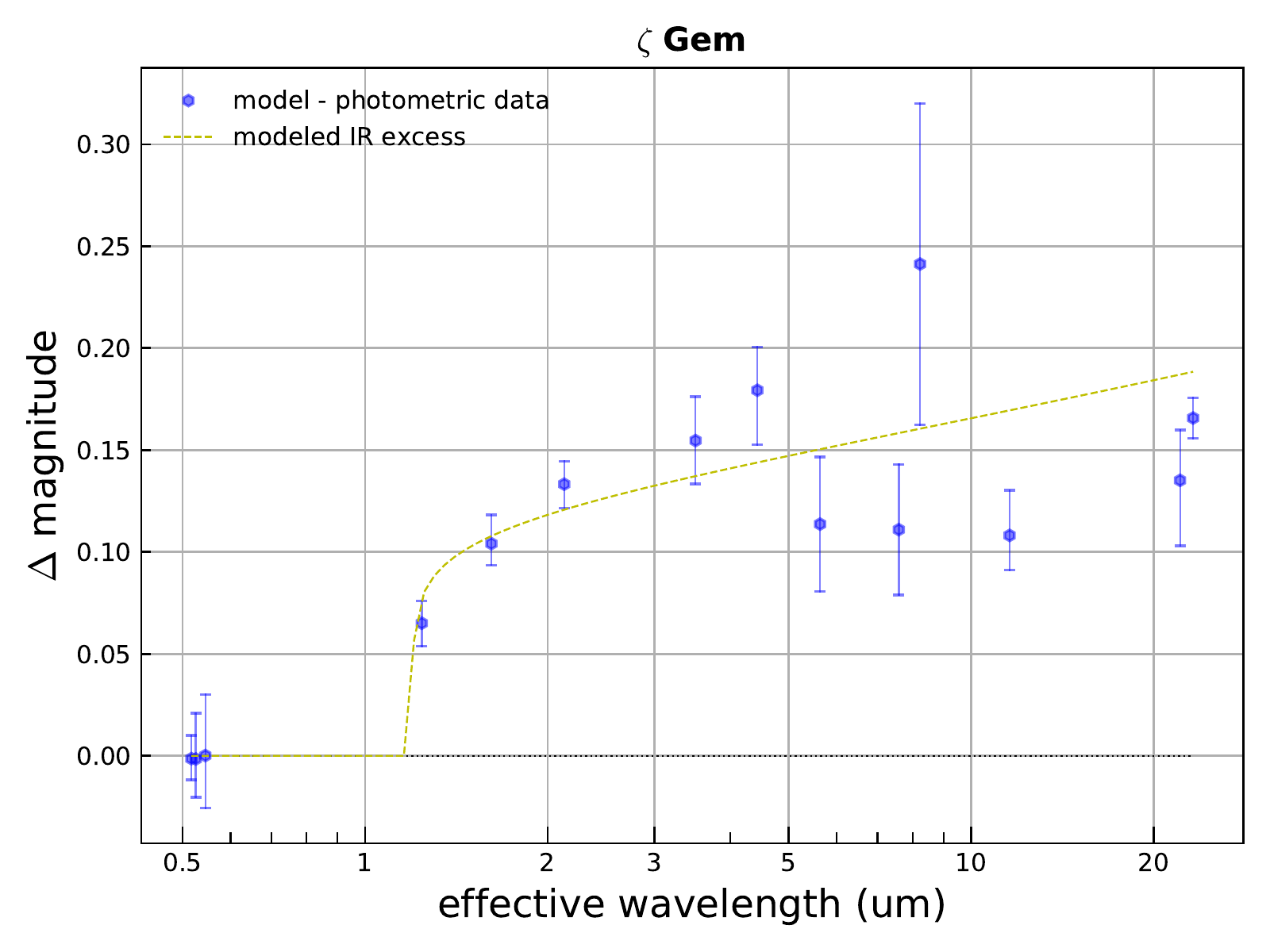}}
			\resizebox{\hsize}{!}{\includegraphics{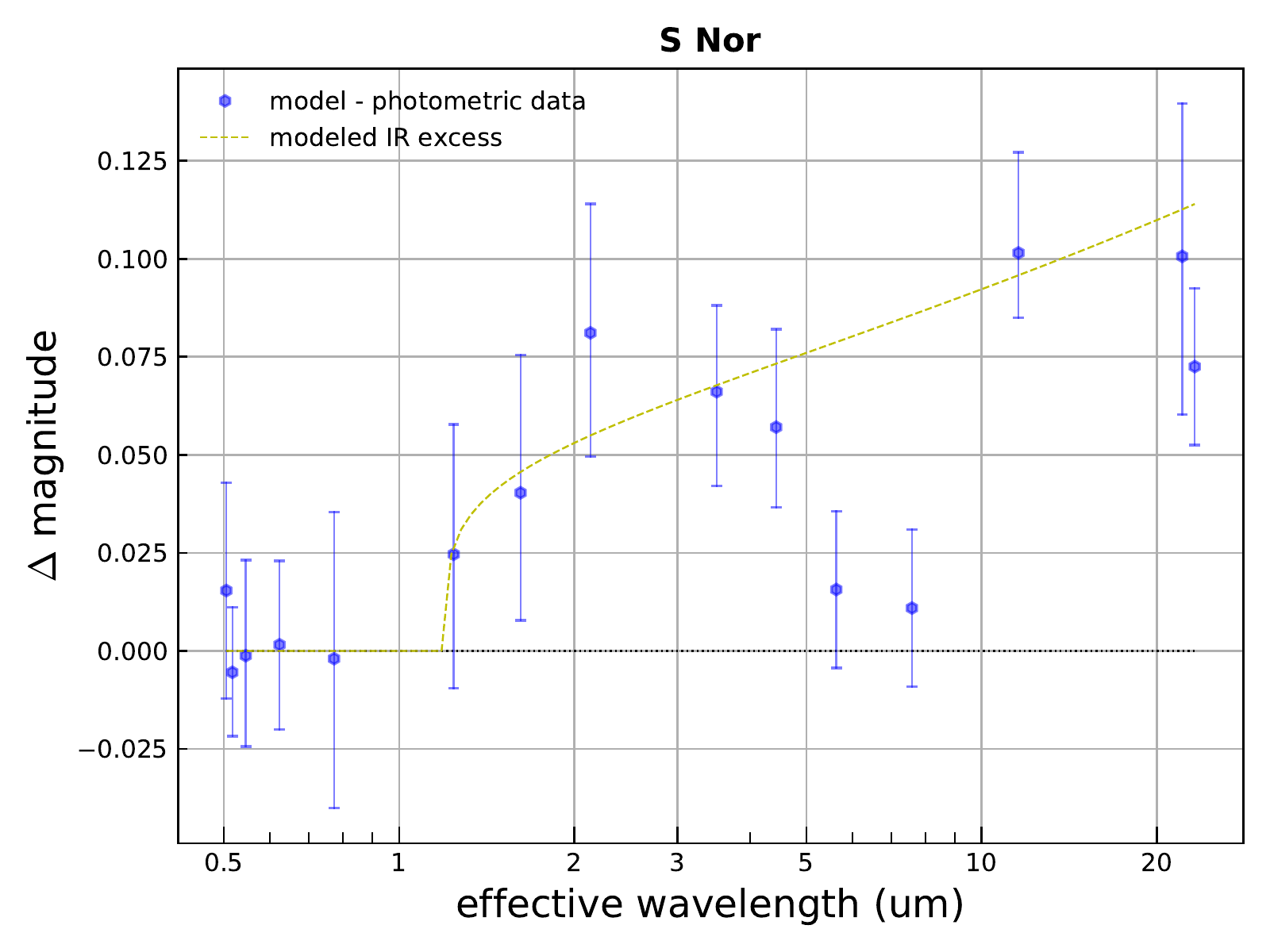}\includegraphics{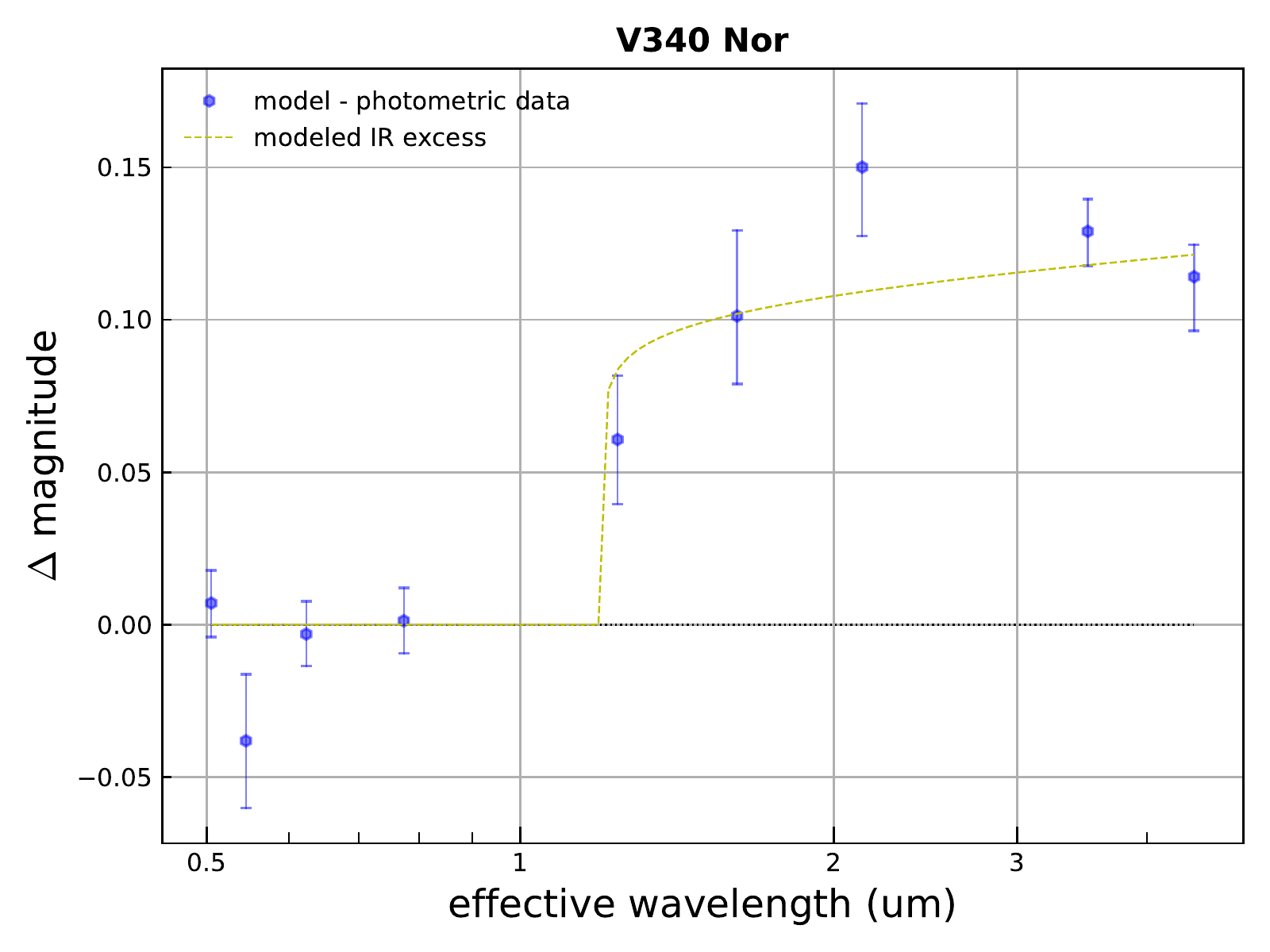}}
			\resizebox{\hsize}{!}{\includegraphics{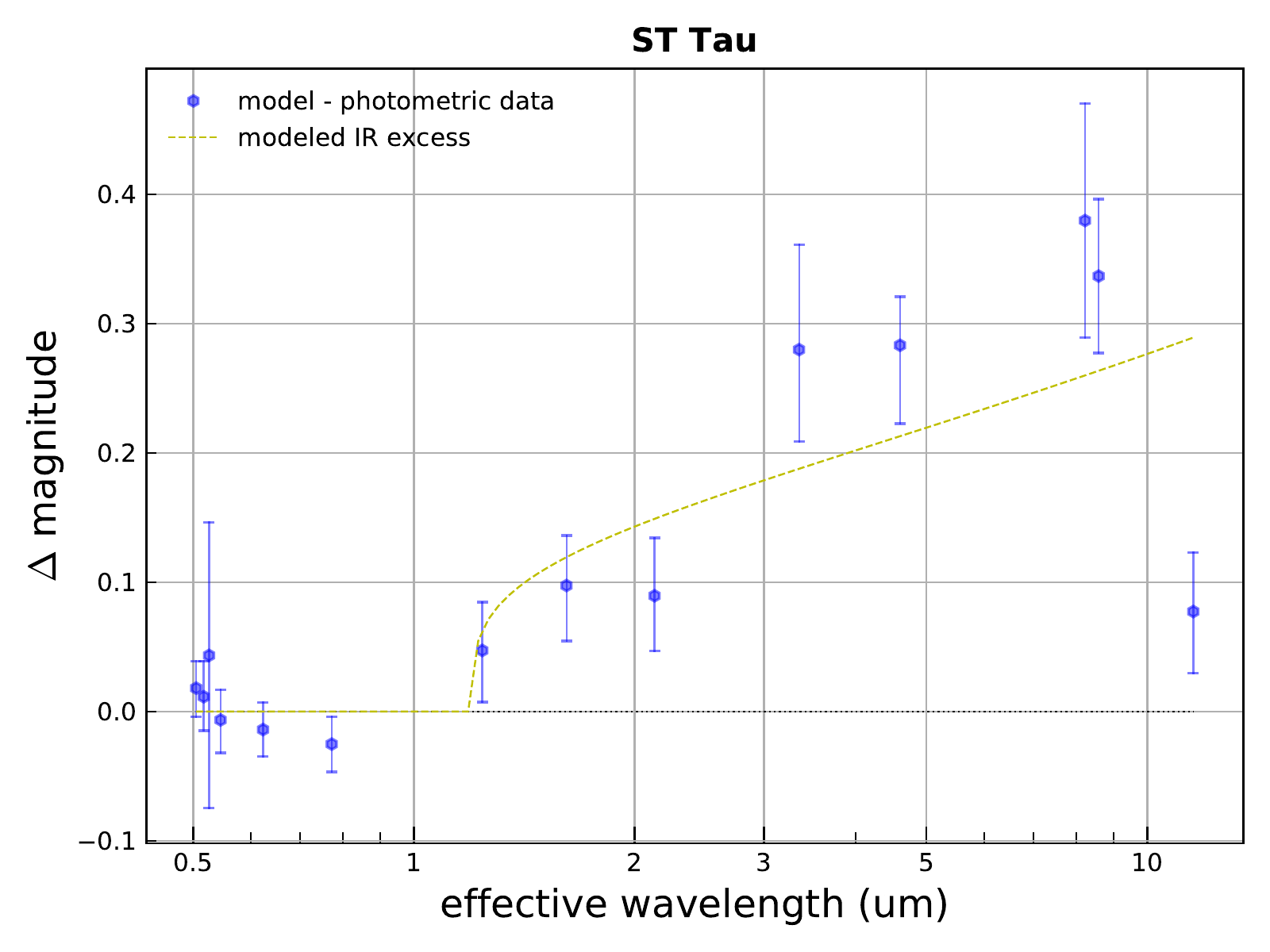}\includegraphics{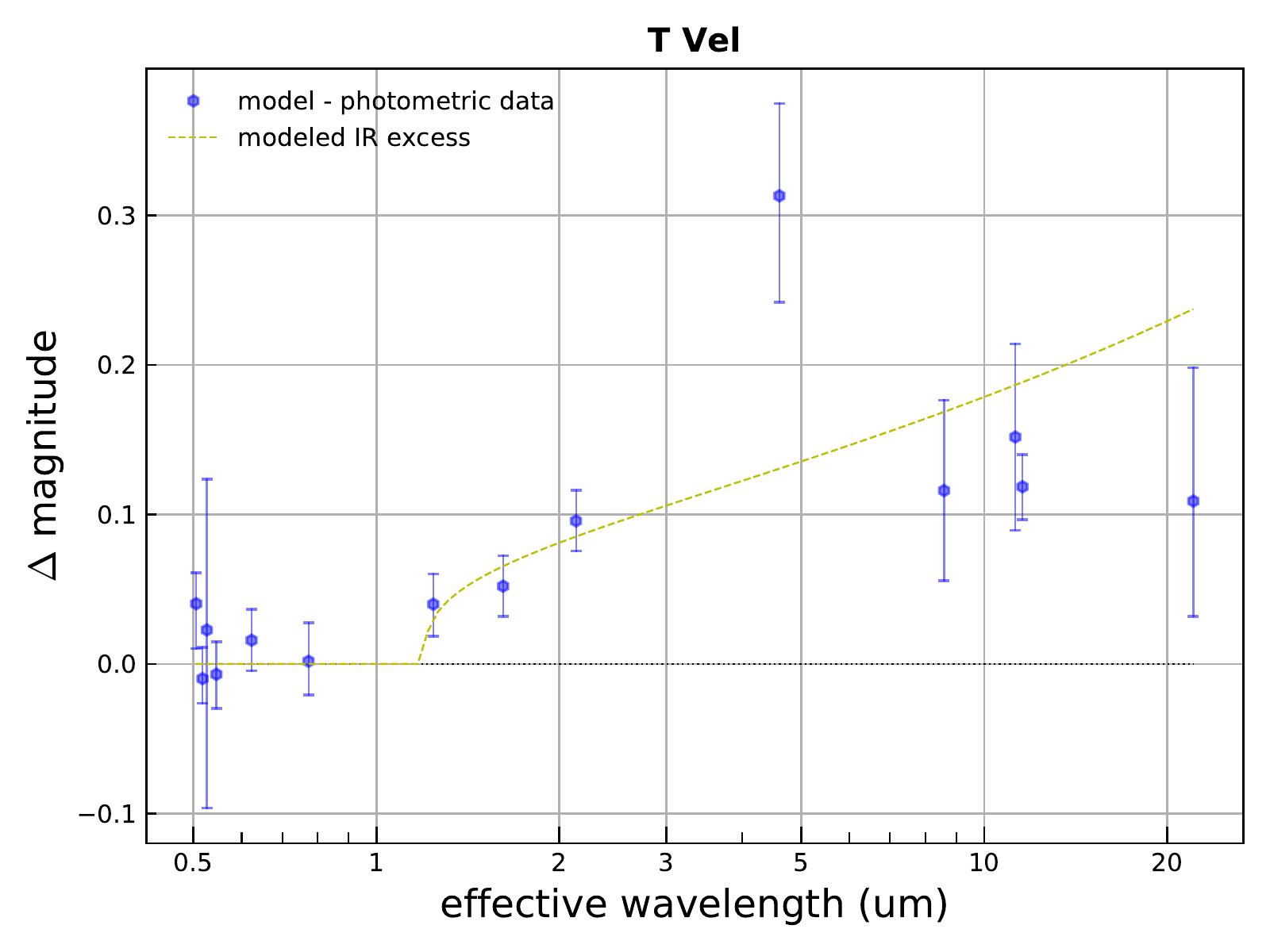}}
			\caption{SPIPS IR-excess model for the Cepheids CD~Cyg,  $\zeta$~Gem, S~Nor, V340~Nor, ST~Tau, and T~Vel, for which we detected a CSE at more than $3\sigma$.}
			\label{image__spips_excess2}
		\end{figure*}
		\begin{figure*}[!h]
		\resizebox{0.5\hsize}{!}{\includegraphics{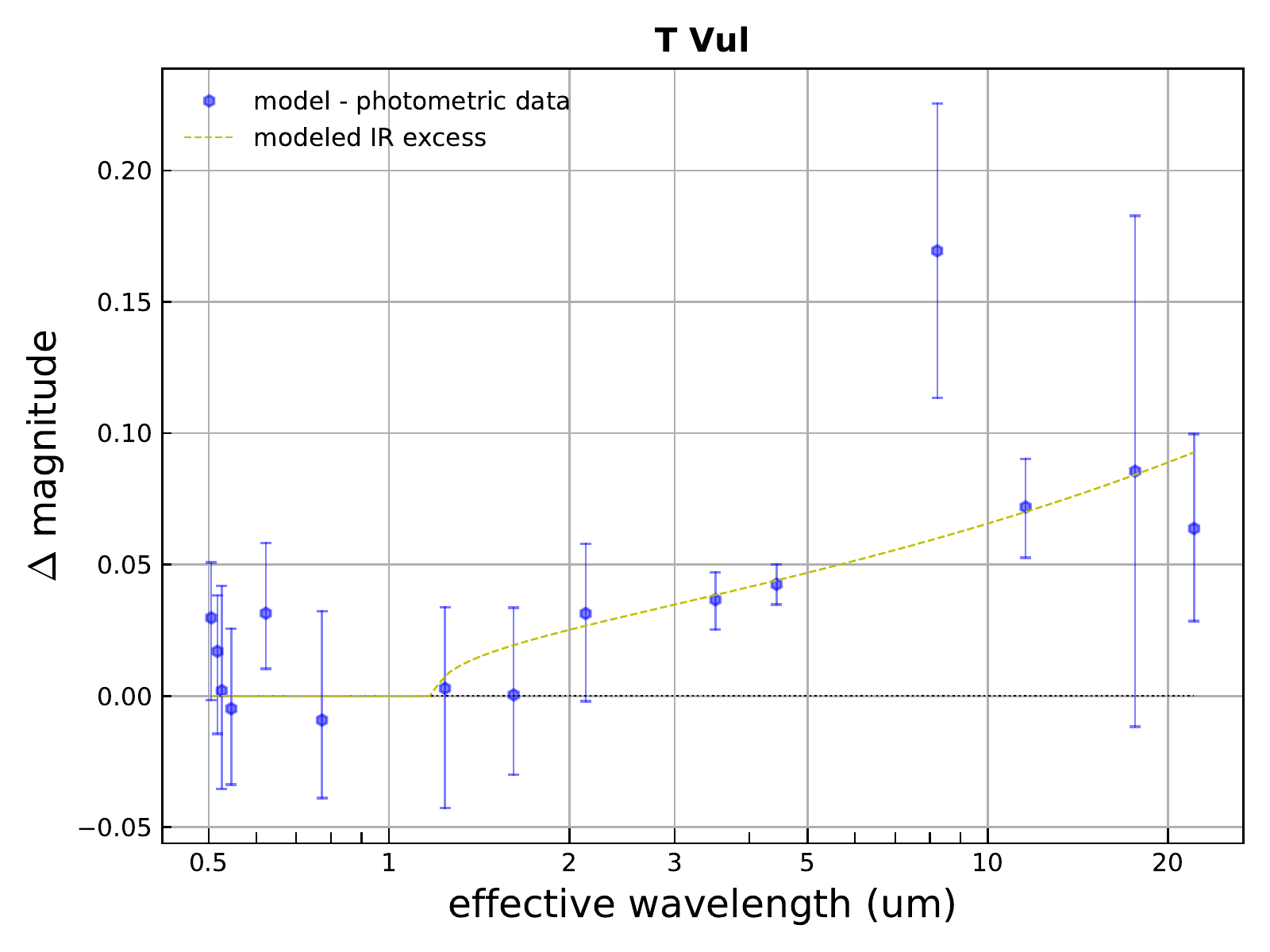}}
		\caption{SPIPS IR-excess model for the Cepheid T~Vul, for which we detected a CSE at more than $3\sigma$.}
		\label{image__spips_excess3}
		\end{figure*}

		\section{Bad SPIPS model for three Cepheids, without fitting a CSE and fixing the colour excess.}
		\label{appendix__spips_model_fixed_reddening}
		
		\begin{figure*}[!h]
			\resizebox{\hsize}{!}{\includegraphics{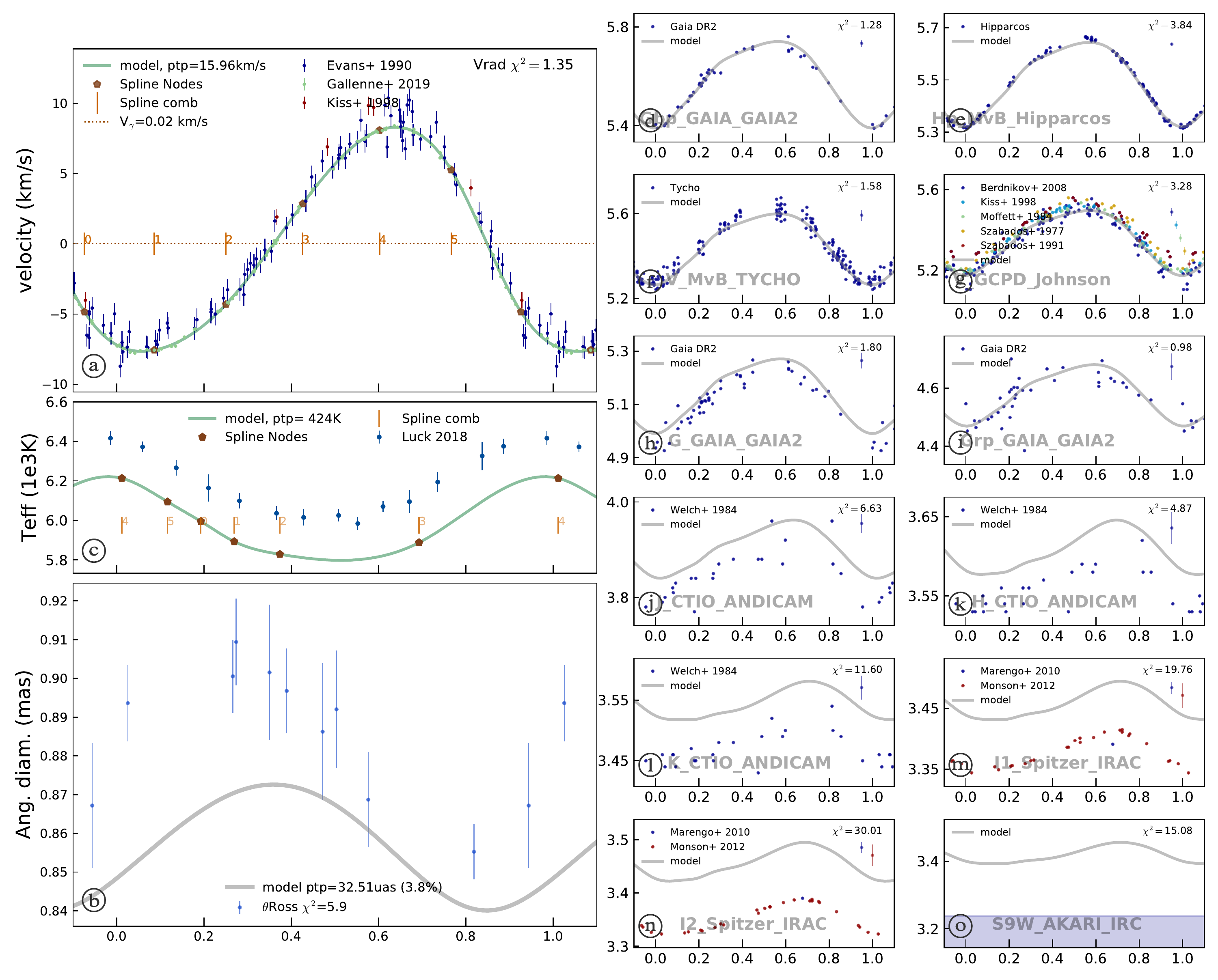}}
			\caption{Bad SPIPS model of FF~Aql without a CSE and with a fixed reddening.}
			\label{image__spips_ffaql_fixed_reddening}
		\end{figure*}
		
		\begin{figure*}[!h]
			\resizebox{\hsize}{!}{\includegraphics{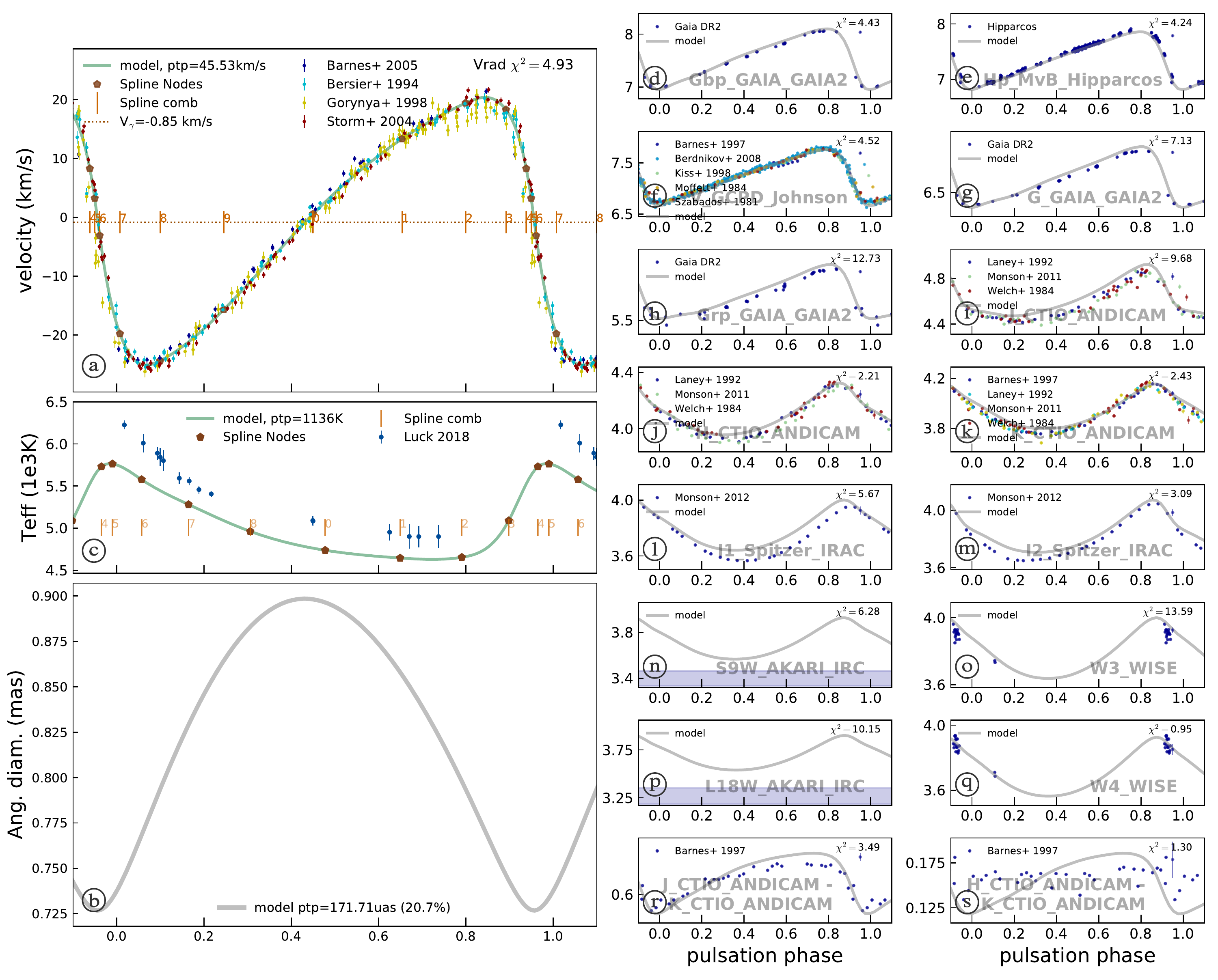}}
			\caption{Bad PIPS model of SV~Vul without a CSE and with a fixed reddening.}
			\label{image__spips_svvul_fixed_reddening}
		\end{figure*}
		
		\begin{figure*}[!h]
			\resizebox{\hsize}{!}{\includegraphics{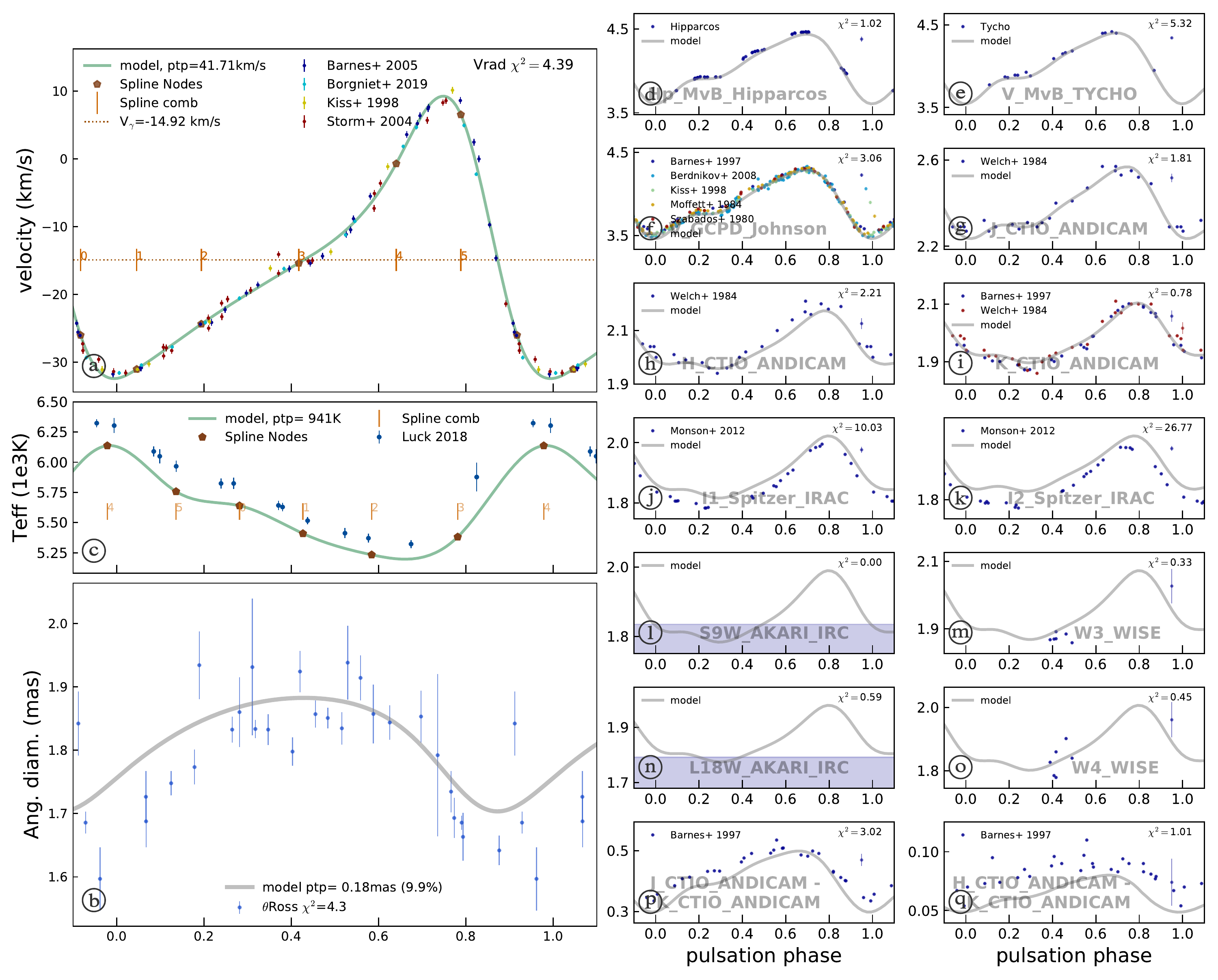}}
			\caption{Bad SPIPS model of $\eta$~Aql without a CSE and with a fixed reddening.}
			\label{image__spips_wsgr_fixed_reddening}
		\end{figure*}

		\section{Diagram temperature - IR excess and colour - IR.}
		\begin{figure*}[!h]
		\resizebox{\hsize}{!}{\includegraphics{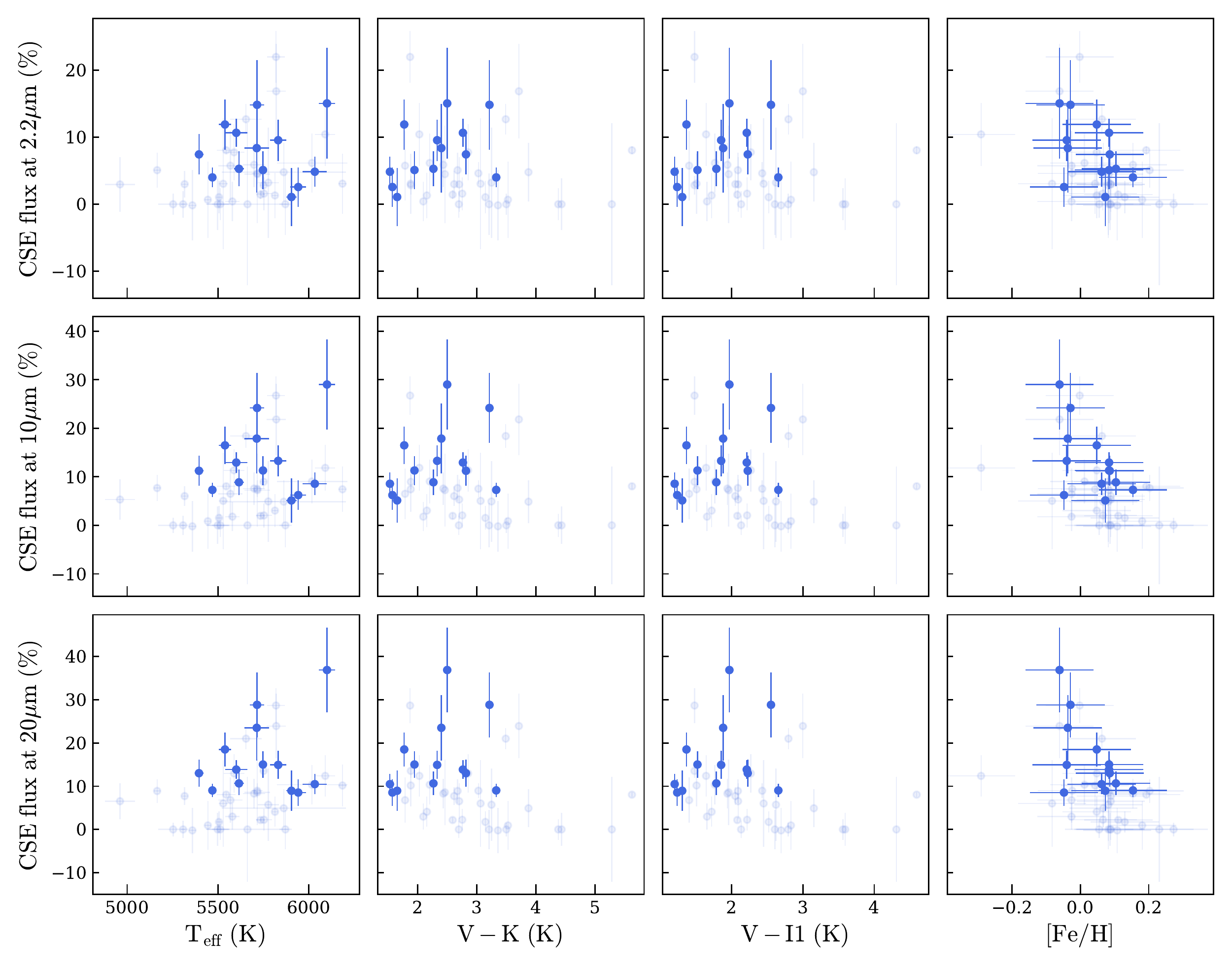}}
		\caption{Diagram $T_\mathrm{eff}$ - IR excess, ($V-K$) - IR excess, and ($V-I1$) - IR excess. The blue dots are the Cepheids for which the IR excess is detected at more than $3\sigma$, and shaded points are detections $<3\sigma$.}
		\label{image__ir_correllation}
		\end{figure*}

%

	\end{appendix}
	
\end{document}